\documentclass[twocolumn]{aastex631}

\newcounter{num}
\shorttitle{Subaru High-$z$ Exploration of Low-Luminosity Quasars (SHELLQs). \Roman{num}}
\shortauthors{Arita et al.}
\graphicspath{{./}{figures/}}

\begin{document}
\setcounter{num}{18}
\title{Subaru High-$z$ Exploration of Low-Luminosity Quasars (SHELLQs). \Roman{num}. The Dark Matter Halo Mass of Quasars at $z\sim6$}

\correspondingauthor{Junya Arita}
\email{jarita@astron.s.u-tokyo.ac.jp}

\author[0009-0007-0864-7094]{Junya Arita}
\affiliation{Department of Astronomy, School of Science, The University of Tokyo, 7-3-1 Hongo, Bunkyo-ku, Tokyo, 113-0033, Japan}

\author[0000-0003-3954-4219]{Nobunari Kashikawa}
\affiliation{Department of Astronomy, School of Science, The University of Tokyo, 7-3-1 Hongo, Bunkyo-ku, Tokyo, 113-0033, Japan}
\affiliation{Research Center for the Early Universe, The University of Tokyo, 7-3-1 Hongo, Bunkyo-ku, Tokyo, 113-0033, Japan}

\author[0000-0001-5063-0340]{Yoshiki Matsuoka}
\affiliation{Research Center for Space and Cosmic Evolution, Ehime University, 2-5 Bunkyo-cho, Matsuyama, Ehime 790-8577, Japan}

\author[0000-0001-7759-6410]{Wanqiu He}
\affil{National Astronomical Observatory of Japan, 2-21-1 Osawa, Mitaka, Tokyo 181-8588, Japan}

\author[0000-0002-9453-0381]{Kei Ito}
\affiliation{Department of Astronomy, School of Science, The University of Tokyo, 7-3-1 Hongo, Bunkyo-ku, Tokyo, 113-0033, Japan}

\author[0000-0002-2725-302X]{Yongming Liang}
\affiliation{Institute for Cosmic Ray Research, The University of Tokyo, 5-1-5 Kashiwanoha, Kashiwa, Chiba 277-8582, Japan}

\author[0000-0002-2134-2902]{Rikako Ishimoto}
\affiliation{Department of Astronomy, School of Science, The University of Tokyo, 7-3-1 Hongo, Bunkyo-ku, Tokyo, 113-0033, Japan}

\author[0000-0002-3800-0554]{Takehiro Yoshioka}
\affiliation{Department of Astronomy, School of Science, The University of Tokyo, 7-3-1 Hongo, Bunkyo-ku, Tokyo, 113-0033, Japan}

\author[0000-0001-7154-3756]{Yoshihiro Takeda}
\affiliation{Department of Astronomy, School of Science, The University of Tokyo, 7-3-1 Hongo, Bunkyo-ku, Tokyo, 113-0033, Japan}

\author[0000-0002-4923-3281]{Kazushi Iwasawa}
\affiliation{Institut de Ci\`{e}ncies del Cosmos (ICCUB), Universitat de Barcelona (IEEC-UB), Mart\'{i} i Franqu\`{e}s, 1, E-08028 Barcelona, Spain}
\affiliation{ICREA, Pg. Llu\'{i}s Companys 23, E-08010 Barcelona, Spain}

\author[0000-0003-2984-6803]{Masafusa Onoue}
\affiliation{Kavli Institute for Astronomy and Astrophysics, Peking University, Beijing 100871, China}
\affiliation{Kavli Institute for the Physics and Mathematics of the Universe (Kavli IPMU, WPI), The University of Tokyo, Chiba 277-8583, Japan}

\author[0000-0002-3531-7863]{Yoshiki Toba}
\affiliation{Research Center for Space and Cosmic Evolution, Ehime University, 2-5 Bunkyo-cho, Matsuyama, Ehime 790-8577, Japan}
\affiliation{National Astronomical Observatory of Japan, 2-21-1 Osawa, Mitaka, Tokyo 181-8588, Japan}
\affiliation{Academia Sinica Institute of Astronomy and Astrophysics, 11F of Astronomy-Mathematics Building, AS/NTU, No.1, Section 4, Roosevelt Road, Taipei 10617, Taiwan}

\author[0000-0001-6186-8792]{Masatoshi Imanishi}
\affiliation{National Astronomical Observatory of Japan, 2-21-1 Osawa, Mitaka, Tokyo 181-8588, Japan}

\begin{abstract}
We present, for the first time, dark matter halo (DMH) mass measurement of quasars at $z\sim6$ based on a clustering analysis of 107 quasars.
Spectroscopically identified quasars are homogeneously extracted from the HSC-SSP wide layer over $891\,\mathrm{deg^2}$.
We evaluate the clustering strength by three different auto-correlation functions: projected correlation function, angular correlation function, and redshift-space correlation function.
The DMH mass of quasars at $z\sim6$ is evaluated as $5.0_{-4.0}^{+7.4}\times10^{12}\,h^{-1}M_\odot$ with the bias parameter $b=20.8\pm8.7$ by the projected correlation function. 
The other two estimators agree with these values, though each uncertainty is large.
The DMH mass of quasars is found to be nearly constant $\sim10^{12.5}\,h^{-1}M_\odot$ throughout cosmic time, suggesting that there is a characteristic DMH mass where quasars are always activated.
As a result, quasars appear in the most massive halos at $z \sim 6$, but in less extreme halos thereafter.
The DMH mass does not appear to exceed the upper limit of $10^{13}\,h^{-1}M_\odot$, which suggests that most quasars reside in DMHs with $M_\mathrm{halo}<10^{13}\,h^{-1}M_\odot$ across most of the cosmic time.
Our results supporting a significant increasing bias with redshift are consistent with the bias evolution model with inefficient AGN feedback at $z\sim6$.
The duty cycle ($f_\mathrm{duty}$) is estimated as $0.019\pm0.008$ by assuming that DMHs in some mass interval can host a quasar.
The average stellar mass is evaluated from stellar-to-halo mass ratio as $M_*=6.5_{-5.2}^{+9.6}\times10^{10}\,h^{-1}M_\odot$, which is found to be consistent with [C \textsc{II}] observational results.
\end{abstract}


\section{Introduction} \label{sec:intro}

According to the current $\mathrm{\Lambda}$CDM theory, the tiny density fluctuation of dark matter in the early universe grows and subsequently collapses into dark matter halos (DMHs).
These halos continuously accrete and hierarchically merge to form high-mass DMHs.
Galaxies are nurtured in the center of DMHs and almost all the galaxies harbor a supermassive black hole (SMBH) in their centers \citep{Kormendy1995}.
Quasars are believed to be powered by gas accretion onto SMBHs \citep{Salpeter1964,Lynden-Bell1969} and outshine in the multiple wavelengths.
Since quasars are one of the most luminous objects in the universe, they are observable even at $z\gtrsim7$ (e.g., \citealp{Mortlock2011,Banados2018,Matsuoka2019,Yang2020,Koptelova2022}).
Quasars are important objects to study the open questions in the early universe; however, it remains unclear how high-$z$ quasars are physically related to the underlying DMHs they inhabit.


One of the important questions is when and how the co-evolution between galaxies and SMBHs manifested, i.e., the masses of which are correlated with those of their host galaxies.
While this relationship in the local universe is well established \citep{Magorrian1998,Tremaine2002,Kormendy2013}, it remains to be elucidated in the early universe.
The parent DMH, which governs both the SMBH and the galaxy, holds the key to unveiling the underlying physical mechanism of their relationship.
The gas accumulated by the gravitational potential of DMHs is consumed to form stars, thus, the relationship between stellar mass and DMH mass is quite natural \citep{White1978}.
It is believed that the gas further loses the angular momentum due to the radiation from active star formation in the DMH and flows into the central SMBH \citep{Laura2002, Granato2004, Shankar2006} to grow more massive.
Otherwise, the steady high density cold gas flow directly from the halo could be responsible for sustaining critical accretion rates leading to rapid growth of $\sim10^9\,M_\odot$ black holes as early as $z\sim7$ (e.g., \citealp{Dimatteo2012}).
Therefore, the mass of a galaxy DMH hosting a SMBH is crucial for understanding their co-evolutionary growth.


The DMH mass is also a key physical quantity to understand the AGN feedback, which is thought to play a significant role in regulating the star formation of the host galaxies \citep{Kormendy2013,Heckman2014}, because it can constrain the duty cycle, the fraction of DMHs that host active quasars (e.g., \citealp{Cole1989,Haiman2001,Martini2001}).
\citet{Hopkins2007} showed that feedback efficiency will greatly change DMH mass evolution at high-$z$.
According to their model, feedback prevents gas accretion against gravity by radiation pressure, and works to stop SMBH growth and eventually defuses the quasar phase.
Thus, if feedback is inefficient to stop the SMBH growth at high-$z$, quasars will live in the highest-mass DMHs.
Since the quasar activity has a huge impact on the host galaxy, unveiling the feedback efficiency helps to advance our understanding of the co-evolution.


The clustering analysis is an effective method to estimate DMH mass.
It quantifies the distribution of objects often through a two-point correlation function.
The two-point correlation function $\xi(r)$ is defined based on the probability $dP$ that an object is observed in the volume element $dV$ apart from the separation $r$ from a given object \citep{Totsuji1969,Peebles1980};
\begin{equation}
    dP = \bar{n}[1+\xi(r)]dV,
\end{equation}
where $\bar{n}$ is the mean number density of the objects.
Quasar host galaxies are believed to reside in the peak of the underlying dark matter density distribution \citep{Dekel1999}.
Using the bias (linear bias) parameter $b$, the relation between two-point correlation functions of quasars $\xi_\mathrm{Q}(r)$ and that of dark matter $\xi_{\mathrm{DM}}(r)$ can be expressed as
\begin{equation}
    \xi_\mathrm{Q}(r) = b^2 \xi_{\mathrm{DM}}(r).
    \label{eq:biasdefinition}
\end{equation}
The bias parameter has been modeled theoretically (e.g., \citealp{Jing1998,SMT2001,Seljak2004,Mandelbaum2005,Tinker_2010}), which gives insight into the DMH mass.


The previous initiative works \citep{Porciani2004,Croom2005,Myers2007} have proved that the quasar bias increases with redshift from today to $z=2\--3$.
However, clustering analyses of quasars at $z>3$, there are a few attempts.
\citet{Shen2007} utilized 4426 spectroscopically identified luminous quasars with $2.9\leq z\leq 5.4$ from the Fifth Data Release of the Sloan Digital Sky Survey (SDSS DR5; \citealt{Schneider2007}) and concluded that quasars typically reside in DMHs with $(2\-- 3)\times 10^{12}\ h^{-1}M_\odot$ at $2.9\leq z\leq 3.5$ and $(4\-- 6)\times 10^{12}\ h^{-1}M_\odot$ at $3.5\leq z\leq 5.4$.
\citet{Eftekharzadeh_2015} measured the clustering signal of quasars from the Baryon Oscillation Spectroscopic Survey (BOSS; \citealt{Smee2013}).
They estimated the DMH mass for quasars at $2.2\leq z\leq 3.4$ as $0.6\--3\times 10^{12}\ h^{-1}M_\odot$.
\citet{He2018} extracted photometrically selected 901 quasars with $\bar{z}_{\mathrm{phot}}\sim3.8$ from the early data release of the Subaru Hyper Suprime-Cam Strategic Survey Program (HSC-SSP; \citealt{Aihara2018a,Aihara2018b}).
They added 342 SDSS quasars \citep{Alam2015} to their sample and evaluated cross-correlation functions (CCF) between the quasars and bright Lyman Break Galaxies (LBGs) from the HSC-SSP.
The typical DMH mass derived from the CCF signal is $1\--2\times 10^{12}\ h^{-1}M_\odot$. 
\citet{Timlin_2018} measured the clustering signal of photometrically selected quasars with $2.9\leq z_{\mathrm{phot}}\leq 5.1$ from SDSS Stripe 82 field \citep{Annis2014,Jiang2014} and presumed that characteristic DMH mass is $1.70\--9.83\times10^{12}\,h^{-1}M_\odot$.
These studies detected significantly large clustering signals, implying that the quasar halo bias rapidly increases beyond $z\sim3$.
In addition, the quasar DMH mass remains approximately constant at $M_{\mathrm{halo}}\sim 10^{12.5}\,h^{-1}M_\odot$ from the present day to $z\sim4$, which will be intriguing to see whether these trends continue to higher-$z$.


Despite intense observational efforts, the clustering measurements have been challenging beyond $z>4$.
This is because clustering analysis requires a quasar sample with sufficient number density, which remarkably decreases towards $z\sim6$ \citep{Fan2022}.
The sample size and the number density of quasars at $z\sim6$ have increased dramatically in the last two decades but the observable quasar population is limited to high-luminosity obtained by ultra wide-field surveys, hindering the increase in their number density.
Increasing the number density of quasars at $z\sim6$ has been a major challenge because of the need for wide and deep observations and expensive spectroscopic observations for fainter quasars.
Hyper Suprime-Cam (HSC; \citealp{Miyazaki_2018,Komiyama_2018,Kawanomoto_2018,Furusawa_2018}) on the Subaru Telescope, which has a large field of view and high sensitivity, has changed the situation.
Utilizing the powerful instrument, wide-field imaging survey program, HSC-SSP, was performed.
From the survey data, Subaru High-$z$ Exploration of Low-Luminosity Quasars (SHELLQs; \citealp{Matsuoka_2016,Matsuoka_2022}) has discovered 162 quasars at $5.66<z<7.07$ over $1200\,\mathrm{deg^2}$, providing high number density to allow for clustering measurements.


In this paper, we, for the first time, present the clustering analysis of quasars at $z\sim6$ by using the SHELLQs sample.
We show the samples for our analysis in Section \ref{sec:data}.
We explain the details of clustering analysis in Section \ref{sec:clustering}.
In Section \ref{sec:discussion}, we derive the important physical quantities from the result in previous section.
Finally, we summarize our results in Section \ref{sec:summary}.
We adopt flat $\Lambda$CDM cosmology with cosmology parameters $(h,\Omega_{m},\Omega_{\Lambda},\sigma_8)=(0.7,0.3,0.7,0.81)$, namely $H_0=70\,\mathrm{km\,s^{-1}\,Mpc^{-1}}$.
All magnitudes in this paper are presented in the AB system \citep{Oke1983}.

\begin{figure}[t]
    \centering
    \includegraphics[width=\columnwidth]{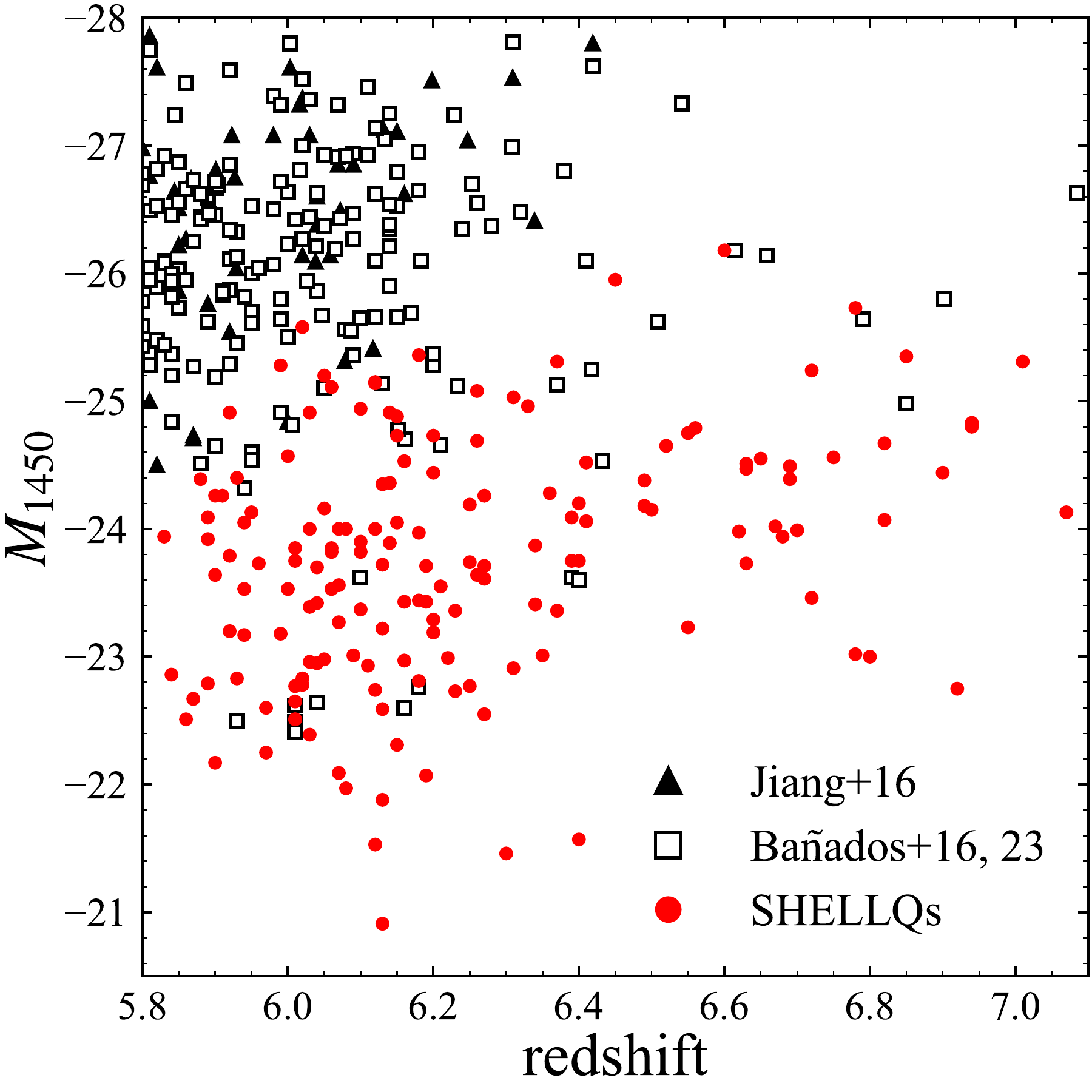}
    \caption{Comparison of $M_{1450}$, the absolute magnitudes at $1450\,\mathrm{\AA}$ of SHELLQs quasars (red circles) and those of known quasars.
    We refer to the $M_{1450}$ from the quasar sample in \citet{Jiang2016} (black triangle) and that of newly identified quasars in \citet{Banados_2016,Banados2023} (black open square).
    }
    \label{fig:comp_absmag}
\end{figure}

\section{Data} \label{sec:data}

\subsection{SHELLQs} \label{subsec:shellqs}
Our main quasar sample is from SHELLQs utilizing HSC-SSP data.
The HSC data are reduced with HSC pipeline, \texttt{hscpipe} \citep{Bosch2018}, which is based on the Large Synoptic Survey Telescope (LSST) pipeline \citep{Juric2017,Bosch2018,Ivezi2019}.
The astrometry and photometric calibration are performed based on the data from Panoramic Survey Telescope and Rapid Response System Data Release 1 (Pan-STARRS1; \citealp{Schlafly2012,Tonry2012,Magnier2013,Chambers2016}).
The SHELLQs quasars are a flux-limited ($m_z<24.5$ for $z\sim 6$ and $m_y<24$ for $z\sim 7$) sample of quasars at $z\sim6\--7$.
These quasars are selected from point sources and by a Bayesian-based probabilistic algorithm, which is applied to the optical HSC-SSP source catalogs.
More details of the sample construction are described in \citet{Matsuoka_2016}.
The spectroscopic observation is performed by utilizing the Faint Object Camera and Spectrograph (FOCAS; \citealp{Kashikawa2002}) on the Subaru Telescope and Optical System for Imaging and low-intermediate-Resolution Integrated Spectroscopy (OSIRIS; \citealp{Cepa2000}) on the Gran Telescopio Canarias.

The advantages of the SHELLQs sample are faintness and high number density thanks to the depth of the HSC-SSP.
Figure \ref{fig:comp_absmag} shows the comparison of the absolute magnitude of quasars detected in SHELLQs, SDSS \citep{Jiang2016}, and Pan-STARRS1 \citep{Banados_2016,Banados2023}, where it is clear that SHELLQs is exploring a unique regime fainter than other surveys.
The SHELLQs have a number density ($0.14\,\mathrm{deg^{-2}}$) of quasars $\sim30$ times more than the SDSS \citep{Jiang2016}, where 52 quasars are detected in $11240\,\mathrm{deg^2}$ at $5.7<z\leq6.4$. 

The original SHELLQs sample consists of 162 spectroscopically confirmed quasars.
We impose the following four criteria to ensure homogeneity, which yields 93 quasars (see Table \ref{tab:quasar_selection}).

\begin{deluxetable}{lc}
\tablehead{\colhead{Requirement}&\colhead{Number}}
\tablecaption{Detail of the sample selection}
\startdata
All quasars identified by SHELLQs & 162\\
1. $z\leq 6.5$ & 132\\
2. Identified in HSC-SSP S20A & 125\\
3. Far from bright star masks and edge regions & 116\\
4. Broad line quasars & 93
\enddata
\tablecomments{We add 14 known quasars into this sample. The additional quasars are listed in Table \ref{tab:bright_quasars}.}
\end{deluxetable}
\label{tab:quasar_selection}

\begin{enumerate}
    \item $z\leq 6.5$
    
    The SHELLQs sample consists of $z\sim6$ quasars selected by $i$-dropouts and $z\sim7$ quasars selected by $z$-dropouts.
    Since the latter has a small sample size and the survey areas of the two do not perfectly match, only the former $z\sim6$ quasars are used in this study to ensure uniformity of the sample.
    The $z\sim6$ quasar sample selection criteria \citep{Matsuoka_2022} is
    \begin{eqnarray}
    m_z<24.5\ \&\ \sigma_z<0.155\ \&\ m_i-m_z>1.5\nonumber\\
    \&\ 0.7<\mu/\mu_\mathrm{PSF}<1.2,
    \end{eqnarray}
    where $\mu$ is the adaptive moment of the source averaged over the two image dimensions and $\mu_\mathrm{PSF}$ is that of the point spread function (PSF) model.
    
    \item Identified in HSC-SSP S20A region
    
    The SHELLQs sample is still growing.
    Optical spectroscopic follow-up observations have been completely executed in the S20A survey area.
    This study uses only SHELLQs quasars spectroscopically confirmed in the S20A region, and quasars added after S21A are removed to account for the uniformity of the sample.

    \item Far from bright star masks and edge regions
    
    We remove quasars in areas with poor data quality, such as near the bright star masks and edges by random points covering HSC-SSP S20A region to preserve sample homogeneity
    \footnote{Specifically, we use the following flags to retrieve random points covering the survey field with a surface number density of $100\,\mathrm{arcmin^{-2}}$: \texttt{\{i,z,y\}\_pixelflags\_edge}, \texttt{\{i,z,y\}\_pixelflags\_saturatedcenter}, \texttt{\{i,z,y\}\_pixelflags\_crcenter}, \texttt{\{i,z,y\}\_pixelflags\_bad}, \texttt{\{i,z,y\}\_mask\_brightstar\_\{halo, ghost, blooming\}}, and impose $\texttt{z\_inputcount\_value}\geq2$.
    We exclude quasars that have no random points within $0.12\,\mathrm{arcmin}$ from the clustering analysis.}.
    
    \item Broad line quasars
    
    According to the unified AGN model \citep{Antonucci1993}, type-I AGNs and type-II AGNs are the same population and the difference purely originates from the inclination angle to observers.
    However, another evolutionary scenario interprets the difference between the two populations in host galaxies.
    The DMH mass measurements by \citet{Hickox2011} found the differences in the DMH mass between obscured and unobscured quasars through clustering analysis.
    \citet{Onoue2021} reported that approximately 20\% of SHELLQs quasars have narrow Ly$\alpha$ emission lines ($\mathrm{FWHM_{Ly\alpha}}<500\,\mathrm{km\,s^{-1}}$) and one of them can be a type-II quasar based on the spectroscopic follow-up.
    Therefore, as long as it is unclear which interpretation is correct, we decide to be conservative in the study to exclude quasars with narrow Ly$\alpha$ emission lines from the sample.
\end{enumerate}

\subsection{Other quasars}

\begin{deluxetable*}{lcccclc}
\tablehead{\colhead{Name}&\colhead{R.A.}&\colhead{Decl.}&\colhead{Redshift}&\colhead{$M_{1450}$ }&\colhead{Survey}&\colhead{Reference}\\
\colhead{}&\colhead{(J2000)}&\colhead{(J2000)}&\colhead{}&\colhead{(mag)}&\colhead{}&\colhead{}}
\tablecaption{Additional quasar sample}
\startdata
SDSS J160254.18$+$422822.9&16:02:54.18&$+$42:28:22.9&6.07&$-26.82$&SDSS &(1)\\
SDSS J000552.33$-$000655.6 &00:05:52.33&$-$00:06:55.6 &5.855&$-26.46$ &SDSS  &(1) \\
CFHQS J021013$-$045620\tablenotemark{a}& 02:10:13.19&$-$04:56:20.9 & 6.44 &$-24.28$  &CFHQS\tablenotemark{b} & (2)\\
CFHQS J021627$-$045534\tablenotemark{a}& 02:16:27.81&$-$04:55:34.1 &6.01&$-22.21$ & CFHQS &(3)\\
CFHQS J022743$-$060530\tablenotemark{a} & 02:27:43.29&$-$06:05:30.2 & 6.20&$-25.03$ & CFHQS & (3)\\
IMS J220417.92$+$011144.8\tablenotemark{a} & 22:04:17.92&$+$01:11:44.8 & 5.944 &$-23.59$ & IMS\tablenotemark{c} &  (4)\\
VIMOS2911001793\tablenotemark{a}&22:19:17.22&$+$01:02:48.9 &6.156&$-23.10$ & Suprime Cam &(5)\\
SDSS J222843.5$+$011032.2\tablenotemark{a} &22:28:43.54&$+$01:10:32.2 &5.95&$-24.53$ & SDSS Stripe82  &(6)\\
SDSS J230735.35$+$003149.4 \tablenotemark{a}&23:07:35.35&$+$00:31:49.4 &5.87&$-24.93$ & SDSS  &(7)\\
SDSS J231546.57$-$002358.1 \tablenotemark{a}& 23:15:46.57&$-$00:23:58.1 &6.117&$-25.38$ & SDSS  &(8)\\
PSO J183.1124$+$05.0926&12:12:26.98&$+$05:05:33.4 &6.439&$-26.99$ & Pan-STARRS1 &(9)\\
VIK J121516.88$+$002324.7 \tablenotemark{a}&12:15:16.88&$+$00:23:24.7 &5.93&$-24.67$ & VIKING\tablenotemark{e} &(10)\\
PSO J184.3389$+$01.5284 \tablenotemark{a}&12:17:21.34&$+$01:31:42.2 &6.20&$-25.37$ & Pan-STARRS1 &(11)\\
PSO J187.3050$+$04.3243  & 12:29:13.21&$+$04:19:27.7 & 5.89 & $-25.4$ & Pan-STARRS1 &(12)
\enddata
\tablenotetext{a}{The quasar is recovered by SHELLQs project \citep{SHELLQs4}.}
\tablenotetext{b}{Canada-France High-$z$ Quasar Survey}
\tablenotetext{c}{Infrared Medium-deep Survey}
\tablenotetext{e}{VISTA Kilo-degree Infrared Galaxy Public Survey}
\tablerefs{(1) \citet{Fan2004}, (2) \citet{Willott_2010}, (3) \citet{Willott_2009}, (4) \citet{Kim_2015}, (5) \citet{Kashikawa2015}, (6) \citet{Zeimann_2011}, (7) \citet{Jiang2009}, (8) \citet{Jiang2008}, (9) \citet{Mazzucchelli_2017}, (10) \citet{Venemans2015}, (11) \citet{Banados_2016}, (12) \citet{Banados2014}.}
\end{deluxetable*}
\label{tab:bright_quasars}

\begin{figure}[ht!]
    \includegraphics[width=\columnwidth,clip]{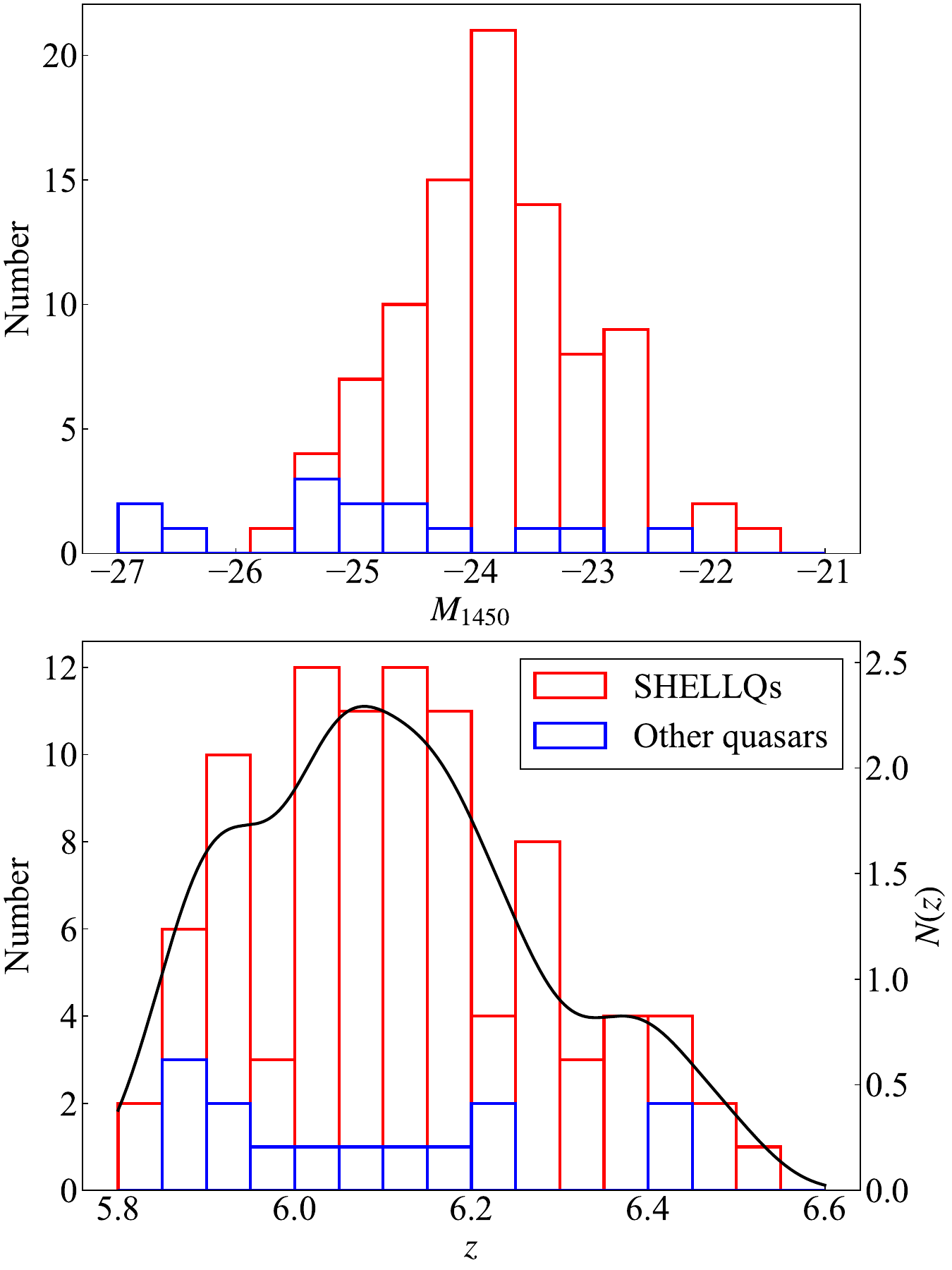}
    \caption{Top: red and blue histograms show 1450 \AA\ absolute magnitude distribution for SHELLQs quasars and other quasars respectively. Bottom: red and blue histograms show the redshift distribution for SHELLQs quasars and other quasars respectively.
    The black line represents $N(z)$, the redshift distribution of SHELLQs quasars estimated by kernel density estimation, which is used in Section \ref{subsec:acf}.}
    \label{fig:mag&z_hist}
\end{figure}

We also add to our sample 14 quasars at $z\sim6$ that were discovered by other surveys (see Table \ref{tab:bright_quasars}).
We select these quasars from the survey, whose area fully covers HSC-SSP S20A field.
Most of them are identified by SDSS (e.g., \citealp{Fan2004}) and Pan-STARRS1 (e.g., \citealp{Banados2014,Banados_2016,Mazzucchelli_2017}), which tend to be brighter than the SHELLQs quasars.
These quasars also satisfy the requirements imposed on SHELLQs quasars.
Ten of the 14 quasars are also detected in the SHELLQs observation but not included in the SHELLQs sample because they had already been found \citep{SHELLQs4}.
We visually inspect the spectra of all these quasars to confirm that they are actually $z\sim6$ quasars.
We assume that clustering strength is independent of quasar brightness, which is confirmed at low-$z$ \citep{Croom2005,Adelberger2006,Myers_2006,Shen2009}.
In fact, when we divide the sample into bright sub-sample ($M_{1450}\leq-24$) and faint sub-sample ($M_{1450}>-24$), the results obtained in Section \ref{sec:clustering} are consistent with each other within their errors.
In summary, our final sample consists of 107 quasars.
The distributions of absolute magnitude and redshift are shown in Figure \ref{fig:mag&z_hist}.

\subsection{Homogeneity of the sample}\label{subsec:homogeneity}
\begin{figure*}
    \centering
    \includegraphics[width=2.2\columnwidth,clip]{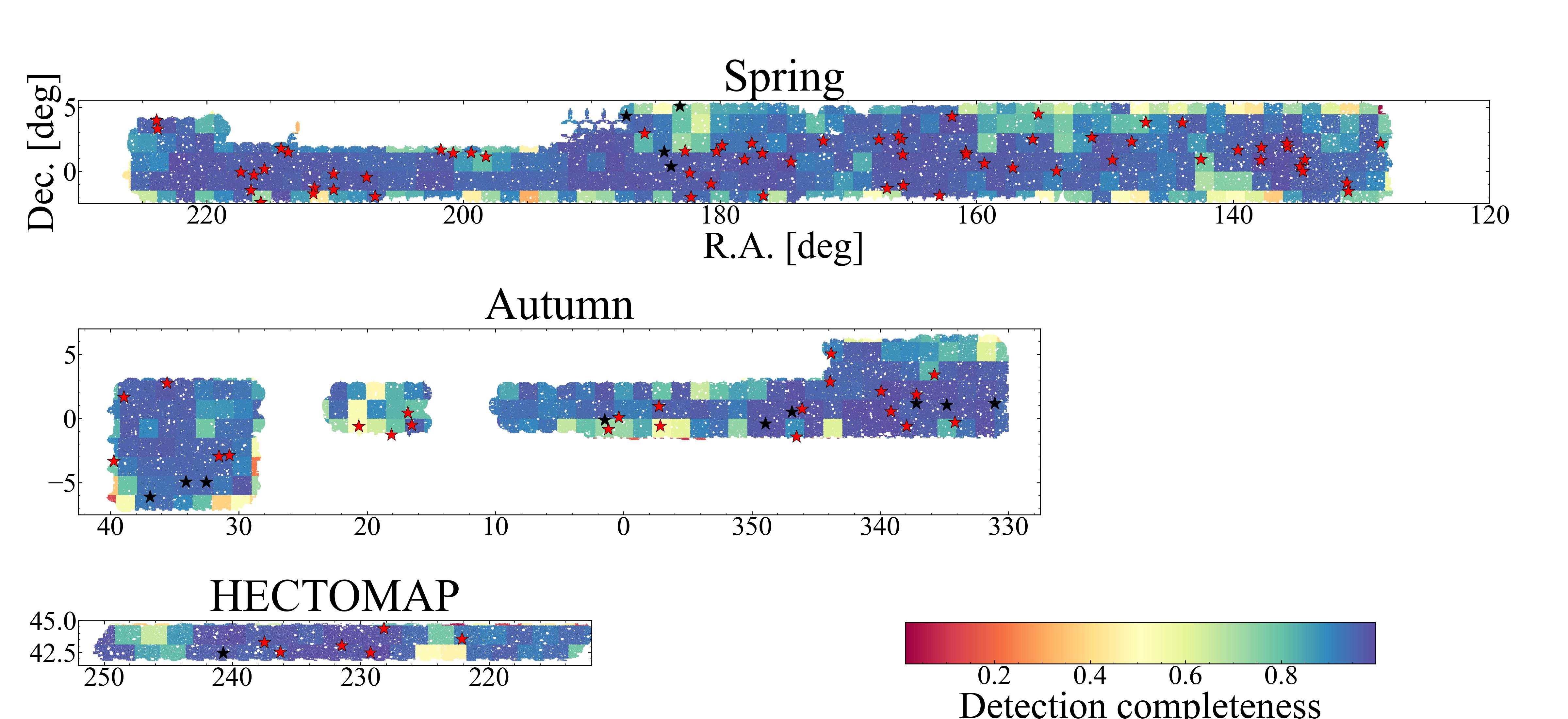}
    \caption{Detection completeness map of the HSC-SSP S20A region.
    The color represents the detection completeness at $m_z=24.5$ of each tract.
    The red and black stars represent the SHELLQs and other quasars, respectively.
    }
    \label{fig:completenessmap}
\end{figure*}

Our sample is distributed over $891\ \mathrm{deg^2}$ of three fields: HECTOMAP, Autumn and Spring.
The sky distribution of our sample quasars is shown in Figure \ref{fig:completenessmap}.
Sample homogeneity is of utmost importance for clustering analysis.
Since the spectroscopy is completely executed for the candidates in S20A, the spatial homogeneity of the photometric data in selecting candidate objects should be examined.
We verify the homogeneity by calculating detection completeness over the survey region.
The detection completeness is defined as the ratio of the number of quasars recovered by \texttt{hscpipe} ver. 8.4\footnote{\url{https://hsc.mtk.nao.ac.jp/pipedoc/pipedoc_8_e/index.html}} to the number of mock quasars scattered at random points on the HSC image in the same manner in \citet{SHELLQs5}.
The PSF of the input mock quasars is generated to be the same as that measured at each image position.
The PSF is modeled by \texttt{PSFEx} \citep{Bertin2011}, which can extract precise PSF model from images processed by \texttt{SExtractor} \citep{Bertin1996}. 
A small region (patch) of $12'\times 12'$ is randomly selected per tract, which consists of 81 patches, over the HSC-SSP region.
We embed more than 3000 mock quasars in the survey field per patch with $m_z=21\--28$ which are randomly spread over the HSC-SSP coadded $z$-band images using \texttt{Balrog} \citep{Suchyta2016}.
We perform photometry on the coadded $z$-band images embedded mock quasars utilizing \texttt{hscpipe} and detect the mock quasars.
The detection completeness is estimated on 662 patches in total.

\begin{figure}
    \centering
    \includegraphics[width=\columnwidth]{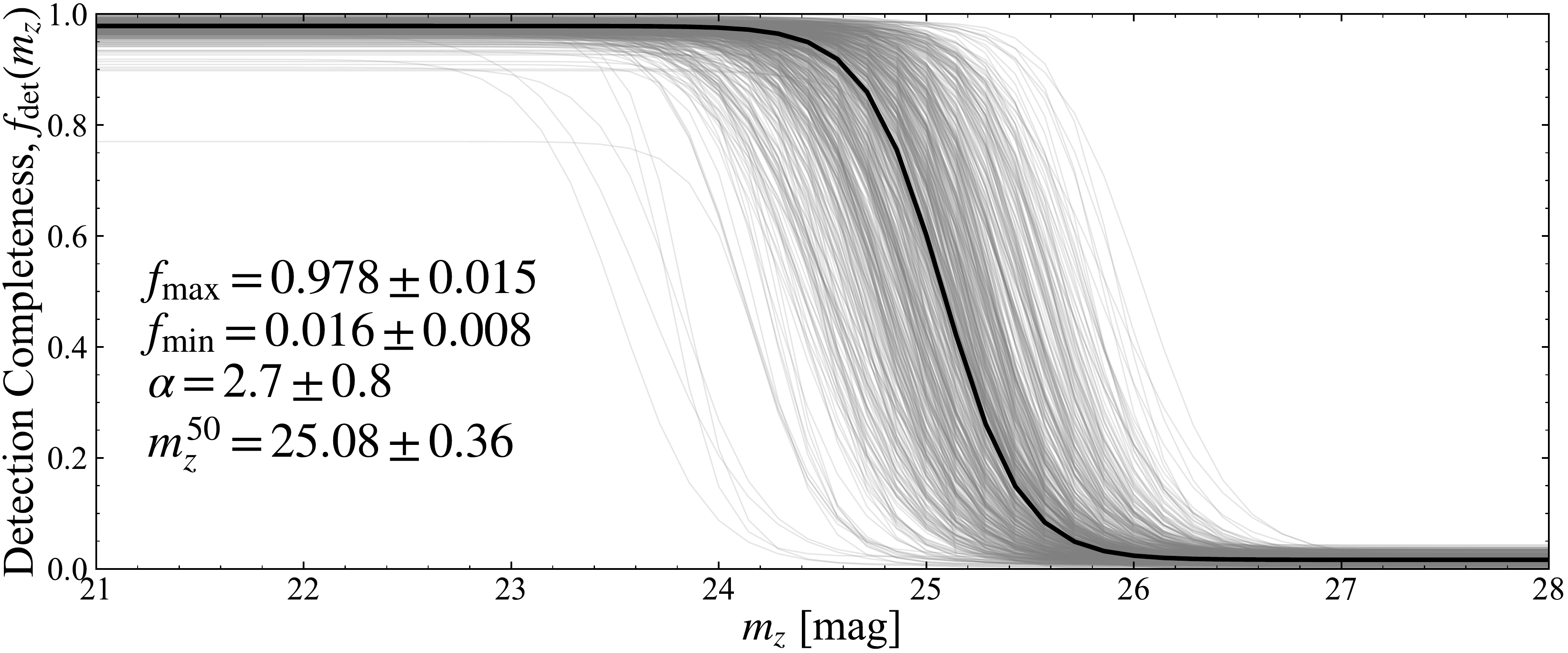}
    \caption{Detection completeness as a function of $z$-band magnitude.
    The black solid line denotes the median completeness of all patches.
    The thin gray solid lines show the detection completeness of each patch.
    Each median parameter and $1\sigma$ error is denoted in the left of the figure.
    }
    \label{fig:detection_completeness}
\end{figure}

The detection completeness is fitted with a function in \citet{Serjeant2000};
\begin{equation}
    f_{\mathrm{det}}(m_z)=\frac{f_\mathrm{max}-f_\mathrm{min}}{2}\{\tanh [\alpha(m_z^{50}-m_z)]+1\}+f_\mathrm{min},
\end{equation}
where $f_\mathrm{max},f_\mathrm{min},\alpha,$ and $m_{z}^{50}$ represent the detection completeness at the brightest magnitude and the faintest magnitude, the sharpness of the function, and the magnitude at which the detection completeness is 50\%, respectively.
Our measurement of each tract is presented in Figure \ref{fig:detection_completeness} and the best-fit parameters with $1\sigma$ error for the median completeness are $f_\mathrm{max}=0.978\pm0.015, f_\mathrm{min}=0.016\pm0.008,\alpha=2.7\pm0.8,m_z^{50}=25.08\pm0.36$, which are also denoted in the figure.
Almost all the functions have similar parameters with a $m_z^{50}$ scatter as small as $\sigma(m_z^{50})=0.36$.
Figure \ref{fig:completenessmap} shows the completeness map at $m_z=24.5$ of the survey region overplotted with the sample quasars.
The completeness holds more than 70\% (80\%) over 85\% (77\%) of the entire survey region, and $>50\%$ over almost all the area at the $z$-band limiting magnitude, and there are few areas of singularly lowered completeness.
It is noted that the following results hardly change when the area with $f_\mathrm{det}<0.5$ is excluded.
Therefore, we conclude that the whole survey area is homogeneous enough to conduct the clustering analysis.

\section{Clustering Analysis} \label{sec:clustering}

\subsection{Auto Correlation Function of the Quasars} \label{subsec:acf}
We first measure a projected correlation function $\omega_p(r_p)$ in Section \ref{subsubsec:projected} that can be directly related to real-space clustering. 
At the same time, to check the robustness of the result, we also measure an angular correlation function $\omega(\theta)$ without redshift information in Section \ref{subsubsec:angular} and a redshift-space correlation function $\xi(s)$, which includes redshift-space distortion, in Section \ref{subsubsec:redshift}.

\subsubsection{Projected correlation function} \label{subsubsec:projected}
We evaluate the projected correlation function $\omega_p$ of the sample.
In this analysis, the comoving distance is calculated from the spectroscopic redshift.
We separate $s$, the three-dimensional distance between two objects, into $r_p$, perpendicular to the line of sight, and $\pi$, parallel with it ($s=\sqrt{r_p^2+\pi^2}$).
We estimate the two-dimensional correlation function $\xi(r_p,\pi)$ from \citet{Landy1993BiasAV};
\begin{equation}
    \xi(r_p,\pi)=\frac{DD(r_p,\pi)-2DR(r_p,\pi)+RR(r_p,\pi)}{RR(r_p,\pi)}
\end{equation}
where $DD(r_p,\pi), DR(r_p,\pi), RR(r_p,\pi)$ represent data-data, data-random, and random-random pair counts within perpendicular distance separation $r_p$ and parallel distance separation $\pi$, respectively.
The survey area is divided into three independent fields, therefore we count the pairs in each field to sum them up before being normalized by all pairs.
The random points are retrieved from the random catalog in HSC-SSP DR3, which has random points scattered over the entire effective survey area, excluding mask areas, at a surface density of $1\,\mathrm{arcmin^{-2}}$.
The total number of random points is 3,209,416.
The redshift of random points is assigned to follow the $N(z)$, which is the redshift distribution of SHELLQs estimated by kernel density estimation (see Figure \ref{fig:mag&z_hist}).
To count pairs, we use \texttt{Corrfunc}\footnote{\url{https://github.com/manodeep/Corrfunc}}, which is a Python package containing routines for clustering analysis.
We also use the package in Section \ref{subsubsec:angular}, Section \ref{subsubsec:redshift}, and Section \ref{subsec:ccf}.
The projected correlation function $\omega_p(r_p)$ is derived by integrating $\xi(r_p,\pi)$ with $\pi$ direction.
\begin{equation}
    \omega_p(r_p)=2\int_{0}^{\pi_{\mathrm{cutoff}}}\xi(r_p,\pi)\ d\pi
    \label{eq:projected}
\end{equation}
where $\pi_{\mathrm{cutoff}}$, which is the optimum limit above which the signal is almost negligible, is fixed to $\pi_{\mathrm{cutoff}}=80\ h^{-1}\mathrm{Mpc}$ after sufficient trial and error.
The redshift distortion is eliminated through the integration \citep{Eftekharzadeh_2015}, though the angular scale of the redshift distortion is much smaller ($<20\,h^{-1}\mathrm{Mpc}$) than the scales of our measurements. 

The uncertainty of $\omega_p(r_p)$ is evaluated by Jackknife resampling \citep{Zehavi_2005}.
In the $k$-th resampling, we exclude $k$-th sub-region, and calculate the correlation function, $\omega_{p,k}(r_p)$.
We divide the survey area into $N=21$ sub-regions using the $k$-means method.
In this case, the covariance matrix is defined as
\begin{eqnarray}
    C_{ij}=\frac{N-1}{N}\sum_{k=1}^{N}(\omega_{p,k}(r_{p,i})-\bar{\omega}_p(r_{p,i}))\nonumber\\
    \times\ (\omega_{p,k}(r_{p,j})-\bar{\omega}_p(r_{p,j}))
    \label{eq:jackknife}
\end{eqnarray}
where $\omega_{p,k}(r_{p,i})$ and $\bar{\omega}_p$ represent the value of $k$-th projected correlation function for $i$-th $r_p$ bin and the mean of the projected correlation function, respectively.
The uncertainty of $\omega_p(r_{p,i})$, $\sigma_i$, is evaluated as $\sigma_i=\sqrt{C_{ii}}$, which is used only for plotting process.

The projected function is related to the real-space correlation function $\xi(r)$ as \citep{Davis1983} 
\begin{equation}
    \omega_p(r_p)=2\int_{r_p}^{\infty}\frac{r\xi(r)}{\sqrt{r^2-r_p^2}}dr.
\end{equation}
Assuming that the real space correlation function is regarded as a power-law function, $\xi(r)=(r/r_0)^{-\gamma}$, the fitted function $\omega_{p,\mathrm{fit}}(r_p)$ is represented as
\begin{equation}
    \frac{\omega_{p,\mathrm{fit}}(r_p)}{r_p}=B\left(\frac{\gamma-1}{2},\frac{1}{2}\right)\left(\frac{r_p}{r_0}\right)^{-\gamma}
\end{equation}
where $\gamma$ is a power-law index of the dark matter correlation function, $B$ represents the beta function, and $r_0$ is the correlation length, which represents the scale of clustering.
In this study, we fit $\gamma$ to a fiducial value ($\gamma=1.8$; \citealp{Peebles1980}).
The black solid line in the top panel of Figure \ref{fig:correlation_functions} represents the power-law function fitted to the projected correlation function based on the $\chi^2$ fit.
Then, we obtain $r_0=23.7\pm 11\ h^{-1}\mathrm{Mpc}$ as listed in Table \ref{tab:DMHmass}.
The goodness-of-fit is evaluated by 
\begin{equation}
    \chi^2=\sum_{i,j}[\omega_p(r_{p,i})-\omega_{p,\mathrm{fit}}(r_{p,i})]C_{ij}^{-1}[\omega_p(r_{p,j})-\omega_{p,\mathrm{fit}}(r_{p,j})].
    \label{eq:goodness-of-fit}
\end{equation}

To see the robustness of the clustering signal, we integrate the real-space correlation function $\xi(r)$ within $r_\mathrm{min}\leq r\leq r_\mathrm{max}$ (e.g., \citealp{Shen2007});
\begin{equation}
    \xi_{100} = \frac{3}{r_\mathrm{max}^3}\int_{r_\mathrm{min}}^{r_\mathrm{max}}\xi(r)r^2\,dr,
    \label{eq:integrate_xi}
\end{equation}
where $r_\mathrm{min}=10\,h^{-1}\mathrm{Mpc}$ and $r_\mathrm{max}=100\,h^{-1}\mathrm{Mpc}$, over which the observed signal is detected in this study.
Since we assume $\xi(r)=(r/r_0)^{-\gamma}$, Equation (\ref{eq:integrate_xi}) reduces to
\begin{equation}
    \xi_{100}=\frac{3r_0^{\gamma}}{(3-\gamma)r_\mathrm{max}^3}(r_\mathrm{max}^{3-\gamma}-r_\mathrm{min}^{3-\gamma}).
    \label{eq:xi100}
\end{equation}
Adopting $r_0=23.7\pm11\,h^{-1}\mathrm{Mpc}$, we obtain $\xi_{100}=0.175\pm0.147$.
Although the uncertainty of each individual data point is large, the overall clustering signal is found to be positive with a significance of more than $1\sigma$.

We also test the robustness in terms of whether the signal can be obtained by chance from a random sample.
We extract the same number of random points as our quasar sample, treat them as data points and evaluate the projected correlation function.
Based on 10000 iteration, we evaluate the probability of obtaining a clustering signal as shown in Figure \ref{fig:correlation_functions}.
Counting the number of the projected correlation function that has positive signals in the same bins in Figure \ref{fig:correlation_functions}, we find that there is only a 4\% probability of obtaining the clustering signal observed in this study.
Hence, we conclude that the signal is not artificial.

\subsubsection{Angular correlation function} \label{subsubsec:angular}
We also evaluate the angular correlation function.
Then, we use the estimator from \citet{Landy1993BiasAV};
\begin{equation}
    \omega(\theta)=\frac{DD(\theta)-2DR(\theta)+RR(\theta)}{RR(\theta)}
    \label{eq:LS_ang}
\end{equation}
where $DD(\theta),DR(\theta),RR(\theta)$ represent the normalized data-data, data-random and random-random pair counts normalized by whole pair counts within an angular separation $\theta$, respectively.
The random points are retrieved from the random catalog in HSC-SSP DR3.

The uncertainty of $\omega(\theta)$ is evaluated by Jackknife resampling in the same manner as the previous section.
We evaluate the uncertainty from the diagonal elements of the covariance matrix derived from Equation (\ref{eq:jackknife}) replacing $\omega_p$ for $\omega$.
The uncertainty of $\omega(\theta_i)$, $\sigma_i$, is evaluated based on the diagonal element of the covariance matrix.

The middle panel of Figure \ref{fig:correlation_functions} represents the result of the angular correlation function.
The black solid line represents the best fit of a single power-law model, $\omega_{\mathrm{true}}(\theta)=A_\omega\theta^{-\beta}$, considering the effect of the limited survey area to the correlation function.
Then, we assume the following function;
\begin{equation}
\omega(\theta)=A_\omega \theta^{-\beta}-\mathrm{IC},
\label{eq:power-law_angular}
\end{equation}
where $A_\omega$ is the amplitude, $\beta$ is the power-law index, and IC is the integral constraint, which is a negative offset as the survey region is limited \citep{Groth1977}.
We fix $\beta$ to $0.8 (=\gamma-1)$ for consistency with the projected correlation function.
We evaluate the integral constraint based on \citet{Woods1997};
\begin{equation}
    \mathrm{IC}=\frac{1}{\Omega^2}\int\!\!\!\int\omega_{\mathrm{true}}(\theta)d\Omega_1d\Omega_2,
\end{equation}
where $\Omega$ represents the solid angle of the survey field.
The integral constraint becomes considerably smaller than the clustering signal in all fields.
Therefore, the integral constraint is ignored in this study.
We assume the error follows the Gaussian function, and evaluate the goodness-of-fit of the fitted function through Equation (\ref{eq:goodness-of-fit}) replacing $\omega_p$ for $\omega$.

We convert the amplitude $A_\omega$ into the correlation length $r_0$ based on \citet{Limber1953}, which formulated
\begin{equation}
    r_0=\left\{A_\omega \frac{c}{H_0 H_\gamma}\frac{[\int N(z)dz]^2}{\int N^2(z)\chi(z)^{1-\gamma}E(z)dz}\right\}^{1/\gamma},
\end{equation}
where 
\begin{equation}
    H_\gamma=B\left(\frac{\gamma-1}{2},\frac{1}{2}\right),
\end{equation}
\begin{equation}
    E(z)= \sqrt{\Omega_{m}(1+z)^3+\Omega_{\lambda}},
\end{equation}
\begin{equation}
    \chi(z)=\frac{1}{H_0}\int_{0}^{z}\frac{1}{E(z')}dz'.
\end{equation}

Finally, we obtain $r_0=27.0\pm8.4\,h^{-1}\mathrm{Mpc}$.
The result is listed in Table \ref{tab:DMHmass}
\footnote{We note that the obtained amplitude is consistent with Shinohara et al. 2023, in preparation, which also evaluate the angular correlation function from 92 quasars at $5.88<z<6.49$ including 81 SHELLQs quasars, although samples are not an exact match.
}.
Adopting $r_0=27.0\pm8.4\,h^{-1}\mathrm{Mpc}$, the Equation (\ref{eq:xi100}) gives $\xi_{100}=0.222\pm0.124$, which suggests that the clustering signal is actually detected.

\subsubsection{Redshift-space correlation function} \label{subsubsec:redshift}
We also evaluate the redshift space correlation function of the quasars.
All redshifts for SHELLQs quasars are measured in Ly$\mathrm{\alpha}$ emission lines, which has an uncertainty up to $\Delta z\sim0.1$, in particular for those without clear Ly$\alpha$ emission \citep{Matsuoka_2022}.
This uncertainty and the redshift distortion due to peculiar velocity induce systematic bias in the redshift-space correlation.
We derive the redshift-space correlation function $\xi(s,\mu)$, where $s$ is the 3D distance and $\mu$ represents the cosine of the angle to the line of sight, utilizing the estimator from \citet{Landy1993BiasAV};
\begin{equation}
\xi(s,\mu)=\frac{DD(s,\mu)-2DR(s,\mu)+RR(s,\mu)}{RR(s,\mu)},    
\end{equation}
where $DD(s,\mu), DR(s,\mu), RR(s,\mu)$ represent data-data, random-random, and random-random pair counts within  a separation $s$ and an angular separation $\arccos\mu$, respectively.
The correlation function of the entire survey are is evaluated by summing the whole pair counts.
As mentioned in Section \ref{subsubsec:projected}, the redshift of random points is assigned to follow $N(z)$, the distribution function in Figure \ref{fig:mag&z_hist}.
The redshift-space correlation function decomposed into multipoles $\xi_l(s)$ is derived by integrating $\xi(s,\mu)$ by $\mu$ \citep{Marin_2015};
\begin{equation}
    \xi_l(s)=\frac{2l+1}{2}\int_{-1}^{1}\xi(s,\mu)L_l(\mu)d\mu,
\end{equation}
where $L_l$ is the Legendre polynomial of order $l$.
We evaluate the mono-pole ($l=0$) of the redshift-space correlation function.

\begin{figure}[ht!]
    \centering
    \includegraphics[width=0.86\columnwidth]{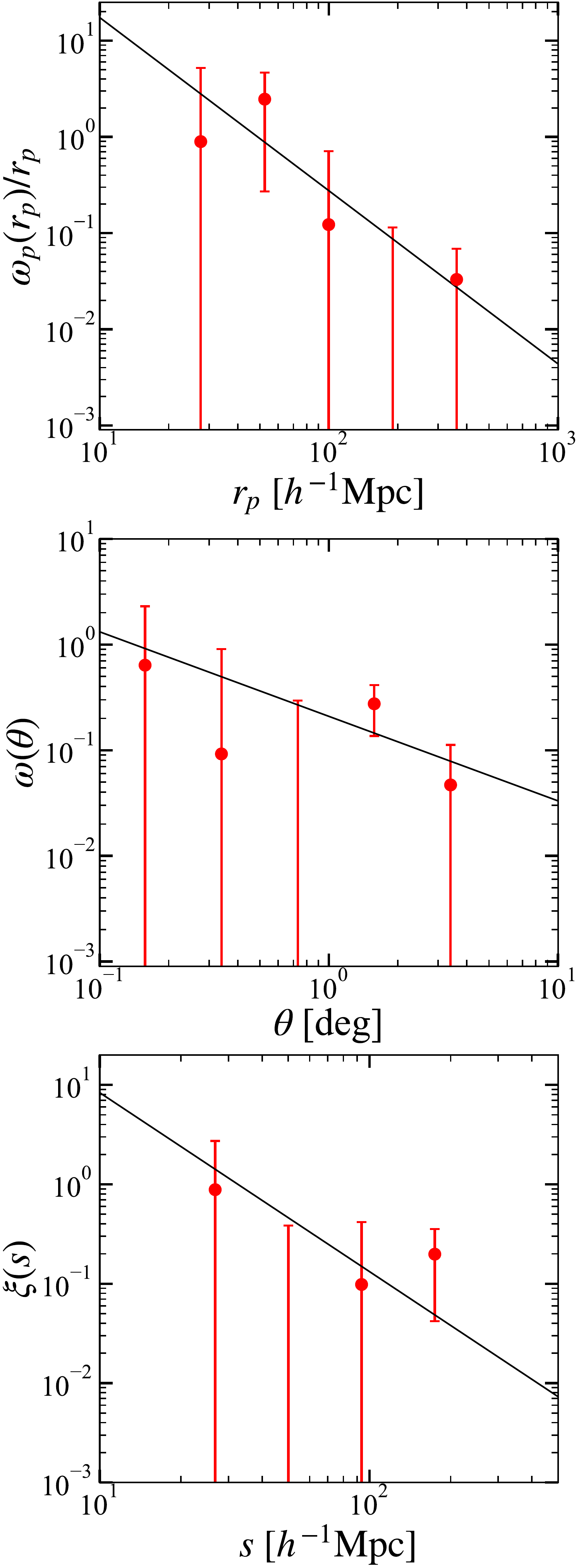}
    \caption{
    The top, middle, and bottom panels show the projected correlation function, the angular correlation function,  and the redshift space correlation function, respectively. 
    The error bars represent the $1\sigma$ error from the Jackknife resampling.
    The black solid lines represent the power-law functions which are fitted by $\chi^2$ fit.
    }
    \label{fig:correlation_functions}
\end{figure}

The bottom panel of Figure \ref{fig:correlation_functions} represents the result of the redshift-space correlation function.
Taking the redshift distortion into account, the redshift-space correlation function $\xi_0(s)$ is related to the real-space correlation function $\xi(r)$ suggested by \citet{Kaiser1987};
\begin{equation}
    \xi_0(s)=\left(b^2+\frac{2}{3}bf+\frac{1}{5}f^2\right)\xi(r),
\end{equation}
where $b$ is the bias parameter which is defined in Equation (\ref{eq:biasdefinition}), $f$ is the gravitational growth factor.
However, the effect of the redshift distortion is negligible because our clustering signal is measured on a large scale, beyond the small scale where redshift distortion can be observed.
We fit the power-law function $\xi(s)=(s/s_0)^{-\gamma}$, black solid line in the bottom panel of Figure \ref{fig:correlation_functions}, to the redshift space correlation function in place of the Kaiser's function by $\chi^2$ fit.
Based on Equation (\ref{eq:goodness-of-fit}) replacing $\omega$ for $\xi$, we obtain $s_0 = 32.5\pm 19\ h^{-1}\mathrm{Mpc}$ as the correlation length in the redshift space, which is almost consistent with that in the real space.
Adopting $s_0 = 32.5\pm 19\ h^{-1}\mathrm{Mpc}$ as the correlation length in the real space, we obtain $\xi_{100}=0.310\pm0.326$, which suggests that the significance of the clustering signal of the redshift-space correlation function is marginal.

As shown in Table \ref{tab:DMHmass}, consistent correlation lengths are obtained using three different correlation functions.
The $\xi_{100}$ shows that the clustering signal is barely detected.
However, the errors for each correlation length are relatively large. 
This is probably due to the sample size not being large enough yet.

\subsection{Cross Correlation Function with Galaxies} \label{subsec:ccf}
We evaluate cross-correlation function (CCF) with our quasars and their neighboring LBGs at the similar redshift.
The LBGs at $z\sim6$ are retrieved from Great Optically Luminous Dropout Research Using Subaru HSC (GOLDRUSH; \citealt{Harikane_2022}).
The LBGs in the wide layer are not suitable for clustering analysis due to their low number density; therefore we use the LBG sample in the Deep and Ultra Deep layer (COSMOS and SXDS) over $8.7\,\mathrm{deg^2}$ of HSC-SSP S18A \citep{Aihara2019} where the SHELLQs quasars reside.
As a result, the number of quasars and LBGs to calculate CCF is limited to 3 and 200, respectively.
The limiting magnitude of LBGs is $m_\mathrm{UV}=25.15$.
They are not spectroscopically confirmed; therefore only angular correlation function can be evaluated.
We use the estimator of CCF from the following equation \citep{Landy1993BiasAV,Cooke_2006};
\begin{equation}
    \omega_{\mathrm{QG}}(\theta)=\frac{QG(\theta)-QR(\theta)-GR(\theta)+RR(\theta)}{RR(\theta)},
    \label{eq:ccf}
\end{equation}
where $QG(\theta), QR(\theta), GR(\theta), RR(\theta)$ represent quasar-galaxy, quasar-random, galaxy-random, and random-random pairs of the given separation normalized by total pairs, respectively.
The random points are retrieved from the random catalog in HSC-SSP DR2 utilizing the same flags of Table 2 in \citet{Harikane_2022} at the surface number density of $1\,\mathrm{arcmin^{-2}}$.

\begin{figure}
    \centering
    \includegraphics[width=0.8\columnwidth]{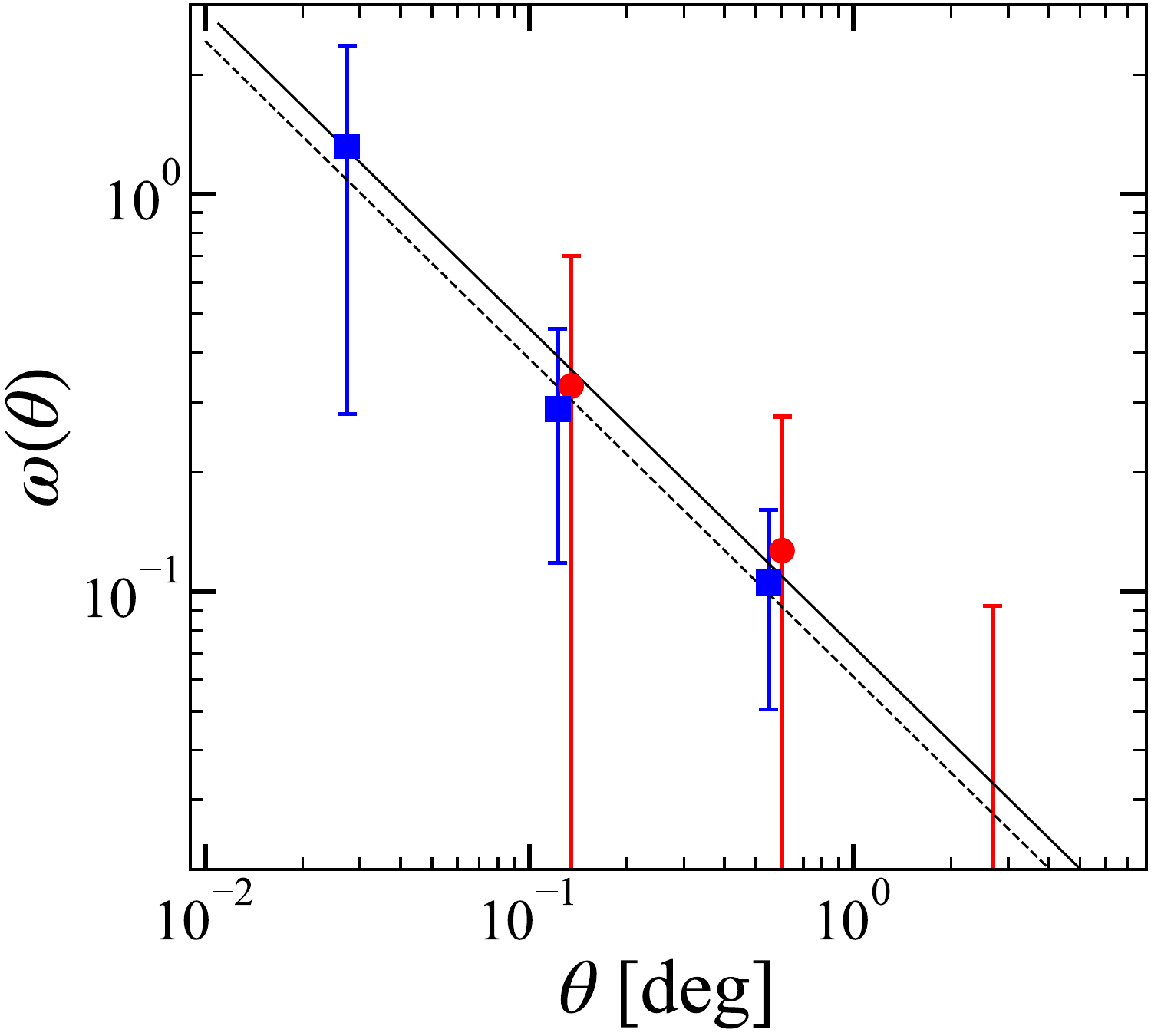}
    \caption{The CCF between quasars and LBGs at $z\sim6$ (red) and ACF of LBGs (blue).
    The solid line and the dashed line show the power-law function fitted to the CCF and the ACF by $\chi^2$ fit, respectively.
    For visibility, CCF plots are offset toward x-axis direction.}
    \label{fig:CCF}
\end{figure}

Figure \ref{fig:CCF} represents the result of CCF (red circles) and auto-correlation function (ACF) of LBGs at $z\sim6$ (blue squares; $\omega_{\mathrm{GG}}$), which is found to be consistent with \citet{Harikane_2022}.
We fit the power-law functions to the CCF and the ACF by $\chi^2$ fit and the results are shown as the solid line and the dashed line, respectively.
We confirme that the angular scale at which we see the CCF signal is large enough to exceed the small scale ($\lesssim 20''$), where the one-halo term is dominant.
The errors are also evaluated by Jackknife resampling mentioned in Section \ref{subsubsec:projected} with $N=5$.
The goodness-of-fit is calculated based on Equation (\ref{eq:goodness-of-fit}).
Our quasar sample has a cross-correlation strength similar to that of the auto-correlation of LBGs at the same redshift.
Although the UV luminosity of the host galaxy, on which the clustering strength of galaxies strongly depends, in our quasar sample is not known, the result seems natural, given that quasars are stochastic processes that all galaxies experience at some period.

The correlation length is derived from the amplitude $A_\omega$ of the power-law function fitted to the CCF.
In CCFs, the Limber's equation is formalized as \citep{Croom_1999}
\begin{equation}
    r_0=\left\{A_\omega \frac{c}{H_0 H_\gamma}\frac{\int N_{\mathrm{Q}}(z)dz\int N_{ \mathrm{G}}(z)dz}{\int N_{\mathrm{Q}}(z)N_{\mathrm{G}}(z)\chi(z)^{1-\gamma}E(z)dz}\right\}^{1/\gamma}.
    \label{eq:correlation_length_CCF}
\end{equation}
The suffix of Q and G in Equation (\ref{eq:correlation_length_CCF}) denote quasars and LBGs, respectively.
The redshift distribution of LBGs is assumed to be the same as \citet{Harikane_2022}.
Finally, we obtain $r_0=17.7\pm8.0\,h^{-1}\mathrm{Mpc}$ as the correlation length of quasars and galaxies.
It should be noted that our LBG sample is photometrically selected and contamination of low-$z$ interlopers, the fraction of which is unknown, reduces the amplitude of cross-correlation.
\citet{Ono2018} concluded that the contamination rate in the $i$-dropout galaxies may be small based on the fact that all 31 spectroscopic $i$-dropout galaxies have $z>5.5$, but it is difficult to know the exact contamination rate in the sample in this study down to the limiting magnitude.

\begin{deluxetable*}{lccccc}
\tablehead{\colhead{Estimator}&\colhead{correlation length}&\colhead{bias}&\colhead{DMH mass}&\colhead{reduced-$\chi^2$}\\
\colhead{} & \colhead{($h^{-1}\mathrm{Mpc}$)} & \colhead{} & \colhead{($10^{12}\ h^{-1}M_\odot$)} &  \colhead{}}
\tablecaption{DMH mass from the clustering analysis}
\startdata
Projected, $\omega_p$ & $23.7\pm11$ &$20.8\pm8.7$ &$5.0_{-4.0}^{+7.4}$ &  0.87 \\ 
Angular, $\omega$ & $27.0\pm8.4$ &$23.4\pm6.6$ &$6.9_{-4.1}^{+6.1}$  & 0.92 \\ 
Redshift space, $\xi$ & $32.5\pm19$ &$27.7\pm15$ &$10.6_{-9.3}^{+17.5}$  & 0.91 \\ 
CCF, $\omega_{\mathrm{QG}}$ & \-- &$19.5\pm16$ &$4.0_{-4.0}^{+14.8}$  & 0.52 
\enddata
\end{deluxetable*}
\label{tab:DMHmass}

\subsection{The Bias Parameter}\label{subsec:correlation_length_bias}
We assume that quasars reside in the peak of DM distribution and trace the distribution of underlying DM \citep{Sheth_1999}.
The bias parameter is derived from the ratio of clustering strength between quasars and underlying DM at a scale of $r=8\ h^{-1}\mathrm{Mpc}$,
\begin{equation}
    b=\sqrt{\frac{\xi(8,z)}{\xi_{\mathrm{DM}}(8,z)}},
    \label{eq:bias}
\end{equation}
assuming that the real space correlation function $\xi(r)$ is approximated by the power-law function.
The correlation function of DM is generated by \texttt{halomod}\footnote{\url{https://github.com/halomod/halomod}} \citep{Murray2013,Murray_2021}, assuming the bias model of \citet{Tinker_2010}, the transfer function of \texttt{CAMB}, and the growth model of \citet{Carroll1992}.
We evaluate the bias parameter as $b=20.8\pm8.7, 23.4\pm6.6$ and, $27.7\pm15$ from the projected correlation function, the angular correlation function, and the redshift space correlation function, respectively.
We also evaluate the bias parameters $b_{\mathrm{QG}}$ and $b_{\mathrm{GG}}$ from the CCF between quasars and LBGs and the ACF of LBGs, respectively.
We derive the bias parameter of quasars $b_{\mathrm{QQ}}$ from \citet{Mountrichas2009},
\begin{equation}
    b_\mathrm{QG}\sim b_\mathrm{QQ}b_\mathrm{GG}.
\end{equation}
We obtain $b_\mathrm{QG}=16.1\pm6.6$ and $b_\mathrm{GG}=13.3\pm2.3$ from the same analysis, yielding $b_\mathrm{QQ}=19.5\pm16$ and they are  summarized in Table \ref{tab:DMHmass}.
The bias parameters derived by four independent methods are consistent with each other within their errors.

\subsection{DMH Mass of $z\sim6$ Quasars}\label{subsec:DMHmass}
We derive typical DMH mass from bias parameters of correlation functions. 
Under the assumption that quasars are the tracer for the underlying DM distribution,
we adopt the bias model in \citet{Tinker_2010}, which is formalized as
\begin{eqnarray}
    b(\nu)=1-A\frac{\nu^a}{\nu^a+\delta^a_c}+B\nu^b+C\nu^c,
    \label{eq:Tinker}
\end{eqnarray}
where $\nu$ is the peak height which is defined as $\nu=\delta_c/\sigma(M)$, $\delta_c$ is the critical density for the collapse of DMHs ($\delta_c=1.686$), and $\sigma(M)$ is the linear matter variance at the radius of each DMH.
We use the other parameters as they are in Table 2 of \citet{Tinker_2010} for $\Delta=200$, which represents the ratio between mean density and background density.
The linear variance is defined as
\begin{equation}
    \sigma^2(M)=\frac{1}{2\pi^2}\int P(k,z)\hat{W}^2(k,R)k^2 dk,
    \label{eq:matter_variance}
\end{equation}
where $P(k,z)$ is the matter power spectrum generated by \texttt{CAMB}\footnote{\url{https://github.com/cmbant/CAMB}} with our cosmology parameters and $\hat{W}$ is the spherical top-hat function defined as 
\begin{equation}
    \hat{W}(k,R)=\frac{3}{(kR)^3}[\sin(kR)-kR\cos(kR)].
\end{equation}
This model is based on the clustering of DMHs in cosmological simulations of the flat $\Lambda$CDM cosmology. 
We obtain the radius of DMH $R_{\mathrm{halo}}$ by solving Equation (\ref{eq:matter_variance}).
Finally, we evaluate the DMH mass $M_\mathrm{halo}$ assuming the spherical DMH;
\begin{equation}
    M_\mathrm{halo}=\frac{4}{3}\pi R_\mathrm{halo}^3\bar{\rho}_m.
\end{equation}
We adopt $\bar{\rho}_m=2.78\times 10^{11}\Omega_m h^2M_\odot$.
Our DMH mass from each estimator is summarized in Table \ref{tab:DMHmass}.
The bias and halo mass of the CCF are slightly smaller than the other three, but this may be due to the contamination of the low-$z$ interlopers to the $z\sim6$ LBG sample (see Section \ref{subsec:ccf}).

The DMH mass derived by four independent methods are consistent with each other within their errors.
However, we note that the DMH mass estimation is sensitive to $\sigma_8$.
For simplicity, the following discussions will use the bias and halo mass obtained from the projected correlation function, but note that there is variation in these evaluations as shown in Table \ref{tab:DMHmass}.

\section{Discussion} \label{sec:discussion}

\subsection{Comparison of DMH mass with other studies} \label{subsec:comparison}

\begin{figure*}
    \centering
    \includegraphics[width=2\columnwidth]{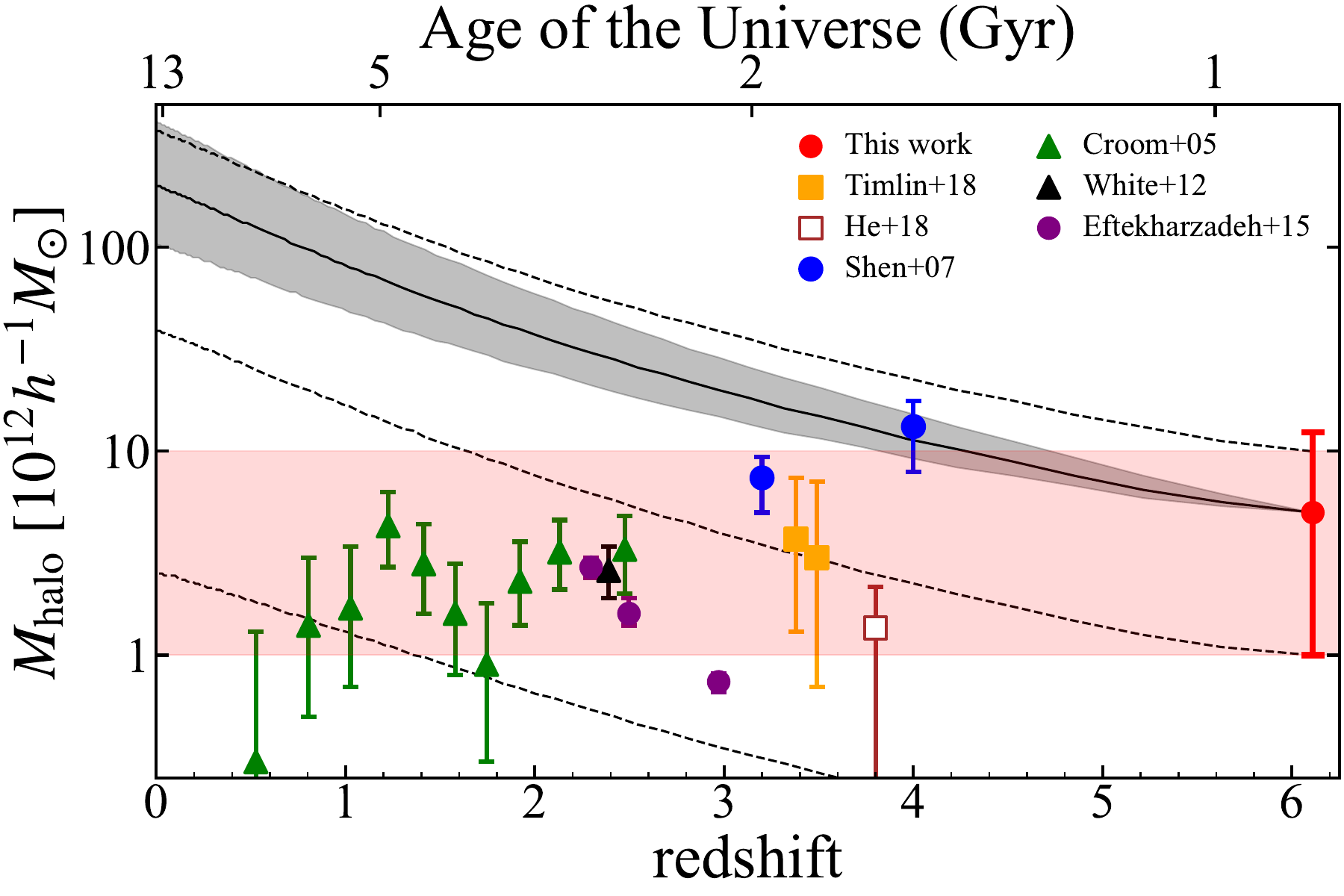}
    \caption{The DMH mass from clustering analysis as a function of redshift.
    Our result from the projected correlation function is represented as a red circle.
    Other symbols represent the DMH masses from previous studies.
    The circle, square, and triangle show studies that derive the projected, angular, and redshift-space correlation function, respectively.
    The filled symbols denote estimates from ACF signal, while others from CCF signal.
    The references are shown in the upper right of the figure.
    These masses are recalculated based on each bias parameter by using the cosmology in the study and the bias model of \citet{Tinker_2010}.
    The solid line with gray shade denotes the DMH mass evolution with its error based on the extended Press-Schechter theory.
    The three dashed lines represent the mass evolution of DMHs with $M_{\mathrm{halo}}=10^{13},10^{12},10^{11}\,h^{-1}M_\odot$ at $z\sim6$ from the top.
    The red-shaded region represents the DMH mass range where most of quasars are expected to reside.}
    \label{fig:Mhalo_summary}
\end{figure*}

This study is the first to obtain the typical DMH mass of quasars at $z\sim6$ from clustering analysis, and not many previous studies have obtained DMH mass at $z\sim6$ using other methods.
\citet{Shimasaku_2019} estimated DMH mass of 49 $z\sim 6$ quasars, assuming that the FWHM of [C \textsc{II}] corresponds to the circular velocity of the DMH.
They estimated that the median DMH mass of the whole samples is $1.2_{-0.6}^{+2.2}\times 10^{12}\ M_\odot$, which is slightly lower than our measurement, though it is consistent within the errors.
\citet{Chen2022} estimated the typical DMH mass of a bright quasar ($M_{1450}<-26.5$) at $z\sim6$ is $2.2_{-1.8}^{+3.4}\times10^{12}\,h^{-1}M_\odot$ by measuring the intergalactic medium density around these luminous quasars, which is also consistent with our result within the error.
Furthermore, a cosmological N-body simulation \citep{Springel2005b} predicts that the virial mass of DMH of quasars at $z=6.2$ is $3.9\times 10^{12}\ h^{-1}M_\odot$, which is consistent with our result.

Figure \ref{fig:Mhalo_summary} shows a compilation of the previous DMH mass measurements based on clustering analysis at lower-$z$.
In the figure, we convert the bias parameter in each previous research into DMH mass adopting our cosmology to reduce the effect of different $\sigma_8$ among this work and the previous research.
Some previous studies use different fitting formulae to infer a DMH mass from clustering, but we confirmed that the difference produces a few percent discrepancy in the DMH mass estimate.
We conclude that the definition does not have a large impact on the DMH mass measurement.
We also plot the mass evolution of DMH with $M_{\mathrm{halo}}=10^{13},10^{12},10^{11}\,h^{-1}M_\odot$ (dotted lines) from the sample mean redshift, $z=6.1$, to $z=0$ based on the extended Press-Schechter theory \citep{Bower1991,Bond1991,Lacey1993}.
Our quasar sample with $M_{\mathrm{halo}}=5.0\times 10^{12}\,h^{-1}M_\odot$ at $z = 6.1$ grows to $2.0_{-1.0}^{+2.2}\times 10^{14}\,h^{-1}M_\odot$ (black solid line) at $z=0$, which is comparable to a rich galaxy cluster at present \citep{Bhattacharya2013}, implying that quasars reside in the most massive DMHs in the early universe.

Interestingly, the DMH mass of quasars has remained almost constant $\sim10^{12.5}\,h^{-1}M_\odot$ across the cosmic time.
Although the errors of each data point and variations even at the same epoch are large, and the DMH mass tends to decrease slightly from $z=1$ to $0$, it appears to remain roughly $M_{\mathrm{halo}}\sim10^{12}\--10^{13}\,h^{-1}M_\odot$.
A quite constant halo mass of quasars as a function of redshift has been suggested up to $z\sim4$ by the previous studies \citep{Trainor2012,Shen2013,Timlin_2018} and this study confirms that the trend continues up to $z\sim6$ for the first time.
This is in clear contrast to the standard growth of DMHs (the dashed lines in Figure \ref{fig:Mhalo_summary}).
\citet{Greiner2021} also concluded from the quasar pair statistics that there is no strong evolution in clustering strength from $z\sim6$ to $z\sim4$. 
\citet{McGreer2016} also used pair statistics to constrain the correlation length at $z\sim5$ as $r_0\gtrsim20\,h^{-1}\mathrm{Mpc}$, which is consistent with the trend.
The observed trend is also consistent with the model (e.g., \citealp{Lidz_2006}) that the characteristic mass of quasar host halos should evolve only weekly with redshift to reproduce the quasar luminosity function, though their constraints are predicted only at $0<z<3$.
Even though quasars at $z=0\--6$ reside in similar host halos of $10^{12.5}\,h^{-1}M_\odot$, this means that, as seen in the next section, higher-$z$ quasars are hosted in DMHs which are more massive (higher bias) for the mass at that time.
In other words, quasars appear in the most massive halos at $z\sim6$, but they appear in less extreme halos at a later time.

Our result that quasars at $z\sim6$ reside in a fairly massive-end halo implies that they could be in overdense regions.
However, observational evidence is far from conclusive, with some studies (e.g., \citealp{Stiavelli2005,Morselli2014,Mignoli2020}) finding quasars in the overdense region and others (e.g., \citealp{Willott2005,Banados2013,Mazzucchelli2017a}) finding no sign of it.
This may be due to differences in the depth and survey area of the overdense regions explored, or different selection criteria for surrounding galaxies, which may have led to a lack of consensus.
Recent James Webb Space Telescope (JWST) observation \citep{Kashino2022}, which assessed the galaxy distribution around quasars at $z\sim6$ on the scale of up to $\sim10\,\mathrm{Mpc}$ in the comoving coordinate, showed a clear overdensity of [O \textsc{iii}] emitters around an ultra-luminous quasar at $z=6.327$.
Another JWST observation by \citet{Wang2023}, which performed an imaging and spectroscopic survey of quasars utilizing NIRCam/WFSS, discovered ten [O \textsc{III}] emitters around a quasar at $z=6.6$ and the galaxy overdensity corresponds to $\delta_{\mathrm{gal}}=12.6_{-5.0}^{+5.9}$ over a $637\,\mathrm{Mpc^3}$ volume in the comoving space.
A large number of such deep and wide observations will provide clearer insights into the large-scale environments of $z\sim6$ quasars.

In the low-mass regime below $M_\mathrm{halo}<10^{12}\,h^{-1}M_\odot$, quasars with small black hole mass,
small stellar mass and extremely low luminosity may not be detected observationally.
In this case, the apparent lower limit of the observed halo mass may be due to observational bias.
In contrast, there may be an upper limit of halo mass rather than a typical halo mass at which quasar activity appears.
In Figure \ref{fig:Mhalo_summary}, there appears to be an upper limit where the quasar DMH mass never exceeds $10^{13}\,h^{-1}M_\odot$, i.e., most quasars reside in DMHs with $M_\mathrm{halo}<10^{13}\,h^{-1}M_\odot$ across most of the cosmic history.
\citet{Fanidakis2013} used \textsc{GALFORM}, a semi-analytic model, and concluded that quasars live in average mass halos and do not reside in the most massive DMHs at any redshift.
In their model, the quasar activity, which is maintained by the cold gas accretion onto a central SMBH, will be suppressed by the radio-mode AGN feedback in a massive halo larger than $10^{13}\,h^{-1}M_\odot$.
If the halo mass of quasars does not exceed $10^{13}\,h^{-1}M_\odot$ at any cosmic time, then such physics may ubiquitously operate.
This is supported by the observation by \citet{Uchiyama2018}, which concluded that few of the most massive protocluster candidates were found around quasars at $z\sim4$.
In other words, at $z\sim 4$, quasars does not exist in overdense regions exceeding $10^{13}\,h^{-1}M_\odot$, but in medium-weight overdense regions below $10^{13}\,h^{-1}M_\odot$.

However, it only appears that the halo mass does not exceed $10^{13}\,h^{-1}M_\odot$ in Figure \ref{fig:Mhalo_summary}, and what is measured from the clustering is the average DMH mass of quasars in each period, and it is therefore strictly inconclusive whether there are no quasars in the halo with a mass exceeding $10^{13}\,h^{-1}M_\odot$.

\subsection{Implication to AGN Feedback}\label{subsec:feedback}

\begin{figure*}
    \centering
    \includegraphics[width=2\columnwidth]{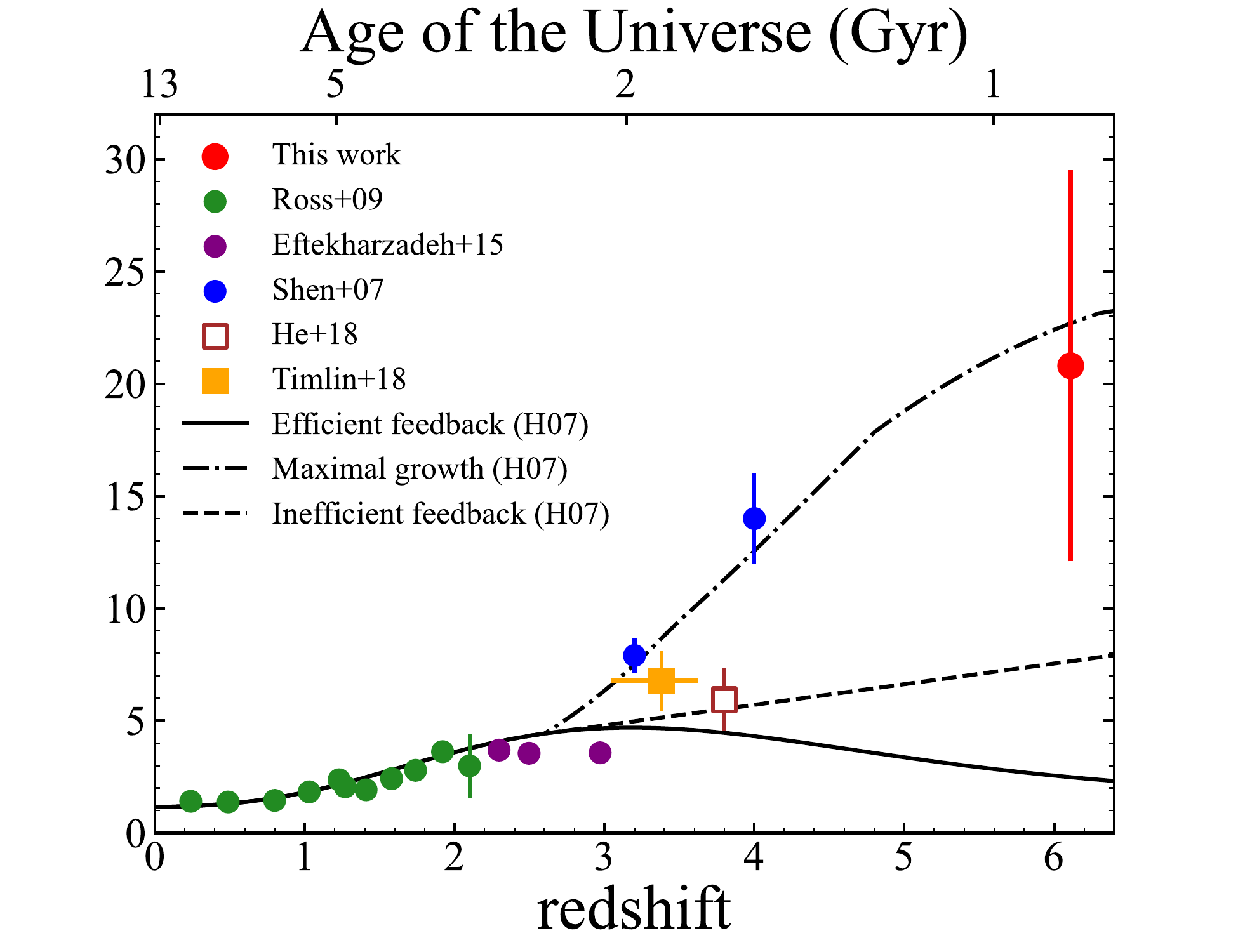}
    \caption{Evolution of the quasar bias parameter.
    This figure is the extension of Figure 12 in \citet{Timlin_2018}.
    The red circle represents our result of the projected correlation function and the other symbols show the result of previous studies.
    Three black lines represent the bias parameters as a function of redshift based on the theoretical models: ``efficient feedback" (solid line), ``inefficient feedback" (dashed line), and ``maximal growth" (dash dotted line).
    The ``efficient feedback" model assumes that the growth of quasars only occurs in their active phase.
    The ``inefficient feedback" model presumes that quasars continue growing periodically even after their active phase until $z\sim2$.
    The ``maximal growth" model premises that quasars keep growing with their host DMH until $z\sim2$.
    In these models, as the feedback becomes inefficient, the DMH mass gets more massive. 
    }
    \label{fig:bias_summary}
\end{figure*}

We compare our bias parameter with theoretical models in \citet{Hopkins2007}, which predicted a bias parameter evolution at $z\gtrsim3$ for three models with simple assumptions: ``efficient feedback," ``inefficient feedback" and ``maximal growth," as shown in Figure \ref{fig:bias_summary}.
In ``efficient feedback" model, quasars only grow during their active phase and the growth thoroughly terminates after the phase.
The bias parameter is predicted to become smaller at higher-$z$ if feedback is efficient.
In ``inefficient feedback" model, quasars and their central SMBHs continue growing periodically even after their active phase until $z\sim2$.
Since their feedback is inefficient, the quasars do not stop growing and shine episodically.
In contrast to the previous model, the quasars tend to reside in more massive DMHs, which makes the bias parameter larger at $z\gtrsim3$.
In the last model, ``maximal growth," quasars keep growing at the same rate with their host DMHs simultaneously until $z\sim2$.
The central SMBHs retain Eddington accretion all the time and their growth is rapid.
The feedback of quasars is less efficient than the second model.
Therefore, the DMHs are the most massive among these models, which is apparent in Figure \ref{fig:bias_summary}.
It should be noted that the model only predicts the evolution of the bias parameter, and no prediction of other observables (e.g., luminosity function, $M\--\sigma$ relation) is given for each assumption.

Our result is most consistent with a large bias parameter, favoring the ``maximal model," which assumes Eddington accretion and the feedback is highly inefficient at $z\sim6$.
This result is consistent with the fact that the Eddington ratio of quasars at $z\sim6$ tends to be higher than that in local \citep{Yang2021}.
However, it is noted that the Eddington ratio of quasars at $z<4$ is usually smaller than unity \citep{Shen2008}, being inconsistent with ``maximal growth" model.
The measurement of bias parameters at $z\sim4$ has not yet been settled, as the results are largely divided into large \citep{Shen2007} and small \citep{He2018,Timlin_2018} values.
Therefore, since Hopkins' models simply attempt to explain the evolution from $z=2$ to $z=6$ with a single physical mechanism, it is not necessary only to support this ``maximal growth" model at $4<z<6$ as it is.
For example, by assuming that the feedback is inefficient at $z\sim6$ while it becomes more efficient until $z\sim4$, an evolution model that the bias keeps low until $z\sim4$ and increases rapidly to $z\sim6$ does not conflict with our observational result.
Alternatively, it could be explained by intermittent BH growth \citep{Inayoshi2022,Li2022}.
To further restrict the models, the measurement of the bias parameter at $z\sim5$ is a key.

\subsection{Duty Cycle}
We also evaluate the duty cycle of quasars which represents the fraction of DMHs that host active quasars.
At first, following the traditional approach \citep{Haiman2001,Martini2001}, we assume that a DMH with more than the threshold $M_\mathrm{min}$ can host a quasar which activates randomly for a certain period.
Under this assumption, the duty cycle $f_\mathrm{duty}$ is defined as the ratio of the number of observed quasars to the number of the whole host halos above $M_\mathrm{min}$.
Therefore, $f_\mathrm{duty}$ is evaluated as
\begin{equation}
    f_\mathrm{duty} =\frac{\int_{L_{\mathrm{min}}}^{\infty}\Phi(L)dL}{\int_{M_\mathrm{min}}^{\infty}n(M)dM},
    \label{eq:dutycycle}
\end{equation}
where $\Phi(L)$ is the quasar luminosity function at $z\sim6$ derived by \citet{SHELLQs5}, $L_\mathrm{min}$ is the minimum luminosity of the quasar sample, $n(M)$ represents the DMH mass function at $z\sim6$ derived by \citet{Sheth_1999}, and $M_\mathrm{min}$ represents the DMH minimum mass to host a quasar.
The quasar luminosity function is evaluated based on the sample almost equivalent to that in this study, by excluding type-II quasars.
We adopt the DMH mass function from \citet{Sheth_1999};
\begin{eqnarray}
    n(M)=-A\sqrt{\frac{2a}{\pi}}\frac{\rho_0}{M}\frac{\delta_c(z)}{\sigma^2(M)}\frac{d\sigma(M)}{dM}\nonumber\\
     \times\left\{1+\left[\frac{\sigma^2(M)}{a\delta_c(z)}\right]^p\right\}\exp\left[-\frac{a\delta_c(z)}{2\sigma^2(M)}\right],
     \label{eq:Sheth99}
\end{eqnarray}
where $A=0.3222,a=0.707,p=0.3,$ and $\delta_c(z)=\delta_c/D(z)$.
The $D(z)$ represents the growth factor from \citet{Carroll1992}.
The minimum mass is estimated from the effective bias which is expressed as
\begin{equation}
    b_{\mathrm{eff}} = \frac{\int_{M_\mathrm{min}}^{\infty}b(M,z)n(M)dM}{\int_{M_\mathrm{min}}^{\infty}n(M)dM},
\end{equation}
where $b(M,z)$ is the bias parameter of the given DMH mass at a given redshift from the model \citep{Tinker_2010}.
Based on the effective bias determined from the clustering analysis, $M_\mathrm{min}$ is evaluated to be $4.5\times 10^{12}\,h^{-1}M_\odot$.
In this case, we obtain $f_\mathrm{duty}=6.3\pm2.7$, exceeding unity, which is unreasonable given its definition.

We consider that it is too simple to assume that all halos above a certain minimum mass can host a quasar, as expressed in Equation (\ref{eq:dutycycle}).
Equation (\ref{eq:dutycycle}) is correct when luminosity and mass are proportional and $L_\mathrm{min}$ corresponds to $M_\mathrm{min}$, but this is not the case of quasars.
In fact, as seen in Figure \ref{fig:Mhalo_summary}, the halo mass of quasars is within a certain narrow range over cosmic time, and there seems to exist an upper limit to the halo mass of quasars.
Although it is difficult to determine the exact mass range, we here simply assume that the DMHs with $12 \leq \log (M_\mathrm{halo}/h^{-1}M_\odot)\leq 13$ can host quasars based on Figure \ref{fig:Mhalo_summary}.
In the case, Equation (\ref{eq:dutycycle}) can be expressed as
\begin{equation}
    f_\mathrm{duty} = \frac{\int_{L_\mathrm{min}}^{\infty}\Phi(L)dL}{\int_{M_1}^{M_2}n(M)dM},
    \label{eq:new_dutycycle}
\end{equation}
where $M_1 = 10^{12}\,h^{-1}M_\odot$ and $M_2 = 10^{13}\,h^{-1}M_\odot$.
This equation gives $f_\mathrm{duty}=0.019\pm0.008$.
The derived $f_\mathrm{duty}$ corresponds to 1.9\% of the age of the universe, namely $\sim1.7\times10^{7}$ yr, as the lifetime of quasars at $z\sim 6$.
While this is consistent with the lifetime obtained from the clustering analysis at low-$z$ (e.g., \citealp{White2012}), it is about equal to the upper limit obtained from the proximity zone size measurements at $z\sim6$ \citep{Eilers2021}.

We derive the duty cycle based on the new definition, which cannot be simply compared with previous results at low-$z$. 
Based on Equation (\ref{eq:new_dutycycle}), we recalculate $f_\mathrm{duty}$ at $z\sim4$ from the luminosity function \citep{Akiyama2018} and obtain $f_\mathrm{duty}=0.012\pm0.001$, which is consistent with the conventional estimate, $f=0.001\--0.06$ \citep{He2018}, and $f_\mathrm{duty}$ at $z\sim6$ by this study.
On the other hand, in the case of \citet{Eftekharzadeh_2015} at $z\sim3$, we obtain $f_\mathrm{duty}=0.0060\pm0.0008$.
Based on \citet{Croom2005}, we obtain $f_\mathrm{duty}=0.0039\pm0.0005, 0.0043\pm0.0005, 0.0042\pm0.0006$ at $z=0.804,1.579,2.475$, respectively.
They are slightly smaller than those at $z>4$.

However, note that there is no justification for the mass range used for integration here.
Unless we know the exact mass distribution of halos that can host quasars, we cannot precisely obtain the denominators in either Equation (\ref{eq:dutycycle}) or (\ref{eq:new_dutycycle}).
Also, the numerator in these equations are the number of quasars observed, and $f_\mathrm{duty}$ will inevitably increase as the limiting magnitude deepens in the future, that is, as $L_\mathrm{min}$ decreases.
Because of this physical discrepancy, $f_\mathrm{duty}$ should be considered to give only very rough estimate.

\subsection{Stellar mass and dynamical mass}\label{subsec:stellar_mass}
We evaluate the stellar mass of host galaxies based on the empirical stellar ($M_*$)-to-halo ($M_\mathrm{halo}$) mass ratio (SHMR) from \citet{Behroozi_2019}.
The SHMR at $z\sim6$ has only evaluated up to $M_\mathrm{halo}^\mathrm{max}=10^{12}\,h^{-1}M_\odot$ and needs to be extended beyond this point to reach the observed halo mass, $M_\mathrm{halo}=5\times10^{12}\,h^{-1}M_\odot$.
However, the pivot mass ($M_\mathrm{halo}^\mathrm{max}$) is just where the slope of this relationship changes, and the slope at the high-mass regime tends to become shallower toward higher-$z$ \citep{Behroozi_2019}; therefore, this extension involves a large uncertainty.
Assuming conservatively here that the ratio above $M_\mathrm{halo}^\mathrm{max}$ does not change from $\mathrm{SHMR}\sim0.013$ at $M_\mathrm{halo}^\mathrm{max}$, the stellar mass is evaluated as $M_*=6.5_{-5.2}^{+9.6}\times10^{10}\,h^{-1}M_\odot$, where the error is estimated from the uncertainty of $M_\mathrm{halo}$ only and does not take into account the uncertainty of the SHMR extrapolation.

On the other hand, the dynamical mass evaluated by the [C \textsc{II}] $158\,\mathrm{\mu m}$ observation is often used as the surrogate of the stellar mass (e.g., \citealp{Willott2015_surrogate,Venemans2016,SHELLQs3_Izumi}) though the dynamical mass is essentially different from the stellar mass.
They estimated the dynamical mass with an assumption of a thin rotation disk with a diameter $D=1\--2\,h^{-1}\mathrm{kpc}$. \citep{Wang2013,Willott2015,Willott2017}.
Stellar mass was evaluated by the [C \textsc{II}] observations of seven SHELLQs quasars as $M_*=(0.91\--20)\times10^{10}\,h^{-1}M_\odot$ \citep{SHELLQs3_Izumi,SHELLQs8_Izumi}.
\citet{Neeleman2021} evaluated the mean stellar mass of 27 brighter quasars as $M_\mathrm{dyn}=(3.5\pm2.5)\times10^{10}\,h^{-1}M_\odot$, which is consistent with \citet{SHELLQs3_Izumi,SHELLQs8_Izumi}.

The stellar mass based on the clustering analysis with the SHMR is consistent with those independently measured from [C \textsc{II}] observation.
It is a bit surprising that they both agree, albeit with large uncertainties: in addition to the SHMR being uncertain at the massive end, there is no guarantee that the quasar hosts will have the same SHMR as the normal galaxy.
On the other hand, there is an implicit assumption that [C \textsc{II}] dynamical mass is a good proxy for the stellar mass of the bulge.

We compare the dynamical mass between the clustering analysis and the [C \textsc{II}] observation.
We estimate the dynamical mass at $D=1\--2\,h^{-1}\mathrm{kpc}$, where [C \textsc{II}] dynamical mass is estimated, from our $M_\mathrm{halo}$ measurement, as follows.
The virial radius $r_\mathrm{vir}$ can be estimated by using the spherical collapse model \citep{Barkana2001};
\begin{eqnarray}
    r_\mathrm{vir}=0.756\left(\frac{M_{\mathrm{halo}}}{10^8\,h^{-1}M_\odot}\right)^{1/3}\left[\frac{\Omega_{m}}{\Omega_m(z)}\frac{\Delta_c}{200}\right]^{-1/3}\nonumber\\
    \left(\frac{1+z}{10}\right)^{-1}\,h^{-1}\mathrm{kpc},
\end{eqnarray}
where 
\begin{equation}
    \Omega_m(z) = \frac{\Omega_m(1+z)^3}{E(z)^2}
\end{equation}
and 
\begin{equation}
    \Delta_c=18\pi^2+82[\Omega_m(z)-1]-39[\Omega_m(z)-1]^2
\end{equation}
is the overdensity at the halo collapse and we obtain $r_\mathrm{vir}=64\,h^{-1}\mathrm{kpc}$, which is much larger than the scale on which the [C \textsc{II}] dynamical mass is estimated.
When assuming a rotation-dominated disk with the flat rotation of DMHs, i.e., rotation velocity does not depend on the radius, the dynamical mass is estimated as $M_\mathrm{dyn}=(0.83\--1.7)\times10^{11}\,h^{-1}M_\odot$ at the scale of $1\--2\,h^{-1}\mathrm{kpc}$, which is larger than the [C \textsc{II}] dynamical mass, and in other words, the [C \textsc{II}] rotation velocity is much slower than the halo circular velocity.
This suggests either that the area where [C \textsc{II}] is detected is substantially central to the halo, where the rotation velocity has not yet reached the maximum halo circular velocity or the rotation of [C \textsc{II}] is independent of the rotation of the halo.
These considerations make it difficult to regard the dynamical mass obtained from [C \textsc{II}] as that of the entire system.
Nevertheless, the stellar masses of both estimates agree, which could be a coincidence due to the large uncertainties in both.

It should be noted that recent direct observation by JWST/NIRCam for host galaxies of a couple of SHELLQs quasars \citep{Ding2022} applies the SED (Spectral Energy Distribution) fitting to derive the stellar mass, which is comparable to that inferred from the halo mass measurement.
\citet{Marshall2023} also used JWST/NIRSpec to detect [O \textsc{III}] $\lambda5008$ emitting regions, which are more extended than the [C \textsc{II}], of the host galaxy, giving a slightly higher dynamical mass.
More observations should be made in the future to increase the number of direct measurements of the stellar mass of quasar host galaxies.
Also, we should keep in mind that the halo mass obtained in this study is still accompanied by a large error.

\section{Summary} \label{sec:summary}
We conduct a clustering analysis of 107 quasars at $z\sim6$, mainly composed of SHELLQs, which have increased the number density of quasars at $z\sim6$ by more than 30 times than SDSS.
This study is the first attempt to measure the DMH mass of quasars at $z\sim6$.
The main results are summarized below.
\begin{enumerate}
    \item 
    The quasars are spectroscopically identified in the HSC-SSP wide layer over $891\,\mathrm{deg^2}$.
    The completeness holds 70\% (80\%) over 85\% (77\%) of the entire survey regions.
    We evaluate the three types of auto-correlation function for our sample: projected correlation function $\omega_p(r_p)$, angular correlation function $\omega(\theta)$, and redshift space correlation function $\xi(s)$.
    We also evaluate the angular cross-correlation function between our quasar sample and LBG sample at $z\sim6$ in the HSC-SSP Deep layer.  
    The DMH mass at $z\sim6$ is evaluated as $5.0_{-4.0}^{+7.4}\times10^{12}\,h^{-1}M_\odot$ with the bias parameter, $b=20.8\pm8.7$ by the projected correlation function.
    The other three estimators agree with these values, though the uncertainties are large due to the small sample size.
    Using extended Press-Schechter theory, we find that the DMH with $5.0\times10^{12}\,h^{-1}M_\odot$ at $z\sim6$ will grow into $2.0_{-1.0}^{+2.2}\times10^{14}\,h^{-1}M_\odot$ at $z=0$, which is comparable to the rich clusters of galaxies today.
    
    \item 
    The DMH mass of quasars is found to be nearly constant $\sim10^{12.5}\,h^{-1}M_\odot$ throughout the cosmic epoch. 
    While there is broad agreement in previous studies that the quasar halo mass remains approximately constant up to $z \sim 4$, this study confirms, for the first time, that this trend continues up to $z \sim 6$.
    This means that there is a characteristic mass of DMH where quasars are always activated.
    As a result, quasars appear in the most massive halos at $z \sim 6$, but in less extreme halos thereafter.
    The mass of the quasar DMH is unlikely to exceed its upper limit of $10^{13}\,h^{-1}M_\odot$.
    This suggests that most quasars reside in DMHs with $M_\mathrm{halo}<10^{13}\,h^{-1}M_\odot$ across most of the cosmic history.
    This is consistent with the model by \citet{Fanidakis2013}.
    In the model, quasar activity, which is maintained by cold gas accretion into the central SMBH, is suppressed by radio-mode AGN feedback for massive halos larger than $10^{13}\,h^{-1}M_\odot$.
    If the quasar halo mass does not exceed $10^{13}\,h^{-1}M_\odot$ at any time, then such physics may be ubiquitously at work.

    \item 
    Our result that the bias parameter $b$ is as large as $b=20.8\pm8.7$ at $z \sim 6$ supports the ``maximum model" proposed by \citet{Hopkins2007}, which assumes that feedback is highly inefficient during $z \sim 4\--6$.
    Without the observational constraint at $z\sim5$, our result along with the previous observations can also be explained by a bias evolution model in which feedback is inefficient at $z\sim6$ but becomes progressively more efficient at $z\sim4$.
    
    \item 
    We estimate the quasar duty cycle $f_\mathrm{duty}$ at $z\sim6$.
    We find that the conventional definition of $f_\mathrm{duty}$ yields an unphysical result as the $f_\mathrm{duty}$ becomes greater than unity.
    We propose a new method to estimate the duty cycle in line with the observational result that DMH mass is nearly constant.
    We assume that DMHs with $12\leq\log(M_\mathrm{halo}/h^{-1}M_\odot)\leq13$ can host quasars.
    Using the number density of DMHs in the mass interval and the quasar luminosity function at $z\sim6$, we achieve $f_\mathrm{duty}=0.019\pm0.008$, which is consistent with that at $z\sim4$.
    
    \item 
    Assuming that the empirical SHMR at $z\sim6$ is constant at $M_\mathrm{halo}>10^{12}\,h^{-1}M_\odot$, the average stellar mass of quasar host galaxies at $z\sim6$ is evaluated from the observed DMH mass to be $M_{*}=6.5_{-5.2}^{+9.6}\times10^{10}\,h^{-1}M_\odot$, which is found to be consistent with those derived from [C \textsc{II}] observations.
\end{enumerate}

The clustering signal measurement utilizing quasar candidates at $z\sim5$ identified by HSC will soon be made, which can constrain the feedback models more rigidly.
More stellar mass measurements from [C \textsc{II}] observations by Atacama Large Millimeter Array (ALMA) will lead to a rigorous comparison with halo masses derived in the study and a constraint on the SHMR at the massive-end at $z\sim6$.
The high sensitivity of JWST will allow us to directly measure the host stellar mass and the dynamical mass of quasars and investigate the environment around quasars (e.g., overdensity).
In the future, more powerful surveys (e.g., Legacy Survey of Space and Time; \citealp{LSST2009}) will contribute to the larger quasar sample at high-$z$, which will lead to the detection of clearer clustering signals.
In addition, promising instruments, such as Nancy Roman Space Telescope and Euclid Satellite, will be expected to identify quasars at $z>7$.
These next-generation instruments will make the sample size deeper and larger, which will have a huge impact on our understanding of the co-evolution in the early universe.

\section*{Acknowledgements}
We appreciate the anonymous referee for constructive comments and suggestions.
We thank Tomo Takahashi, Takumi Shinohara, and Teruaki Suyama for fruitful discussions and Kazuhiro Shimasaku, and Yuichi Harikane for useful suggestions.

J.A. is supported by the International Graduate Program for Excellence in Earth-Space Science (IGPEES).
N.K. was supported by the Japan Society for the Promotion of Science through Grant-in-Aid for Scientific Research 21H04490.
Y.M. was supported by the Japan Society for the Promotion of Science KAKENHI grant No. JP17H04830 and No. 21H04494.
K.I. acknowledges support by grant PID2019-105510GB-C33 funded by MCIN/AEI/10.13039/501100011033 and ``Unit of excellence Mar\'ia de Maeztu 2020-2023'' awarded to ICCUB (CEX2019-000918-M).
M.O. is supported by the National Natural Science Foundation of China (12150410307). 

The Hyper Suprime-Cam (HSC) collaboration includes the astronomical communities of Japan and Taiwan, and Princeton University.  The HSC instrumentation and software were developed by the National Astronomical Observatory of Japan (NAOJ), the Kavli Institute for the Physics and Mathematics of the Universe (Kavli IPMU), the University of Tokyo, the High Energy Accelerator Research Organization (KEK), the Academia Sinica Institute for Astronomy and Astrophysics in Taiwan (ASIAA), and Princeton University.  Funding was contributed by the FIRST program from the Japanese Cabinet Office, the Ministry of Education, Culture, Sports, Science and Technology (MEXT), the Japan Society for the Promotion of Science (JSPS), Japan Science and Technology Agency  (JST), the Toray Science  Foundation, NAOJ, Kavli IPMU, KEK, ASIAA, and Princeton University. 
This paper is based [in part] on data collected at the Subaru Telescope and retrieved from the HSC data archive system, which is operated by Subaru Telescope and Astronomy Data Center (ADC) at NAOJ. Data analysis was in part carried out with the cooperation of Center for Computational Astrophysics (CfCA) at NAOJ.  We are honored and grateful for the opportunity of observing the Universe from Maunakea, which has the cultural, historical and natural significance in Hawaii.

This paper makes use of software developed for Vera C. Rubin Observatory. We thank the Rubin Observatory for making their code available as free software at \url{http://pipelines.lsst.io/}. 

The Pan-STARRS1 Surveys (PS1) and the PS1 public science archive have been made possible through contributions by the Institute for Astronomy, the University of Hawaii, the Pan-STARRS Project Office, the Max Planck Society and its participating institutes, the Max Planck Institute for Astronomy, Heidelberg, and the Max Planck Institute for Extraterrestrial Physics, Garching, The Johns Hopkins University, Durham University, the University of Edinburgh, the Queen’s University Belfast, the Harvard-Smithsonian Center for Astrophysics, the Las Cumbres Observatory Global Telescope Network Incorporated, the National Central University of Taiwan, the Space Telescope Science Institute, the National Aeronautics and Space Administration under grant No. NNX08AR22G issued through the Planetary Science Division of the NASA Science Mission Directorate, the National Science Foundation grant No. AST-1238877, the University of Maryland, Eotvos Lorand University (ELTE), the Los Alamos National Laboratory, and the Gordon and Betty Moore Foundation.

\vspace{5mm}
\facilities{Subaru}

\software{astropy \citep{2013A&A...558A..33A,2018AJ....156..123A},  
          CAMB \citep{Lewis2011},
          Corrfunc \citep{10.1007/978-981-13-7729-7_1,2020MNRAS.491.3022S},
          halomod \citep{Murray2013,Murray_2021},
          hscpipe \citep{Bosch2018}
          Numpy \citep{harris2020array},
          Matplotlib \citep{Hunter:2007},
          Pandas \citep{mckinney2010data},
          Scipy \citep{2020SciPy-NMeth}
          }

\bibliography{main}{}

\begin{thebibliography}{}
\expandafter\ifx\csname natexlab\endcsname\relax\def\natexlab#1{#1}\fi
\providecommand{\url}[1]{\href{#1}{#1}}
\providecommand{\dodoi}[1]{doi:~\href{http://doi.org/#1}{\nolinkurl{#1}}}
\providecommand{\doeprint}[1]{\href{http://ascl.net/#1}{\nolinkurl{http://ascl.net/#1}}}
\providecommand{\doarXiv}[1]{\href{https://arxiv.org/abs/#1}{\nolinkurl{https://arxiv.org/abs/#1}}}

\bibitem[{{Adelberger} {et~al.}(2006){Adelberger}, {Steidel}, {Kollmeier}, \&
  {Reddy}}]{Adelberger2006}
{Adelberger}, K.~L., {Steidel}, C.~C., {Kollmeier}, J.~A., \& {Reddy}, N.~A.
  2006, \apj, 637, 74, \dodoi{10.1086/497896}

\bibitem[{{Aihara} {et~al.}(2018{\natexlab{a}}){Aihara}, {Arimoto},
  {Armstrong}, {Arnouts}, {Bahcall}, {Bickerton}, {Bosch}, {Bundy}, {Capak},
  {Chan}, {Chiba}, {Coupon}, {Egami}, {Enoki}, {Finet}, {Fujimori}, {Fujimoto},
  {Furusawa}, {Furusawa}, {Goto}, {Goulding}, {Greco}, {Greene}, {Gunn},
  {Hamana}, {Harikane}, {Hashimoto}, {Hattori}, {Hayashi}, {Hayashi},
  {He{\l}miniak}, {Higuchi}, {Hikage}, {Ho}, {Hsieh}, {Huang}, {Huang},
  {Ikeda}, {Imanishi}, {Inoue}, {Iwasawa}, {Iwata}, {Jaelani}, {Jian},
  {Kamata}, {Karoji}, {Kashikawa}, {Katayama}, {Kawanomoto}, {Kayo}, {Koda},
  {Koike}, {Kojima}, {Komiyama}, {Konno}, {Koshida}, {Koyama}, {Kusakabe},
  {Leauthaud}, {Lee}, {Lin}, {Lin}, {Lupton}, {Mandelbaum}, {Matsuoka},
  {Medezinski}, {Mineo}, {Miyama}, {Miyatake}, {Miyazaki}, {Momose}, {More},
  {More}, {Moritani}, {Moriya}, {Morokuma}, {Mukae}, {Murata}, {Murayama},
  {Nagao}, {Nakata}, {Niida}, {Niikura}, {Nishizawa}, {Obuchi}, {Oguri},
  {Oishi}, {Okabe}, {Okamoto}, {Okura}, {Ono}, {Onodera}, {Onoue}, {Osato},
  {Ouchi}, {Price}, {Pyo}, {Sako}, {Sawicki}, {Shibuya}, {Shimasaku},
  {Shimono}, {Shirasaki}, {Silverman}, {Simet}, {Speagle}, {Spergel},
  {Strauss}, {Sugahara}, {Sugiyama}, {Suto}, {Suyu}, {Suzuki}, {Tait},
  {Takada}, {Takata}, {Tamura}, {Tanaka}, {Tanaka}, {Tanaka}, {Tanaka},
  {Terai}, {Terashima}, {Toba}, {Tominaga}, {Toshikawa}, {Turner}, {Uchida},
  {Uchiyama}, {Umetsu}, {Uraguchi}, {Urata}, {Usuda}, {Utsumi}, {Wang}, {Wang},
  {Wong}, {Yabe}, {Yamada}, {Yamanoi}, {Yasuda}, {Yeh}, {Yonehara}, \&
  {Yuma}}]{Aihara2018a}
{Aihara}, H., {Arimoto}, N., {Armstrong}, R., {et~al.} 2018{\natexlab{a}},
  \pasj, 70, S4, \dodoi{10.1093/pasj/psx066}

\bibitem[{{Aihara} {et~al.}(2018{\natexlab{b}}){Aihara}, {Armstrong},
  {Bickerton}, {Bosch}, {Coupon}, {Furusawa}, {Hayashi}, {Ikeda}, {Kamata},
  {Karoji}, {Kawanomoto}, {Koike}, {Komiyama}, {Lang}, {Lupton}, {Mineo},
  {Miyatake}, {Miyazaki}, {Morokuma}, {Obuchi}, {Oishi}, {Okura}, {Price},
  {Takata}, {Tanaka}, {Tanaka}, {Tanaka}, {Uchida}, {Uraguchi}, {Utsumi},
  {Wang}, {Yamada}, {Yamanoi}, {Yasuda}, {Arimoto}, {Chiba}, {Finet},
  {Fujimori}, {Fujimoto}, {Furusawa}, {Goto}, {Goulding}, {Gunn}, {Harikane},
  {Hattori}, {Hayashi}, {He{\l}miniak}, {Higuchi}, {Hikage}, {Ho}, {Hsieh},
  {Huang}, {Huang}, {Imanishi}, {Iwata}, {Jaelani}, {Jian}, {Kashikawa},
  {Katayama}, {Kojima}, {Konno}, {Koshida}, {Kusakabe}, {Leauthaud}, {Lee},
  {Lin}, {Lin}, {Mandelbaum}, {Matsuoka}, {Medezinski}, {Miyama}, {Momose},
  {More}, {More}, {Mukae}, {Murata}, {Murayama}, {Nagao}, {Nakata}, {Niida},
  {Niikura}, {Nishizawa}, {Oguri}, {Okabe}, {Ono}, {Onodera}, {Onoue}, {Ouchi},
  {Pyo}, {Shibuya}, {Shimasaku}, {Simet}, {Speagle}, {Spergel}, {Strauss},
  {Sugahara}, {Sugiyama}, {Suto}, {Suzuki}, {Tait}, {Takada}, {Terai}, {Toba},
  {Turner}, {Uchiyama}, {Umetsu}, {Urata}, {Usuda}, {Yeh}, \&
  {Yuma}}]{Aihara2018b}
{Aihara}, H., {Armstrong}, R., {Bickerton}, S., {et~al.} 2018{\natexlab{b}},
  \pasj, 70, S8, \dodoi{10.1093/pasj/psx081}

\bibitem[{Aihara {et~al.}(2019)Aihara, AlSayyad, Ando, Armstrong, Bosch, Egami,
  Furusawa, Furusawa, Goulding, Harikane, Hikage, Ho, Hsieh, Huang, Ikeda,
  Imanishi, Ito, Iwata, Jaelani, Kakuma, Kawana, Kikuta, Kobayashi, Koike,
  Komiyama, Li, Liang, Lin, Luo, Lupton, Lust, MacArthur, Matsuoka, Mineo,
  Miyatake, Miyazaki, More, Murata, Namiki, Nishizawa, Oguri, Okabe, Okamoto,
  Okura, Ono, Onodera, Onoue, Osato, Ouchi, Shibuya, Strauss, Sugiyama, Suto,
  Takada, Takagi, Takata, Takita, Tanaka, Terai, Toba, Uchiyama, Utsumi, Wang,
  Wang, \& Yamada}]{Aihara2019}
Aihara, H., AlSayyad, Y., Ando, M., {et~al.} 2019, Publications of the
  Astronomical Society of Japan, 71, \dodoi{10.1093/pasj/psz103}

\bibitem[{{Akiyama} {et~al.}(2018){Akiyama}, {He}, {Ikeda}, {Niida}, {Nagao},
  {Bosch}, {Coupon}, {Enoki}, {Imanishi}, {Kashikawa}, {Kawaguchi}, {Komiyama},
  {Lee}, {Matsuoka}, {Miyazaki}, {Nishizawa}, {Oguri}, {Ono}, {Onoue}, {Ouchi},
  {Schulze}, {Silverman}, {Tanaka}, {Tanaka}, {Terashima}, {Toba}, \&
  {Ueda}}]{Akiyama2018}
{Akiyama}, M., {He}, W., {Ikeda}, H., {et~al.} 2018, \pasj, 70, S34,
  \dodoi{10.1093/pasj/psx091}

\bibitem[{{Alam} {et~al.}(2015){Alam}, {Albareti}, {Allende Prieto}, {Anders},
  {Anderson}, {Anderton}, {Andrews}, {Armengaud}, {Aubourg}, {Bailey}, {Basu},
  {Bautista}, {Beaton}, {Beers}, {Bender}, {Berlind}, {Beutler}, {Bhardwaj},
  {Bird}, {Bizyaev}, {Blake}, {Blanton}, {Blomqvist}, {Bochanski}, {Bolton},
  {Bovy}, {Shelden Bradley}, {Brandt}, {Brauer}, {Brinkmann}, {Brown},
  {Brownstein}, {Burden}, {Burtin}, {Busca}, {Cai}, {Capozzi}, {Carnero
  Rosell}, {Carr}, {Carrera}, {Chambers}, {Chaplin}, {Chen}, {Chiappini},
  {Chojnowski}, {Chuang}, {Clerc}, {Comparat}, {Covey}, {Croft}, {Cuesta},
  {Cunha}, {da Costa}, {Da Rio}, {Davenport}, {Dawson}, {De Lee}, {Delubac},
  {Deshpande}, {Dhital}, {Dutra-Ferreira}, {Dwelly}, {Ealet}, {Ebelke},
  {Edmondson}, {Eisenstein}, {Ellsworth}, {Elsworth}, {Epstein}, {Eracleous},
  {Escoffier}, {Esposito}, {Evans}, {Fan}, {Fern{\'a}ndez-Alvar}, {Feuillet},
  {Filiz Ak}, {Finley}, {Finoguenov}, {Flaherty}, {Fleming}, {Font-Ribera},
  {Foster}, {Frinchaboy}, {Galbraith-Frew}, {Garc{\'\i}a},
  {Garc{\'\i}a-Hern{\'a}ndez}, {Garc{\'\i}a P{\'e}rez}, {Gaulme}, {Ge},
  {G{\'e}nova-Santos}, {Georgakakis}, {Ghezzi}, {Gillespie}, {Girardi},
  {Goddard}, {Gontcho}, {Gonz{\'a}lez Hern{\'a}ndez}, {Grebel}, {Green},
  {Grieb}, {Grieves}, {Gunn}, {Guo}, {Harding}, {Hasselquist}, {Hawley},
  {Hayden}, {Hearty}, {Hekker}, {Ho}, {Hogg}, {Holley-Bockelmann}, {Holtzman},
  {Honscheid}, {Huber}, {Huehnerhoff}, {Ivans}, {Jiang}, {Johnson},
  {Kinemuchi}, {Kirkby}, {Kitaura}, {Klaene}, {Knapp}, {Kneib}, {Koenig},
  {Lam}, {Lan}, {Lang}, {Laurent}, {Le Goff}, {Leauthaud}, {Lee}, {Lee},
  {Licquia}, {Liu}, {Long}, {L{\'o}pez-Corredoira}, {Lorenzo-Oliveira},
  {Lucatello}, {Lundgren}, {Lupton}, {Mack}, {Mahadevan}, {Maia}, {Majewski},
  {Malanushenko}, {Malanushenko}, {Manchado}, {Manera}, {Mao}, {Maraston},
  {Marchwinski}, {Margala}, {Martell}, {Martig}, {Masters}, {Mathur},
  {McBride}, {McGehee}, {McGreer}, {McMahon}, {M{\'e}nard}, {Menzel},
  {Merloni}, {M{\'e}sz{\'a}ros}, {Miller}, {Miralda-Escud{\'e}}, {Miyatake},
  {Montero-Dorta}, {More}, {Morganson}, {Morice-Atkinson}, {Morrison},
  {Mosser}, {Muna}, {Myers}, {Nandra}, {Newman}, {Neyrinck}, {Nguyen},
  {Nichol}, {Nidever}, {Noterdaeme}, {Nuza}, {O'Connell}, {O'Connell},
  {O'Connell}, {Ogando}, {Olmstead}, {Oravetz}, {Oravetz}, {Osumi}, {Owen},
  {Padgett}, {Padmanabhan}, {Paegert}, {Palanque-Delabrouille}, {Pan},
  {Parejko}, {P{\^a}ris}, {Park}, {Pattarakijwanich}, {Pellejero-Ibanez},
  {Pepper}, {Percival}, {P{\'e}rez-Fournon}, {P{\'e}rez-R{\`a}fols},
  {Petitjean}, {Pieri}, {Pinsonneault}, {Porto de Mello}, {Prada}, {Prakash},
  {Price-Whelan}, {Protopapas}, {Raddick}, {Rahman}, {Reid}, {Rich}, {Rix},
  {Robin}, {Rockosi}, {Rodrigues}, {Rodr{\'\i}guez-Torres}, {Roe}, {Ross},
  {Ross}, {Rossi}, {Ruan}, {Rubi{\~n}o-Mart{\'\i}n}, {Rykoff},
  {Salazar-Albornoz}, {Salvato}, {Samushia}, {S{\'a}nchez}, {Santiago},
  {Sayres}, {Schiavon}, {Schlegel}, {Schmidt}, {Schneider}, {Schultheis},
  {Schwope}, {Sc{\'o}ccola}, {Scott}, {Sellgren}, {Seo}, {Serenelli}, {Shane},
  {Shen}, {Shetrone}, {Shu}, {Silva Aguirre}, {Sivarani}, {Skrutskie},
  {Slosar}, {Smith}, {Sobreira}, {Souto}, {Stassun}, {Steinmetz}, {Stello},
  {Strauss}, {Streblyanska}, {Suzuki}, {Swanson}, {Tan}, {Tayar}, {Terrien},
  {Thakar}, {Thomas}, {Thomas}, {Thompson}, {Tinker}, {Tojeiro}, {Troup},
  {Vargas-Maga{\~n}a}, {Vazquez}, {Verde}, {Viel}, {Vogt}, {Wake}, {Wang},
  {Weaver}, {Weinberg}, {Weiner}, {White}, {Wilson}, {Wisniewski},
  {Wood-Vasey}, {Ye`che}, {York}, {Zakamska}, {Zamora}, {Zasowski}, {Zehavi},
  {Zhao}, {Zheng}, {Zhou}, {Zhou}, {Zou}, \& {Zhu}}]{Alam2015}
{Alam}, S., {Albareti}, F.~D., {Allende Prieto}, C., {et~al.} 2015, \apjs, 219,
  12, \dodoi{10.1088/0067-0049/219/1/12}

\bibitem[{{Annis} {et~al.}(2014){Annis}, {Soares-Santos}, {Strauss}, {Becker},
  {Dodelson}, {Fan}, {Gunn}, {Hao}, {Ivezi{\'c}}, {Jester}, {Jiang},
  {Johnston}, {Kubo}, {Lampeitl}, {Lin}, {Lupton}, {Miknaitis}, {Seo}, {Simet},
  \& {Yanny}}]{Annis2014}
{Annis}, J., {Soares-Santos}, M., {Strauss}, M.~A., {et~al.} 2014, \apj, 794,
  120, \dodoi{10.1088/0004-637X/794/2/120}

\bibitem[{Antonucci(1993)}]{Antonucci1993}
Antonucci, R. 1993, Annual Review of Astronomy and Astrophysics, 31, 473,
  \dodoi{10.1146/annurev.aa.31.090193.002353}

\bibitem[{{Astropy Collaboration} {et~al.}(2013){Astropy Collaboration},
  {Robitaille}, {Tollerud}, {Greenfield}, {Droettboom}, {Bray}, {Aldcroft},
  {Davis}, {Ginsburg}, {Price-Whelan}, {Kerzendorf}, {Conley}, {Crighton},
  {Barbary}, {Muna}, {Ferguson}, {Grollier}, {Parikh}, {Nair}, {Unther},
  {Deil}, {Woillez}, {Conseil}, {Kramer}, {Turner}, {Singer}, {Fox}, {Weaver},
  {Zabalza}, {Edwards}, {Azalee Bostroem}, {Burke}, {Casey}, {Crawford},
  {Dencheva}, {Ely}, {Jenness}, {Labrie}, {Lim}, {Pierfederici}, {Pontzen},
  {Ptak}, {Refsdal}, {Servillat}, \& {Streicher}}]{2013A&A...558A..33A}
{Astropy Collaboration}, {Robitaille}, T.~P., {Tollerud}, E.~J., {et~al.} 2013,
  \aap, 558, A33, \dodoi{10.1051/0004-6361/201322068}

\bibitem[{{Astropy Collaboration} {et~al.}(2018){Astropy Collaboration},
  {Price-Whelan}, {Sip{\H{o}}cz}, {G{\"u}nther}, {Lim}, {Crawford}, {Conseil},
  {Shupe}, {Craig}, {Dencheva}, {Ginsburg}, {VanderPlas}, {Bradley},
  {P{\'e}rez-Su{\'a}rez}, {de Val-Borro}, {Aldcroft}, {Cruz}, {Robitaille},
  {Tollerud}, {Ardelean}, {Babej}, {Bach}, {Bachetti}, {Bakanov}, {Bamford},
  {Barentsen}, {Barmby}, {Baumbach}, {Berry}, {Biscani}, {Boquien}, {Bostroem},
  {Bouma}, {Brammer}, {Bray}, {Breytenbach}, {Buddelmeijer}, {Burke},
  {Calderone}, {Cano Rodr{\'\i}guez}, {Cara}, {Cardoso}, {Cheedella}, {Copin},
  {Corrales}, {Crichton}, {D'Avella}, {Deil}, {Depagne}, {Dietrich}, {Donath},
  {Droettboom}, {Earl}, {Erben}, {Fabbro}, {Ferreira}, {Finethy}, {Fox},
  {Garrison}, {Gibbons}, {Goldstein}, {Gommers}, {Greco}, {Greenfield},
  {Groener}, {Grollier}, {Hagen}, {Hirst}, {Homeier}, {Horton}, {Hosseinzadeh},
  {Hu}, {Hunkeler}, {Ivezi{\'c}}, {Jain}, {Jenness}, {Kanarek}, {Kendrew},
  {Kern}, {Kerzendorf}, {Khvalko}, {King}, {Kirkby}, {Kulkarni}, {Kumar},
  {Lee}, {Lenz}, {Littlefair}, {Ma}, {Macleod}, {Mastropietro}, {McCully},
  {Montagnac}, {Morris}, {Mueller}, {Mumford}, {Muna}, {Murphy}, {Nelson},
  {Nguyen}, {Ninan}, {N{\"o}the}, {Ogaz}, {Oh}, {Parejko}, {Parley}, {Pascual},
  {Patil}, {Patil}, {Plunkett}, {Prochaska}, {Rastogi}, {Reddy Janga},
  {Sabater}, {Sakurikar}, {Seifert}, {Sherbert}, {Sherwood-Taylor}, {Shih},
  {Sick}, {Silbiger}, {Singanamalla}, {Singer}, {Sladen}, {Sooley},
  {Sornarajah}, {Streicher}, {Teuben}, {Thomas}, {Tremblay}, {Turner},
  {Terr{\'o}n}, {van Kerkwijk}, {de la Vega}, {Watkins}, {Weaver}, {Whitmore},
  {Woillez}, {Zabalza}, \& {Astropy Contributors}}]{2018AJ....156..123A}
{Astropy Collaboration}, {Price-Whelan}, A.~M., {Sip{\H{o}}cz}, B.~M., {et~al.}
  2018, \aj, 156, 123, \dodoi{10.3847/1538-3881/aabc4f}

\bibitem[{{Ba{\~{n}}ados} {et~al.}(2013){Ba{\~{n}}ados}, {Venemans}, {Walter},
  {Kurk}, {Overzier}, \& {Ouchi}}]{Banados2013}
{Ba{\~{n}}ados}, E., {Venemans}, B., {Walter}, F., {et~al.} 2013, \apj, 773,
  178, \dodoi{10.1088/0004-637X/773/2/178}

\bibitem[{{Ba{\~{n}}ados} {et~al.}(2014){Ba{\~{n}}ados}, {Venemans},
  {Morganson}, {Decarli}, {Walter}, {Chambers}, {Rix}, {Farina}, {Fan},
  {Jiang}, {McGreer}, {De Rosa}, {Simcoe}, {Wei{\ss}}, {Price}, {Morgan},
  {Burgett}, {Greiner}, {Kaiser}, {Kudritzki}, {Magnier}, {Metcalfe}, {Stubbs},
  {Sweeney}, {Tonry}, {Wainscoat}, \& {Waters}}]{Banados2014}
{Ba{\~{n}}ados}, E., {Venemans}, B.~P., {Morganson}, E., {et~al.} 2014, \aj,
  148, 14, \dodoi{10.1088/0004-6256/148/1/14}

\bibitem[{{Ba{\~{n}}ados} {et~al.}(2016){Ba{\~{n}}ados}, Venemans, Decarli,
  Farina, Mazzucchelli, Walter, Fan, Stern, Schlafly, Chambers, Rix, Jiang,
  McGreer, Simcoe, Wang, Yang, Morganson, Rosa, Greiner, Balokovi{\'{c}},
  Burgett, Cooper, Draper, Flewelling, Hodapp, Jun, Kaiser, Kudritzki, Magnier,
  Metcalfe, Miller, Schindler, Tonry, Wainscoat, Waters, \&
  Yang}]{Banados_2016}
{Ba{\~{n}}ados}, E., Venemans, B.~P., Decarli, R., {et~al.} 2016, The
  Astrophysical Journal Supplement Series, 227, 11,
  \dodoi{10.3847/0067-0049/227/1/11}

\bibitem[{{Ba{\~n}ados} {et~al.}(2018){Ba{\~n}ados}, {Venemans},
  {Mazzucchelli}, {Farina}, {Walter}, {Wang}, {Decarli}, {Stern}, {Fan},
  {Davies}, {Hennawi}, {Simcoe}, {Turner}, {Rix}, {Yang}, {Kelson}, {Rudie}, \&
  {Winters}}]{Banados2018}
{Ba{\~n}ados}, E., {Venemans}, B.~P., {Mazzucchelli}, C., {et~al.} 2018, \nat,
  553, 473, \dodoi{10.1038/nature25180}

\bibitem[{{Ba{\~{n}}ados} {et~al.}(2023){Ba{\~{n}}ados}, {Schindler},
  {Venemans}, {Connor}, {Decarli}, {Farina}, {Mazzucchelli}, {Meyer}, {Stern},
  {Walter}, {Fan}, {Hennawi}, {Khusanova}, {Morrell}, {Nanni}, {Noirot},
  {Pensabene}, {Rix}, {Simon}, {Verdoes Kleijn}, {Xie}, {Yang}, \&
  {Connor}}]{Banados2023}
{Ba{\~{n}}ados}, E., {Schindler}, J.-T., {Venemans}, B.~P., {et~al.} 2023,
  \apjs, 265, 29, \dodoi{10.3847/1538-4365/acb3c7}

\bibitem[{{Barkana} \& {Loeb}(2001)}]{Barkana2001}
{Barkana}, R., \& {Loeb}, A. 2001, \physrep, 349, 125,
  \dodoi{10.1016/S0370-1573(01)00019-9}

\bibitem[{Behroozi {et~al.}(2019)Behroozi, Wechsler, Hearin, \&
  Conroy}]{Behroozi_2019}
Behroozi, P., Wechsler, R.~H., Hearin, A.~P., \& Conroy, C. 2019, Monthly
  Notices of the Royal Astronomical Society, 488, 3143,
  \dodoi{10.1093/mnras/stz1182}

\bibitem[{{Bertin}(2011)}]{Bertin2011}
{Bertin}, E. 2011, in Astronomical Society of the Pacific Conference Series,
  Vol. 442, Astronomical Data Analysis Software and Systems XX, ed. I.~N.
  {Evans}, A.~{Accomazzi}, D.~J. {Mink}, \& A.~H. {Rots}, 435

\bibitem[{{Bertin} \& {Arnouts}(1996)}]{Bertin1996}
{Bertin}, E., \& {Arnouts}, S. 1996, \aaps, 117, 393,
  \dodoi{10.1051/aas:1996164}

\bibitem[{{Bhattacharya} {et~al.}(2013){Bhattacharya}, {Habib}, {Heitmann}, \&
  {Vikhlinin}}]{Bhattacharya2013}
{Bhattacharya}, S., {Habib}, S., {Heitmann}, K., \& {Vikhlinin}, A. 2013, \apj,
  766, 32, \dodoi{10.1088/0004-637X/766/1/32}

\bibitem[{{Bond} {et~al.}(1991){Bond}, {Cole}, {Efstathiou}, \&
  {Kaiser}}]{Bond1991}
{Bond}, J.~R., {Cole}, S., {Efstathiou}, G., \& {Kaiser}, N. 1991, \apj, 379,
  440, \dodoi{10.1086/170520}

\bibitem[{{Bosch} {et~al.}(2018){Bosch}, {Armstrong}, {Bickerton}, {Furusawa},
  {Ikeda}, {Koike}, {Lupton}, {Mineo}, {Price}, {Takata}, {Tanaka}, {Yasuda},
  {AlSayyad}, {Becker}, {Coulton}, {Coupon}, {Garmilla}, {Huang}, {Krughoff},
  {Lang}, {Leauthaud}, {Lim}, {Lust}, {MacArthur}, {Mandelbaum}, {Miyatake},
  {Miyazaki}, {Murata}, {More}, {Okura}, {Owen}, {Swinbank}, {Strauss},
  {Yamada}, \& {Yamanoi}}]{Bosch2018}
{Bosch}, J., {Armstrong}, R., {Bickerton}, S., {et~al.} 2018, \pasj, 70, S5,
  \dodoi{10.1093/pasj/psx080}

\bibitem[{{Bower}(1991)}]{Bower1991}
{Bower}, R.~G. 1991, \mnras, 248, 332, \dodoi{10.1093/mnras/248.2.332}

\bibitem[{{Carroll} {et~al.}(1992){Carroll}, {Press}, \&
  {Turner}}]{Carroll1992}
{Carroll}, S.~M., {Press}, W.~H., \& {Turner}, E.~L. 1992, \araa, 30, 499,
  \dodoi{10.1146/annurev.aa.30.090192.002435}

\bibitem[{{Cepa} {et~al.}(2000){Cepa}, {Aguiar}, {Escalera},
  {Gonzalez-Serrano}, {Joven-Alvarez}, {Peraza}, {Rasilla}, {Rodriguez-Ramos},
  {Gonzalez}, {Cobos Duenas}, {Sanchez}, {Tejada}, {Bland-Hawthorn},
  {Militello}, \& {Rosa}}]{Cepa2000}
{Cepa}, J., {Aguiar}, M., {Escalera}, V.~G., {et~al.} 2000, in Society of
  Photo-Optical Instrumentation Engineers (SPIE) Conference Series, Vol. 4008,
  Optical and IR Telescope Instrumentation and Detectors, ed. M.~{Iye} \& A.~F.
  {Moorwood}, 623--631, \dodoi{10.1117/12.395520}

\bibitem[{{Chambers} {et~al.}(2016){Chambers}, {Magnier}, {Metcalfe},
  {Flewelling}, {Huber}, {Waters}, {Denneau}, {Draper}, {Farrow}, {Finkbeiner},
  {Holmberg}, {Koppenhoefer}, {Price}, {Rest}, {Saglia}, {Schlafly}, {Smartt},
  {Sweeney}, {Wainscoat}, {Burgett}, {Chastel}, {Grav}, {Heasley}, {Hodapp},
  {Jedicke}, {Kaiser}, {Kudritzki}, {Luppino}, {Lupton}, {Monet}, {Morgan},
  {Onaka}, {Shiao}, {Stubbs}, {Tonry}, {White}, {Ba{\~n}ados}, {Bell},
  {Bender}, {Bernard}, {Boegner}, {Boffi}, {Botticella}, {Calamida},
  {Casertano}, {Chen}, {Chen}, {Cole}, {Deacon}, {Frenk}, {Fitzsimmons},
  {Gezari}, {Gibbs}, {Goessl}, {Goggia}, {Gourgue}, {Goldman}, {Grant},
  {Grebel}, {Hambly}, {Hasinger}, {Heavens}, {Heckman}, {Henderson}, {Henning},
  {Holman}, {Hopp}, {Ip}, {Isani}, {Jackson}, {Keyes}, {Koekemoer}, {Kotak},
  {Le}, {Liska}, {Long}, {Lucey}, {Liu}, {Martin}, {Masci}, {McLean}, {Mindel},
  {Misra}, {Morganson}, {Murphy}, {Obaika}, {Narayan}, {Nieto-Santisteban},
  {Norberg}, {Peacock}, {Pier}, {Postman}, {Primak}, {Rae}, {Rai}, {Riess},
  {Riffeser}, {Rix}, {R{\"o}ser}, {Russel}, {Rutz}, {Schilbach}, {Schultz},
  {Scolnic}, {Strolger}, {Szalay}, {Seitz}, {Small}, {Smith}, {Soderblom},
  {Taylor}, {Thomson}, {Taylor}, {Thakar}, {Thiel}, {Thilker}, {Unger},
  {Urata}, {Valenti}, {Wagner}, {Walder}, {Walter}, {Watters}, {Werner},
  {Wood-Vasey}, \& {Wyse}}]{Chambers2016}
{Chambers}, K.~C., {Magnier}, E.~A., {Metcalfe}, N., {et~al.} 2016, arXiv
  e-prints, arXiv:1612.05560, \dodoi{10.48550/arXiv.1612.05560}

\bibitem[{{Chen} {et~al.}(2022){Chen}, {Eilers}, {Bosman}, {Gnedin}, {Fan},
  {Wang}, {Yang}, {D'Odorico}, {Becker}, {Bischetti}, {Mazzucchelli},
  {Mesinger}, \& {Pallottini}}]{Chen2022}
{Chen}, H., {Eilers}, A.-C., {Bosman}, S. E.~I., {et~al.} 2022, \apj, 931, 29,
  \dodoi{10.3847/1538-4357/ac658d}

\bibitem[{{Cole} \& {Kaiser}(1989)}]{Cole1989}
{Cole}, S., \& {Kaiser}, N. 1989, \mnras, 237, 1127,
  \dodoi{10.1093/mnras/237.4.1127}

\bibitem[{Cooke {et~al.}(2006)Cooke, Wolfe, Gawiser, \& Prochaska}]{Cooke_2006}
Cooke, J., Wolfe, A.~M., Gawiser, E., \& Prochaska, J.~X. 2006, The
  Astrophysical Journal, 652, 994, \dodoi{10.1086/507476}

\bibitem[{Croom \& Shanks(1999)}]{Croom_1999}
Croom, S.~M., \& Shanks, T. 1999, Monthly Notices of the Royal Astronomical
  Society, 303, 411, \dodoi{10.1046/j.1365-8711.1999.02232.x}

\bibitem[{{Croom} {et~al.}(2005){Croom}, {Boyle}, {Shanks}, {Smith}, {Miller},
  {Outram}, {Loaring}, {Hoyle}, \& {da {\^A}ngela}}]{Croom2005}
{Croom}, S.~M., {Boyle}, B.~J., {Shanks}, T., {et~al.} 2005, \mnras, 356, 415,
  \dodoi{10.1111/j.1365-2966.2004.08379.x}

\bibitem[{{Davis} \& {Peebles}(1983)}]{Davis1983}
{Davis}, M., \& {Peebles}, P.~J.~E. 1983, \apj, 267, 465,
  \dodoi{10.1086/160884}

\bibitem[{{Dekel} \& {Lahav}(1999)}]{Dekel1999}
{Dekel}, A., \& {Lahav}, O. 1999, \apj, 520, 24, \dodoi{10.1086/307428}

\bibitem[{{Di Matteo} {et~al.}(2012){Di Matteo}, {Khandai}, {DeGraf}, {Feng},
  {Croft}, {Lopez}, \& {Springel}}]{Dimatteo2012}
{Di Matteo}, T., {Khandai}, N., {DeGraf}, C., {et~al.} 2012, \apjl, 745, L29,
  \dodoi{10.1088/2041-8205/745/2/L29}

\bibitem[{{Ding} {et~al.}(2022){Ding}, {Onoue}, {Silverman}, {Matsuoka},
  {Izumi}, {Strauss}, {Jahnke}, {Taufik Andika}, {Aoki}, {Baba}, {Bieri},
  {Bosman}, {Eilers}, {Fujimoto}, {Habouzit}, {Haiman}, {Imanishi}, {Inayoshi},
  {Iwasawa}, {Kashikawa}, {Kawaguchi}, {Kohno}, {Lee}, {Li}, {Lupi}, {Lyu},
  {Nagao}, {Overzier}, {Phillips}, {Schindler}, {Schramm}, {Shimasaku}, {Toba},
  {Trakhtenbrot}, {Trebitsch}, {Treu}, {Umehata}, {Venemans}, {Vestergaard},
  {Volonteri}, {Walter}, {Wang}, \& {Yang}}]{Ding2022}
{Ding}, X., {Onoue}, M., {Silverman}, J.~D., {et~al.} 2022, arXiv e-prints,
  arXiv:2211.14329.
\newblock \doarXiv{2211.14329}

\bibitem[{Eftekharzadeh {et~al.}(2015)Eftekharzadeh, Myers, White, Weinberg,
  Schneider, Shen, Font-Ribera, Ross, Paris, \&
  Streblyanska}]{Eftekharzadeh_2015}
Eftekharzadeh, S., Myers, A.~D., White, M., {et~al.} 2015, Monthly Notices of
  the Royal Astronomical Society, 453, 2779, \dodoi{10.1093/mnras/stv1763}

\bibitem[{{Eilers} {et~al.}(2021){Eilers}, {Hennawi}, {Davies}, \&
  {Simcoe}}]{Eilers2021}
{Eilers}, A.-C., {Hennawi}, J.~F., {Davies}, F.~B., \& {Simcoe}, R.~A. 2021,
  \apj, 917, 38, \dodoi{10.3847/1538-4357/ac0a76}

\bibitem[{Fan {et~al.}(2022)Fan, Banados, \& Simcoe}]{Fan2022}
Fan, X., Banados, E., \& Simcoe, R.~A. 2022, Quasars and the Intergalactic
  Medium at Cosmic Dawn,  arXiv, \dodoi{10.48550/ARXIV.2212.06907}

\bibitem[{{Fan} {et~al.}(2004){Fan}, {Hennawi}, {Richards}, {Strauss},
  {Schneider}, {Donley}, {Young}, {Annis}, {Lin}, {Lampeitl}, {Lupton}, {Gunn},
  {Knapp}, {Brandt}, {Anderson}, {Bahcall}, {Brinkmann}, {Brunner}, {Fukugita},
  {Szalay}, {Szokoly}, \& {York}}]{Fan2004}
{Fan}, X., {Hennawi}, J.~F., {Richards}, G.~T., {et~al.} 2004, \aj, 128, 515,
  \dodoi{10.1086/422434}

\bibitem[{{Fanidakis} {et~al.}(2013){Fanidakis}, {Macci{\`o}}, {Baugh},
  {Lacey}, \& {Frenk}}]{Fanidakis2013}
{Fanidakis}, N., {Macci{\`o}}, A.~V., {Baugh}, C.~M., {Lacey}, C.~G., \&
  {Frenk}, C.~S. 2013, \mnras, 436, 315, \dodoi{10.1093/mnras/stt1567}

\bibitem[{{Ferrarese}(2002)}]{Laura2002}
{Ferrarese}, L. 2002, \apj, 578, 90, \dodoi{10.1086/342308}

\bibitem[{Furusawa {et~al.}(2017)Furusawa, Koike, Takata, Okura, Miyatake,
  Lupton, Bickerton, Price, Bosch, Yasuda, Mineo, Yamada, Miyazaki, Nakata,
  Koshida, Komiyama, Utsumi, Kawanomoto, Jeschke, Noumaru, Schubert, Iwata,
  Finet, Fujiyoshi, Tajitsu, Terai, \& Lee}]{Furusawa_2018}
Furusawa, H., Koike, M., Takata, T., {et~al.} 2017, Publications of the
  Astronomical Society of Japan, 70, \dodoi{10.1093/pasj/psx079}

\bibitem[{{Granato} {et~al.}(2004){Granato}, {De Zotti}, {Silva}, {Bressan}, \&
  {Danese}}]{Granato2004}
{Granato}, G.~L., {De Zotti}, G., {Silva}, L., {Bressan}, A., \& {Danese}, L.
  2004, \apj, 600, 580, \dodoi{10.1086/379875}

\bibitem[{{Greiner} {et~al.}(2021){Greiner}, {Bolmer}, {Yates}, {Habouzit},
  {Ba{\~n}ados}, {Afonso}, \& {Schady}}]{Greiner2021}
{Greiner}, J., {Bolmer}, J., {Yates}, R.~M., {et~al.} 2021, \aap, 654, A79,
  \dodoi{10.1051/0004-6361/202140790}

\bibitem[{{Groth} \& {Peebles}(1977)}]{Groth1977}
{Groth}, E.~J., \& {Peebles}, P.~J.~E. 1977, \apj, 217, 385,
  \dodoi{10.1086/155588}

\bibitem[{{Haiman} \& {Hui}(2001)}]{Haiman2001}
{Haiman}, Z., \& {Hui}, L. 2001, \apj, 547, 27, \dodoi{10.1086/318330}

\bibitem[{Harikane {et~al.}(2022)Harikane, Ono, Ouchi, Liu, Sawicki, Shibuya,
  Behroozi, He, Shimasaku, Arnouts, Coupon, Fujimoto, Gwyn, Huang, Inoue,
  Kashikawa, Komiyama, Matsuoka, \& Willott}]{Harikane_2022}
Harikane, Y., Ono, Y., Ouchi, M., {et~al.} 2022, The Astrophysical Journal
  Supplement Series, 259, 20, \dodoi{10.3847/1538-4365/ac3dfc}

\bibitem[{Harris {et~al.}(2020)Harris, Millman, van~der Walt, Gommers,
  Virtanen, Cournapeau, Wieser, Taylor, Berg, Smith, Kern, Picus, Hoyer, van
  Kerkwijk, Brett, Haldane, del R{\'{i}}o, Wiebe, Peterson,
  G{\'{e}}rard-Marchant, Sheppard, Reddy, Weckesser, Abbasi, Gohlke, \&
  Oliphant}]{harris2020array}
Harris, C.~R., Millman, K.~J., van~der Walt, S.~J., {et~al.} 2020, Nature, 585,
  357, \dodoi{10.1038/s41586-020-2649-2}

\bibitem[{{He} {et~al.}(2018){He}, {Akiyama}, {Bosch}, {Enoki}, {Harikane},
  {Ikeda}, {Kashikawa}, {Kawaguchi}, {Komiyama}, {Lee}, {Matsuoka}, {Miyazaki},
  {Nagao}, {Nagashima}, {Niida}, {Nishizawa}, {Oguri}, {Onoue}, {Oogi},
  {Ouchi}, {Schulze}, {Shirasaki}, {Silverman}, {Tanaka}, {Tanaka}, {Toba},
  {Uchiyama}, \& {Yamashita}}]{He2018}
{He}, W., {Akiyama}, M., {Bosch}, J., {et~al.} 2018, \pasj, 70, S33,
  \dodoi{10.1093/pasj/psx129}

\bibitem[{{Heckman} \& {Best}(2014)}]{Heckman2014}
{Heckman}, T.~M., \& {Best}, P.~N. 2014, \araa, 52, 589,
  \dodoi{10.1146/annurev-astro-081913-035722}

\bibitem[{{Hickox} {et~al.}(2011){Hickox}, {Myers}, {Brodwin}, {Alexander},
  {Forman}, {Jones}, {Murray}, {Brown}, {Cool}, {Kochanek}, {Dey}, {Jannuzi},
  {Eisenstein}, {Assef}, {Eisenhardt}, {Gorjian}, {Stern}, {Le Floc'h},
  {Caldwell}, {Goulding}, \& {Mullaney}}]{Hickox2011}
{Hickox}, R.~C., {Myers}, A.~D., {Brodwin}, M., {et~al.} 2011, \apj, 731, 117,
  \dodoi{10.1088/0004-637X/731/2/117}

\bibitem[{{Hopkins} {et~al.}(2007){Hopkins}, {Lidz}, {Hernquist}, {Coil},
  {Myers}, {Cox}, \& {Spergel}}]{Hopkins2007}
{Hopkins}, P.~F., {Lidz}, A., {Hernquist}, L., {et~al.} 2007, \apj, 662, 110,
  \dodoi{10.1086/517512}

\bibitem[{Hunter(2007)}]{Hunter:2007}
Hunter, J.~D. 2007, Computing in Science \& Engineering, 9, 90,
  \dodoi{10.1109/MCSE.2007.55}

\bibitem[{{Inayoshi} {et~al.}(2022){Inayoshi}, {Nakatani}, {Toyouchi},
  {Hosokawa}, {Kuiper}, \& {Onoue}}]{Inayoshi2022}
{Inayoshi}, K., {Nakatani}, R., {Toyouchi}, D., {et~al.} 2022, \apj, 927, 237,
  \dodoi{10.3847/1538-4357/ac4751}

\bibitem[{{Ivezi{\'c}} {et~al.}(2019){Ivezi{\'c}}, {Kahn}, {Tyson}, {Abel},
  {Acosta}, {Allsman}, {Alonso}, {AlSayyad}, {Anderson}, {Andrew}, \&
  et~al.}]{Ivezi2019}
{Ivezi{\'c}}, {\v Z}., {Kahn}, S.~M., {Tyson}, J.~A., {et~al.} 2019, \apj, 873,
  111, \dodoi{10.3847/1538-4357/ab042c}

\bibitem[{{Izumi} {et~al.}(2018){Izumi}, {Onoue}, {Shirakata}, {Nagao},
  {Kohno}, {Matsuoka}, {Imanishi}, {Strauss}, {Kashikawa}, {Schulze},
  {Silverman}, {Fujimoto}, {Harikane}, {Toba}, {Umehata}, {Nakanishi},
  {Greene}, {Tamura}, {Taniguchi}, {Yamaguchi}, {Goto}, {Hashimoto},
  {Ikarashi}, {Iono}, {Iwasawa}, {Lee}, {Makiya}, {Minezaki}, \&
  {Tang}}]{SHELLQs3_Izumi}
{Izumi}, T., {Onoue}, M., {Shirakata}, H., {et~al.} 2018, \pasj, 70, 36,
  \dodoi{10.1093/pasj/psy026}

\bibitem[{{Izumi} {et~al.}(2019){Izumi}, {Onoue}, {Matsuoka}, {Nagao},
  {Strauss}, {Imanishi}, {Kashikawa}, {Fujimoto}, {Kohno}, {Toba}, {Umehata},
  {Goto}, {Ueda}, {Shirakata}, {Silverman}, {Greene}, {Harikane}, {Hashimoto},
  {Ikarashi}, {Iono}, {Iwasawa}, {Lee}, {Minezaki}, {Nakanishi}, {Tamura},
  {Tang}, \& {Taniguchi}}]{SHELLQs8_Izumi}
{Izumi}, T., {Onoue}, M., {Matsuoka}, Y., {et~al.} 2019, \pasj, 71, 111,
  \dodoi{10.1093/pasj/psz096}

\bibitem[{{Jiang} {et~al.}(2008){Jiang}, {Fan}, {Annis}, {Becker}, {White},
  {Chiu}, {Lin}, {Lupton}, {Richards}, {Strauss}, {Jester}, \&
  {Schneider}}]{Jiang2008}
{Jiang}, L., {Fan}, X., {Annis}, J., {et~al.} 2008, \aj, 135, 1057,
  \dodoi{10.1088/0004-6256/135/3/1057}

\bibitem[{{Jiang} {et~al.}(2009){Jiang}, {Fan}, {Bian}, {Annis}, {Chiu},
  {Jester}, {Lin}, {Lupton}, {Richards}, {Strauss}, {Malanushenko},
  {Malanushenko}, \& {Schneider}}]{Jiang2009}
{Jiang}, L., {Fan}, X., {Bian}, F., {et~al.} 2009, \aj, 138, 305,
  \dodoi{10.1088/0004-6256/138/1/305}

\bibitem[{{Jiang} {et~al.}(2014){Jiang}, {Fan}, {Bian}, {McGreer}, {Strauss},
  {Annis}, {Buck}, {Green}, {Hodge}, {Myers}, {Rafiee}, \&
  {Richards}}]{Jiang2014}
---. 2014, \apjs, 213, 12, \dodoi{10.1088/0067-0049/213/1/12}

\bibitem[{{Jiang} {et~al.}(2016){Jiang}, {McGreer}, {Fan}, {Strauss},
  {Ba{\~n}ados}, {Becker}, {Bian}, {Farnsworth}, {Shen}, {Wang}, {Wang},
  {Wang}, {White}, {Wu}, {Wu}, {Yang}, \& {Yang}}]{Jiang2016}
{Jiang}, L., {McGreer}, I.~D., {Fan}, X., {et~al.} 2016, \apj, 833, 222,
  \dodoi{10.3847/1538-4357/833/2/222}

\bibitem[{{Jing}(1998)}]{Jing1998}
{Jing}, Y.~P. 1998, \apjl, 503, L9, \dodoi{10.1086/311530}

\bibitem[{{Juri{\'c}} {et~al.}(2017){Juri{\'c}}, {Kantor}, {Lim}, {Lupton},
  {Dubois-Felsmann}, {Jenness}, {Axelrod}, {Aleksi{\'c}}, {Allsman},
  {AlSayyad}, {Alt}, {Armstrong}, {Basney}, {Becker}, {Becla}, {Biswas},
  {Bosch}, {Boutigny}, {Kind}, {Ciardi}, {Connolly}, {Daniel}, {Daues},
  {Economou}, {Chiang}, {Fausti}, {Fisher-Levine}, {Freemon}, {Gris},
  {Hernandez}, {Hoblitt}, {Ivezi{\'c}}, {Jammes}, {Jevremovi{\'c}}, {Jones},
  {Kalmbach}, {Kasliwal}, {Krughoff}, {Lurie}, {Lust}, {MacArthur}, {Melchior},
  {Moeyens}, {Nidever}, {Owen}, {Parejko}, {Peterson}, {Petravick},
  {Pietrowicz}, {Price}, {Reiss}, {Shaw}, {Sick}, {Slater}, {Strauss},
  {Sullivan}, {Swinbank}, {Van Dyk}, {Vuj{\v c}i{\'c}}, {Withers}, \&
  {Yoachim}}]{Juric2017}
{Juri{\'c}}, M., {Kantor}, J., {Lim}, K.-T., {et~al.} 2017, in Astronomical
  Society of the Pacific Conference Series, Vol. 512, Astronomical Data
  Analysis Software and Systems XXV, ed. N.~P.~F. {Lorente}, K.~{Shortridge},
  \& R.~{Wayth}, 279.
\newblock \doarXiv{1512.07914}

\bibitem[{Kaiser(1987)}]{Kaiser1987}
Kaiser, N. 1987, Monthly Notices of the Royal Astronomical Society, 227, 1,
  \dodoi{10.1093/mnras/227.1.1}

\bibitem[{{Kashikawa} {et~al.}(2002){Kashikawa}, {Aoki}, {Asai}, {Ebizuka},
  {Inata}, {Iye}, {Kawabata}, {Kosugi}, {Ohyama}, {Okita}, {Ozawa}, {Saito},
  {Sasaki}, {Sekiguchi}, {Shimizu}, {Taguchi}, {Takata}, {Yadoumaru}, \&
  {Yoshida}}]{Kashikawa2002}
{Kashikawa}, N., {Aoki}, K., {Asai}, R., {et~al.} 2002, \pasj, 54, 819,
  \dodoi{10.1093/pasj/54.6.819}

\bibitem[{{Kashikawa} {et~al.}(2015){Kashikawa}, {Ishizaki}, {Willott},
  {Onoue}, {Im}, {Furusawa}, {Toshikawa}, {Ishikawa}, {Niino}, {Shimasaku},
  {Ouchi}, \& {Hibon}}]{Kashikawa2015}
{Kashikawa}, N., {Ishizaki}, Y., {Willott}, C.~J., {et~al.} 2015, \apj, 798,
  28, \dodoi{10.1088/0004-637X/798/1/28}

\bibitem[{{Kashino} {et~al.}(2022){Kashino}, {Lilly}, {Matthee}, {Eilers},
  {Mackenzie}, {Bordoloi}, \& {Simcoe}}]{Kashino2022}
{Kashino}, D., {Lilly}, S.~J., {Matthee}, J., {et~al.} 2022, arXiv e-prints,
  arXiv:2211.08254, \dodoi{10.48550/arXiv.2211.08254}

\bibitem[{Kawanomoto {et~al.}(2018)Kawanomoto, Uraguchi, Komiyama, Miyazaki,
  Furusawa, Finet, Hattori, Wang, Yasuda, \& Suzuki}]{Kawanomoto_2018}
Kawanomoto, S., Uraguchi, F., Komiyama, Y., {et~al.} 2018, Publications of the
  Astronomical Society of Japan, 70, \dodoi{10.1093/pasj/psy056}

\bibitem[{Kim {et~al.}(2015)Kim, Im, Jeon, Kim, Choi, Hong, Hyun, Jun,
  Karouzos, Kim, Kim, Kim, Kim, Lee, Pak, Park, Taak, \& Yoon}]{Kim_2015}
Kim, Y., Im, M., Jeon, Y., {et~al.} 2015, The Astrophysical Journal, 813, L35,
  \dodoi{10.1088/2041-8205/813/2/l35}

\bibitem[{Komiyama {et~al.}(2017)Komiyama, Obuchi, Nakaya, Kamata, Kawanomoto,
  Utsumi, Miyazaki, Uraguchi, Furusawa, Morokuma, Uchida, Miyatake, Mineo,
  Fujimori, Aihara, Karoji, Gunn, \& Wang}]{Komiyama_2018}
Komiyama, Y., Obuchi, Y., Nakaya, H., {et~al.} 2017, Publications of the
  Astronomical Society of Japan, 70, \dodoi{10.1093/pasj/psx069}

\bibitem[{Koptelova \& Hwang(2022)}]{Koptelova2022}
Koptelova, E., \& Hwang, C.-Y. 2022, Dense nitrogen-enriched circumnuclear
  region of the new high-redshift quasar ULAS J0816+2134 at z=7.46,  arXiv,
  \dodoi{10.48550/ARXIV.2212.05862}

\bibitem[{{Kormendy} \& {Ho}(2013)}]{Kormendy2013}
{Kormendy}, J., \& {Ho}, L.~C. 2013, \araa, 51, 511,
  \dodoi{10.1146/annurev-astro-082708-101811}

\bibitem[{{Kormendy} \& {Richstone}(1995)}]{Kormendy1995}
{Kormendy}, J., \& {Richstone}, D. 1995, \araa, 33, 581,
  \dodoi{10.1146/annurev.aa.33.090195.003053}

\bibitem[{{Lacey} \& {Cole}(1993)}]{Lacey1993}
{Lacey}, C., \& {Cole}, S. 1993, \mnras, 262, 627,
  \dodoi{10.1093/mnras/262.3.627}

\bibitem[{Landy \& Szalay(1993)}]{Landy1993BiasAV}
Landy, S.~D., \& Szalay, A.~S. 1993, The Astrophysical Journal, 412, 64

\bibitem[{{Lewis} \& {Challinor}(2011)}]{Lewis2011}
{Lewis}, A., \& {Challinor}, A. 2011, {CAMB: Code for Anisotropies in the
  Microwave Background}, Astrophysics Source Code Library, record
  ascl:1102.026.
\newblock \doeprint{1102.026}

\bibitem[{{Li} {et~al.}(2022){Li}, {Inayoshi}, {Onoue}, \& {Toyouchi}}]{Li2022}
{Li}, W., {Inayoshi}, K., {Onoue}, M., \& {Toyouchi}, D. 2022, arXiv e-prints,
  arXiv:2210.02308, \dodoi{10.48550/arXiv.2210.02308}

\bibitem[{Lidz {et~al.}(2006)Lidz, Hopkins, Cox, Hernquist, \&
  Robertson}]{Lidz_2006}
Lidz, A., Hopkins, P.~F., Cox, T.~J., Hernquist, L., \& Robertson, B. 2006, The
  Astrophysical Journal, 641, 41, \dodoi{10.1086/500444}

\bibitem[{{Limber}(1953)}]{Limber1953}
{Limber}, D.~N. 1953, \apj, 117, 134, \dodoi{10.1086/145672}

\bibitem[{{LSST Science Collaboration} {et~al.}(2009){LSST Science
  Collaboration}, {Abell}, {Allison}, {Anderson}, {Andrew}, {Angel}, {Armus},
  {Arnett}, {Asztalos}, {Axelrod}, {Bailey}, {Ballantyne}, {Bankert},
  {Barkhouse}, {Barr}, {Barrientos}, {Barth}, {Bartlett}, {Becker}, {Becla},
  {Beers}, {Bernstein}, {Biswas}, {Blanton}, {Bloom}, {Bochanski}, {Boeshaar},
  {Borne}, {Bradac}, {Brandt}, {Bridge}, {Brown}, {Brunner}, {Bullock},
  {Burgasser}, {Burge}, {Burke}, {Cargile}, {Chandrasekharan}, {Chartas},
  {Chesley}, {Chu}, {Cinabro}, {Claire}, {Claver}, {Clowe}, {Connolly}, {Cook},
  {Cooke}, {Cooray}, {Covey}, {Culliton}, {de Jong}, {de Vries}, {Debattista},
  {Delgado}, {Dell'Antonio}, {Dhital}, {Di Stefano}, {Dickinson}, {Dilday},
  {Djorgovski}, {Dobler}, {Donalek}, {Dubois-Felsmann}, {Durech},
  {Eliasdottir}, {Eracleous}, {Eyer}, {Falco}, {Fan}, {Fassnacht}, {Ferguson},
  {Fernandez}, {Fields}, {Finkbeiner}, {Figueroa}, {Fox}, {Francke}, {Frank},
  {Frieman}, {Fromenteau}, {Furqan}, {Galaz}, {Gal-Yam}, {Garnavich},
  {Gawiser}, {Geary}, {Gee}, {Gibson}, {Gilmore}, {Grace}, {Green}, {Gressler},
  {Grillmair}, {Habib}, {Haggerty}, {Hamuy}, {Harris}, {Hawley}, {Heavens},
  {Hebb}, {Henry}, {Hileman}, {Hilton}, {Hoadley}, {Holberg}, {Holman},
  {Howell}, {Infante}, {Ivezic}, {Jacoby}, {Jain}, {R}, {Jedicke}, {Jee},
  {Garrett Jernigan}, {Jha}, {Johnston}, {Jones}, {Juric}, {Kaasalainen},
  {Styliani}, {Kafka}, {Kahn}, {Kaib}, {Kalirai}, {Kantor}, {Kasliwal},
  {Keeton}, {Kessler}, {Knezevic}, {Kowalski}, {Krabbendam}, {Krughoff},
  {Kulkarni}, {Kuhlman}, {Lacy}, {Lepine}, {Liang}, {Lien}, {Lira}, {Long},
  {Lorenz}, {Lotz}, {Lupton}, {Lutz}, {Macri}, {Mahabal}, {Mandelbaum},
  {Marshall}, {May}, {McGehee}, {Meadows}, {Meert}, {Milani}, {Miller},
  {Miller}, {Mills}, {Minniti}, {Monet}, {Mukadam}, {Nakar}, {Neill}, {Newman},
  {Nikolaev}, {Nordby}, {O'Connor}, {Oguri}, {Oliver}, {Olivier}, {Olsen},
  {Olsen}, {Olszewski}, {Oluseyi}, {Padilla}, {Parker}, {Pepper}, {Peterson},
  {Petry}, {Pinto}, {Pizagno}, {Popescu}, {Prsa}, {Radcka}, {Raddick},
  {Rasmussen}, {Rau}, {Rho}, {Rhoads}, {Richards}, {Ridgway}, {Robertson},
  {Roskar}, {Saha}, {Sarajedini}, {Scannapieco}, {Schalk}, {Schindler},
  {Schmidt}, {Schmidt}, {Schneider}, {Schumacher}, {Scranton}, {Sebag},
  {Seppala}, {Shemmer}, {Simon}, {Sivertz}, {Smith}, {Allyn Smith}, {Smith},
  {Spitz}, {Stanford}, {Stassun}, {Strader}, {Strauss}, {Stubbs}, {Sweeney},
  {Szalay}, {Szkody}, {Takada}, {Thorman}, {Trilling}, {Trimble}, {Tyson}, {Van
  Berg}, {Vanden Berk}, {VanderPlas}, {Verde}, {Vrsnak}, {Walkowicz},
  {Wandelt}, {Wang}, {Wang}, {Warner}, {Wechsler}, {West}, {Wiecha},
  {Williams}, {Willman}, {Wittman}, {Wolff}, {Wood-Vasey}, {Wozniak}, {Young},
  {Zentner}, \& {Zhan}}]{LSST2009}
{LSST Science Collaboration}, {Abell}, P.~A., {Allison}, J., {et~al.} 2009,
  arXiv e-prints, arXiv:0912.0201, \dodoi{10.48550/arXiv.0912.0201}

\bibitem[{{Lynden-Bell}(1969)}]{Lynden-Bell1969}
{Lynden-Bell}, D. 1969, \nat, 223, 690, \dodoi{10.1038/223690a0}

\bibitem[{{Magnier} {et~al.}(2013){Magnier}, {Schlafly}, {Finkbeiner}, {Juric},
  {Tonry}, {Burgett}, {Chambers}, {Flewelling}, {Kaiser}, {Kudritzki},
  {Morgan}, {Price}, {Sweeney}, \& {Stubbs}}]{Magnier2013}
{Magnier}, E.~A., {Schlafly}, E., {Finkbeiner}, D., {et~al.} 2013, \apjs, 205,
  20, \dodoi{10.1088/0067-0049/205/2/20}

\bibitem[{{Magorrian} {et~al.}(1998){Magorrian}, {Tremaine}, {Richstone},
  {Bender}, {Bower}, {Dressler}, {Faber}, {Gebhardt}, {Green}, {Grillmair},
  {Kormendy}, \& {Lauer}}]{Magorrian1998}
{Magorrian}, J., {Tremaine}, S., {Richstone}, D., {et~al.} 1998, \aj, 115,
  2285, \dodoi{10.1086/300353}

\bibitem[{{Mandelbaum} {et~al.}(2005){Mandelbaum}, {Tasitsiomi}, {Seljak},
  {Kravtsov}, \& {Wechsler}}]{Mandelbaum2005}
{Mandelbaum}, R., {Tasitsiomi}, A., {Seljak}, U., {Kravtsov}, A.~V., \&
  {Wechsler}, R.~H. 2005, \mnras, 362, 1451,
  \dodoi{10.1111/j.1365-2966.2005.09417.x}

\bibitem[{Mar{\'{\i} }n {et~al.}(2015)Mar{\'{\i} }n, Beutler, Blake, Koda,
  Kazin, \& Schneider}]{Marin_2015}
Mar{\'{\i} }n, F.~A., Beutler, F., Blake, C., {et~al.} 2015, Monthly Notices of
  the Royal Astronomical Society, 455, 4046, \dodoi{10.1093/mnras/stv2502}

\bibitem[{{Marshall} {et~al.}(2023){Marshall}, {Perna}, {Willott}, {Maiolino},
  {Scholtz}, {{\"U}bler}, {Carniani}, {Arribas}, {L{\"u}tzgendorf}, {Bunker},
  {Charlot}, {Ferruit}, {Jakobsen}, {Rodr{\'\i}guez Del Pino}, {B{\"o}ker},
  {Cameron}, {Cresci}, {Curtis-Lake}, {Jones}, {Kumari}, \&
  {P{\'e}rez-Gonz{\'a}lez}}]{Marshall2023}
{Marshall}, M.~A., {Perna}, M., {Willott}, C.~J., {et~al.} 2023, arXiv
  e-prints, arXiv:2302.04795, \dodoi{10.48550/arXiv.2302.04795}

\bibitem[{{Martini} \& {Weinberg}(2001)}]{Martini2001}
{Martini}, P., \& {Weinberg}, D.~H. 2001, \apj, 547, 12, \dodoi{10.1086/318331}

\bibitem[{{Matsuoka} {et~al.}(2016){Matsuoka}, {Onoue}, {Kashikawa}, {Iwasawa},
  {Strauss}, {Nagao}, {Imanishi}, {Niida}, {Toba}, {Akiyama}, {Asami}, {Bosch},
  {Foucaud}, {Furusawa}, {Goto}, {Gunn}, {Harikane}, {Ikeda}, {Kawaguchi},
  {Kikuta}, {Komiyama}, {Lupton}, {Minezaki}, {Miyazaki}, {Morokuma},
  {Murayama}, {Nishizawa}, {Ono}, {Ouchi}, {Price}, {Sameshima}, {Silverman},
  {Sugiyama}, {Tait}, {Takada}, {Takata}, {Tanaka}, {Tang}, \&
  {Utsumi}}]{Matsuoka_2016}
{Matsuoka}, Y., {Onoue}, M., {Kashikawa}, N., {et~al.} 2016, \apj, 828, 26,
  \dodoi{10.3847/0004-637X/828/1/26}

\bibitem[{{Matsuoka} {et~al.}(2018{\natexlab{a}}){Matsuoka}, {Iwasawa},
  {Onoue}, {Kashikawa}, {Strauss}, {Lee}, {Imanishi}, {Nagao}, {Akiyama},
  {Asami}, {Bosch}, {Furusawa}, {Goto}, {Gunn}, {Harikane}, {Ikeda}, {Izumi},
  {Kawaguchi}, {Kato}, {Kikuta}, {Kohno}, {Komiyama}, {Lupton}, {Minezaki},
  {Miyazaki}, {Morokuma}, {Murayama}, {Niida}, {Nishizawa}, {Oguri}, {Ono},
  {Ouchi}, {Price}, {Sameshima}, {Schulze}, {Shirakata}, {Silverman},
  {Sugiyama}, {Tait}, {Takada}, {Takata}, {Tanaka}, {Tang}, {Toba}, {Utsumi},
  {Wang}, \& {Yamashita}}]{SHELLQs4}
{Matsuoka}, Y., {Iwasawa}, K., {Onoue}, M., {et~al.} 2018{\natexlab{a}}, \apjs,
  237, 5, \dodoi{10.3847/1538-4365/aac724}

\bibitem[{{Matsuoka} {et~al.}(2018{\natexlab{b}}){Matsuoka}, {Strauss},
  {Kashikawa}, {Onoue}, {Iwasawa}, {Tang}, {Lee}, {Imanishi}, {Nagao},
  {Akiyama}, {Asami}, {Bosch}, {Furusawa}, {Goto}, {Gunn}, {Harikane}, {Ikeda},
  {Izumi}, {Kawaguchi}, {Kato}, {Kikuta}, {Kohno}, {Komiyama}, {Lupton},
  {Minezaki}, {Miyazaki}, {Murayama}, {Niida}, {Nishizawa}, {Noboriguchi},
  {Oguri}, {Ono}, {Ouchi}, {Price}, {Sameshima}, {Schulze}, {Shirakata},
  {Silverman}, {Sugiyama}, {Tait}, {Takada}, {Takata}, {Tanaka}, {Toba},
  {Utsumi}, {Wang}, \& {Yamashita}}]{SHELLQs5}
{Matsuoka}, Y., {Strauss}, M.~A., {Kashikawa}, N., {et~al.} 2018{\natexlab{b}},
  \apj, 869, 150, \dodoi{10.3847/1538-4357/aaee7a}

\bibitem[{{Matsuoka} {et~al.}(2019){Matsuoka}, {Onoue}, {Kashikawa}, {Strauss},
  {Iwasawa}, {Lee}, {Imanishi}, {Nagao}, {Akiyama}, {Asami}, {Bosch},
  {Furusawa}, {Goto}, {Gunn}, {Harikane}, {Ikeda}, {Izumi}, {Kawaguchi},
  {Kato}, {Kikuta}, {Kohno}, {Komiyama}, {Koyama}, {Lupton}, {Minezaki},
  {Miyazaki}, {Murayama}, {Niida}, {Nishizawa}, {Noboriguchi}, {Oguri}, {Ono},
  {Ouchi}, {Price}, {Sameshima}, {Schulze}, {Shirakata}, {Silverman},
  {Sugiyama}, {Tait}, {Takada}, {Takata}, {Tanaka}, {Tang}, {Toba}, {Utsumi},
  {Wang}, \& {Yamashita}}]{Matsuoka2019}
{Matsuoka}, Y., {Onoue}, M., {Kashikawa}, N., {et~al.} 2019, \apjl, 872, L2,
  \dodoi{10.3847/2041-8213/ab0216}

\bibitem[{{Matsuoka} {et~al.}(2022){Matsuoka}, Iwasawa, Onoue, Izumi,
  Kashikawa, Strauss, Imanishi, Nagao, Akiyama, Silverman, Asami, Bosch,
  Furusawa, Goto, Gunn, Harikane, Ikeda, Ishimoto, Kawaguchi, Kato, Kikuta,
  Kohno, Komiyama, Lee, Lupton, Minezaki, Miyazaki, Murayama, Nishizawa, Oguri,
  Ono, Ouchi, Price, Sameshima, Sugiyama, Tait, Takada, Takahashi, Takata,
  Tanaka, Toba, Utsumi, Wang, \& Yamashita}]{Matsuoka_2022}
{Matsuoka}, Y., Iwasawa, K., Onoue, M., {et~al.} 2022, The Astrophysical
  Journal Supplement Series, 259, 18, \dodoi{10.3847/1538-4365/ac3d31}

\bibitem[{{Mazzucchelli} {et~al.}(2017){Mazzucchelli}, {Ba{\~n}ados},
  {Decarli}, {Farina}, {Venemans}, {Walter}, \& {Overzier}}]{Mazzucchelli2017a}
{Mazzucchelli}, C., {Ba{\~n}ados}, E., {Decarli}, R., {et~al.} 2017, \apj, 834,
  83, \dodoi{10.3847/1538-4357/834/1/83}

\bibitem[{Mazzucchelli {et~al.}(2017)Mazzucchelli, Ba{\~{n} }ados, Venemans,
  Decarli, Farina, Walter, Eilers, Rix, Simcoe, Stern, Fan, Schlafly, Rosa,
  Hennawi, Chambers, Greiner, Burgett, Draper, Kaiser, Kudritzki, Magnier,
  Metcalfe, Waters, \& Wainscoat}]{Mazzucchelli_2017}
Mazzucchelli, C., Ba{\~{n} }ados, E., Venemans, B.~P., {et~al.} 2017, The
  Astrophysical Journal, 849, 91, \dodoi{10.3847/1538-4357/aa9185}

\bibitem[{{McGreer} {et~al.}(2016){McGreer}, {Eftekharzadeh}, {Myers}, \&
  {Fan}}]{McGreer2016}
{McGreer}, I.~D., {Eftekharzadeh}, S., {Myers}, A.~D., \& {Fan}, X. 2016, \aj,
  151, 61, \dodoi{10.3847/0004-6256/151/3/61}

\bibitem[{McKinney {et~al.}(2010)}]{mckinney2010data}
McKinney, W., {et~al.} 2010, in Proceedings of the 9th Python in Science
  Conference, Vol. 445, Austin, TX, 51--56

\bibitem[{{Mignoli} {et~al.}(2020){Mignoli}, {Gilli}, {Decarli}, {Vanzella},
  {Balmaverde}, {Cappelluti}, {Cassar{\`a}}, {Comastri}, {Cusano}, {Iwasawa},
  {Marchesi}, {Prandoni}, {Vignali}, {Vito}, {Zamorani}, {Chiaberge}, \&
  {Norman}}]{Mignoli2020}
{Mignoli}, M., {Gilli}, R., {Decarli}, R., {et~al.} 2020, \aap, 642, L1,
  \dodoi{10.1051/0004-6361/202039045}

\bibitem[{{Miyazaki} {et~al.}(2018){Miyazaki}, {Komiyama}, {Kawanomoto}, {Doi},
  {Furusawa}, {Hamana}, {Hayashi}, {Ikeda}, {Kamata}, {Karoji}, {Koike},
  {Kurakami}, {Miyama}, {Morokuma}, {Nakata}, {Namikawa}, {Nakaya}, {Nariai},
  {Obuchi}, {Oishi}, {Okada}, {Okura}, {Tait}, {Takata}, {Tanaka}, {Tanaka},
  {Terai}, {Tomono}, {Uraguchi}, {Usuda}, {Utsumi}, {Yamada}, {Yamanoi},
  {Aihara}, {Fujimori}, {Mineo}, {Miyatake}, {Oguri}, {Uchida}, {Tanaka},
  {Yasuda}, {Takada}, {Murayama}, {Nishizawa}, {Sugiyama}, {Chiba}, {Futamase},
  {Wang}, {Chen}, {Ho}, {Liaw}, {Chiu}, {Ho}, {Lai}, {Lee}, {Jeng}, {Iwamura},
  {Armstrong}, {Bickerton}, {Bosch}, {Gunn}, {Lupton}, {Loomis}, {Price},
  {Smith}, {Strauss}, {Turner}, {Suzuki}, {Miyazaki}, {Muramatsu}, {Yamamoto},
  {Endo}, {Ezaki}, {Ito}, {Kawaguchi}, {Sofuku}, {Taniike}, {Akutsu}, {Dojo},
  {Kasumi}, {Matsuda}, {Imoto}, {Miwa}, {Suzuki}, {Takeshi}, \&
  {Yokota}}]{Miyazaki_2018}
{Miyazaki}, S., {Komiyama}, Y., {Kawanomoto}, S., {et~al.} 2018, \pasj, 70, S1,
  \dodoi{10.1093/pasj/psx063}

\bibitem[{{Morselli} {et~al.}(2014){Morselli}, {Mignoli}, {Gilli}, {Vignali},
  {Comastri}, {Sani}, {Cappelluti}, {Zamorani}, {Brusa}, {Gallozzi}, \&
  {Vanzella}}]{Morselli2014}
{Morselli}, L., {Mignoli}, M., {Gilli}, R., {et~al.} 2014, \aap, 568, A1,
  \dodoi{10.1051/0004-6361/201423853}

\bibitem[{{Mortlock} {et~al.}(2011){Mortlock}, {Warren}, {Venemans}, {Patel},
  {Hewett}, {McMahon}, {Simpson}, {Theuns}, {Gonz{\'a}les-Solares}, {Adamson},
  {Dye}, {Hambly}, {Hirst}, {Irwin}, {Kuiper}, {Lawrence}, \&
  {R{\"o}ttgering}}]{Mortlock2011}
{Mortlock}, D.~J., {Warren}, S.~J., {Venemans}, B.~P., {et~al.} 2011, \nat,
  474, 616, \dodoi{10.1038/nature10159}

\bibitem[{{Mountrichas} {et~al.}(2009){Mountrichas}, {Sawangwit}, {Shanks},
  {Croom}, {Schneider}, {Myers}, \& {Pimbblet}}]{Mountrichas2009}
{Mountrichas}, G., {Sawangwit}, U., {Shanks}, T., {et~al.} 2009, \mnras, 394,
  2050, \dodoi{10.1111/j.1365-2966.2009.14456.x}

\bibitem[{Murray {et~al.}(2021)Murray, Diemer, Chen, Neuhold, Schnapp, Peruzzi,
  Blevins, \& Engelman}]{Murray_2021}
Murray, S., Diemer, B., Chen, Z., {et~al.} 2021, Astronomy and Computing, 36,
  100487, \dodoi{10.1016/j.ascom.2021.100487}

\bibitem[{Murray {et~al.}(2013)Murray, Power, \& Robotham}]{Murray2013}
Murray, S., Power, C., \& Robotham, A. 2013, HMFcalc: An Online Tool for
  Calculating Dark Matter Halo Mass Functions,  arXiv,
  \dodoi{10.48550/ARXIV.1306.6721}

\bibitem[{{Myers} {et~al.}(2007){Myers}, {Brunner}, {Richards}, {Nichol},
  {Schneider}, \& {Bahcall}}]{Myers2007}
{Myers}, A.~D., {Brunner}, R.~J., {Richards}, G.~T., {et~al.} 2007, \apj, 658,
  99, \dodoi{10.1086/511520}

\bibitem[{Myers {et~al.}(2006)Myers, Brunner, Richards, Nichol, Schneider,
  Berk, Scranton, Gray, \& Brinkmann}]{Myers_2006}
Myers, A.~D., Brunner, R.~J., Richards, G.~T., {et~al.} 2006, The Astrophysical
  Journal, 638, 622, \dodoi{10.1086/499093}

\bibitem[{{Neeleman} {et~al.}(2021){Neeleman}, {Novak}, {Venemans}, {Walter},
  {Decarli}, {Kaasinen}, {Schindler}, {Ba{\~n}ados}, {Carilli}, {Drake}, {Fan},
  \& {Rix}}]{Neeleman2021}
{Neeleman}, M., {Novak}, M., {Venemans}, B.~P., {et~al.} 2021, \apj, 911, 141,
  \dodoi{10.3847/1538-4357/abe70f}

\bibitem[{{Oke} \& {Gunn}(1983)}]{Oke1983}
{Oke}, J.~B., \& {Gunn}, J.~E. 1983, \apj, 266, 713, \dodoi{10.1086/160817}

\bibitem[{{Ono} {et~al.}(2018){Ono}, {Ouchi}, {Harikane}, {Toshikawa}, {Rauch},
  {Yuma}, {Sawicki}, {Shibuya}, {Shimasaku}, {Oguri}, {Willott}, {Akhlaghi},
  {Akiyama}, {Coupon}, {Kashikawa}, {Komiyama}, {Konno}, {Lin}, {Matsuoka},
  {Miyazaki}, {Nagao}, {Nakajima}, {Silverman}, {Tanaka}, {Taniguchi}, \&
  {Wang}}]{Ono2018}
{Ono}, Y., {Ouchi}, M., {Harikane}, Y., {et~al.} 2018, \pasj, 70, S10,
  \dodoi{10.1093/pasj/psx103}

\bibitem[{{Onoue} {et~al.}(2021){Onoue}, {Matsuoka}, {Kashikawa}, {Strauss},
  {Iwasawa}, {Izumi}, {Nagao}, {Asami}, {Fujimoto}, {Harikane}, {Hashimoto},
  {Imanishi}, {Lee}, {Shibuya}, \& {Toba}}]{Onoue2021}
{Onoue}, M., {Matsuoka}, Y., {Kashikawa}, N., {et~al.} 2021, \apj, 919, 61,
  \dodoi{10.3847/1538-4357/ac0f07}

\bibitem[{{Peebles}(1980)}]{Peebles1980}
{Peebles}, P.~J.~E. 1980, {The large-scale structure of the universe}

\bibitem[{{Porciani} {et~al.}(2004){Porciani}, {Magliocchetti}, \&
  {Norberg}}]{Porciani2004}
{Porciani}, C., {Magliocchetti}, M., \& {Norberg}, P. 2004, \mnras, 355, 1010,
  \dodoi{10.1111/j.1365-2966.2004.08408.x}

\bibitem[{{Salpeter}(1964)}]{Salpeter1964}
{Salpeter}, E.~E. 1964, \apj, 140, 796, \dodoi{10.1086/147973}

\bibitem[{{Schlafly} {et~al.}(2012){Schlafly}, {Finkbeiner}, {Juri{\'c}},
  {Magnier}, {Burgett}, {Chambers}, {Grav}, {Hodapp}, {Kaiser}, {Kudritzki},
  {Martin}, {Morgan}, {Price}, {Rix}, {Stubbs}, {Tonry}, \&
  {Wainscoat}}]{Schlafly2012}
{Schlafly}, E.~F., {Finkbeiner}, D.~P., {Juri{\'c}}, M., {et~al.} 2012, \apj,
  756, 158, \dodoi{10.1088/0004-637X/756/2/158}

\bibitem[{{Schneider} {et~al.}(2007){Schneider}, {Hall}, {Richards}, {Strauss},
  {Vanden Berk}, {Anderson}, {Brandt}, {Fan}, {Jester}, {Gray}, {Gunn},
  {SubbaRao}, {Thakar}, {Stoughton}, {Szalay}, {Yanny}, {York}, {Bahcall},
  {Barentine}, {Blanton}, {Brewington}, {Brinkmann}, {Brunner}, {Castander},
  {Csabai}, {Frieman}, {Fukugita}, {Harvanek}, {Hogg}, {Ivezi{\'c}}, {Kent},
  {Kleinman}, {Knapp}, {Kron}, {Krzesi{\'n}ski}, {Long}, {Lupton}, {Nitta},
  {Pier}, {Saxe}, {Shen}, {Snedden}, {Weinberg}, \& {Wu}}]{Schneider2007}
{Schneider}, D.~P., {Hall}, P.~B., {Richards}, G.~T., {et~al.} 2007, \aj, 134,
  102, \dodoi{10.1086/518474}

\bibitem[{{Seljak} \& {Warren}(2004)}]{Seljak2004}
{Seljak}, U., \& {Warren}, M.~S. 2004, \mnras, 355, 129,
  \dodoi{10.1111/j.1365-2966.2004.08297.x}

\bibitem[{{Serjeant} {et~al.}(2000){Serjeant}, {Oliver}, {Rowan-Robinson},
  {Crockett}, {Missoulis}, {Sumner}, {Gruppioni}, {Mann}, {Eaton}, {Elbaz},
  {Clements}, {Baker}, {Efstathiou}, {Cesarsky}, {Danese}, {Franceschini},
  {Genzel}, {Lawrence}, {Lemke}, {McMahon}, {Miley}, {Puget}, \&
  {Rocca-Volmerange}}]{Serjeant2000}
{Serjeant}, S., {Oliver}, S., {Rowan-Robinson}, M., {et~al.} 2000, \mnras, 316,
  768, \dodoi{10.1046/j.1365-8711.2000.03551.x}

\bibitem[{{Shankar} {et~al.}(2006){Shankar}, {Lapi}, {Salucci}, {De Zotti}, \&
  {Danese}}]{Shankar2006}
{Shankar}, F., {Lapi}, A., {Salucci}, P., {De Zotti}, G., \& {Danese}, L. 2006,
  \apj, 643, 14, \dodoi{10.1086/502794}

\bibitem[{{Shen} {et~al.}(2008){Shen}, {Greene}, {Strauss}, {Richards}, \&
  {Schneider}}]{Shen2008}
{Shen}, Y., {Greene}, J.~E., {Strauss}, M.~A., {Richards}, G.~T., \&
  {Schneider}, D.~P. 2008, \apj, 680, 169, \dodoi{10.1086/587475}

\bibitem[{{Shen} {et~al.}(2007){Shen}, {Strauss}, {Oguri}, {Hennawi}, {Fan},
  {Richards}, {Hall}, {Gunn}, {Schneider}, {Szalay}, {Thakar}, {Vanden Berk},
  {Anderson}, {Bahcall}, {Connolly}, \& {Knapp}}]{Shen2007}
{Shen}, Y., {Strauss}, M.~A., {Oguri}, M., {et~al.} 2007, \aj, 133, 2222,
  \dodoi{10.1086/513517}

\bibitem[{{Shen} {et~al.}(2009){Shen}, {Strauss}, {Ross}, {Hall}, {Lin},
  {Richards}, {Schneider}, {Weinberg}, {Connolly}, {Fan}, {Hennawi}, {Shankar},
  {Vanden Berk}, {Bahcall}, \& {Brunner}}]{Shen2009}
{Shen}, Y., {Strauss}, M.~A., {Ross}, N.~P., {et~al.} 2009, \apj, 697, 1656,
  \dodoi{10.1088/0004-637X/697/2/1656}

\bibitem[{{Shen} {et~al.}(2013){Shen}, {McBride}, {White}, {Zheng}, {Myers},
  {Guo}, {Kirkpatrick}, {Padmanabhan}, {Parejko}, {Ross}, {Schlegel},
  {Schneider}, {Streblyanska}, {Swanson}, {Zehavi}, {Pan}, {Bizyaev},
  {Brewington}, {Ebelke}, {Malanushenko}, {Malanushenko}, {Oravetz}, {Simmons},
  \& {Snedden}}]{Shen2013}
{Shen}, Y., {McBride}, C.~K., {White}, M., {et~al.} 2013, \apj, 778, 98,
  \dodoi{10.1088/0004-637X/778/2/98}

\bibitem[{{Sheth} {et~al.}(2001){Sheth}, {Mo}, \& {Tormen}}]{SMT2001}
{Sheth}, R.~K., {Mo}, H.~J., \& {Tormen}, G. 2001, \mnras, 323, 1,
  \dodoi{10.1046/j.1365-8711.2001.04006.x}

\bibitem[{Sheth \& Tormen(1999)}]{Sheth_1999}
Sheth, R.~K., \& Tormen, G. 1999, Monthly Notices of the Royal Astronomical
  Society, 308, 119, \dodoi{10.1046/j.1365-8711.1999.02692.x}

\bibitem[{Shimasaku \& Izumi(2019)}]{Shimasaku_2019}
Shimasaku, K., \& Izumi, T. 2019, The Astrophysical Journal, 872, L29,
  \dodoi{10.3847/2041-8213/ab053f}

\bibitem[{{Sinha} \& {Garrison}(2019)}]{10.1007/978-981-13-7729-7_1}
{Sinha}, M., \& {Garrison}, L.~H. 2019, in Software Challenges to Exascale
  Computing, ed. A.~Majumdar \& R.~Arora (Singapore: Springer Singapore),
  3--20.
\newblock \url{https://doi.org/10.1007/978-981-13-7729-7_1}

\bibitem[{{Sinha} \& {Garrison}(2020)}]{2020MNRAS.491.3022S}
{Sinha}, M., \& {Garrison}, L.~H. 2020, \mnras, 491, 3022,
  \dodoi{10.1093/mnras/stz3157}

\bibitem[{{Smee} {et~al.}(2013){Smee}, {Gunn}, {Uomoto}, {Roe}, {Schlegel},
  {Rockosi}, {Carr}, {Leger}, {Dawson}, {Olmstead}, {Brinkmann}, {Owen},
  {Barkhouser}, {Honscheid}, {Harding}, {Long}, {Lupton}, {Loomis}, {Anderson},
  {Annis}, {Bernardi}, {Bhardwaj}, {Bizyaev}, {Bolton}, {Brewington}, {Briggs},
  {Burles}, {Burns}, {Castander}, {Connolly}, {Davenport}, {Ebelke}, {Epps},
  {Feldman}, {Friedman}, {Frieman}, {Heckman}, {Hull}, {Knapp}, {Lawrence},
  {Loveday}, {Mannery}, {Malanushenko}, {Malanushenko}, {Merrelli}, {Muna},
  {Newman}, {Nichol}, {Oravetz}, {Pan}, {Pope}, {Ricketts}, {Shelden},
  {Sandford}, {Siegmund}, {Simmons}, {Smith}, {Snedden}, {Schneider},
  {SubbaRao}, {Tremonti}, {Waddell}, \& {York}}]{Smee2013}
{Smee}, S.~A., {Gunn}, J.~E., {Uomoto}, A., {et~al.} 2013, \aj, 146, 32,
  \dodoi{10.1088/0004-6256/146/2/32}

\bibitem[{{Springel} {et~al.}(2005){Springel}, {White}, {Jenkins}, {Frenk},
  {Yoshida}, {Gao}, {Navarro}, {Thacker}, {Croton}, {Helly}, {Peacock}, {Cole},
  {Thomas}, {Couchman}, {Evrard}, {Colberg}, \& {Pearce}}]{Springel2005b}
{Springel}, V., {White}, S. D.~M., {Jenkins}, A., {et~al.} 2005, \nat, 435,
  629, \dodoi{10.1038/nature03597}

\bibitem[{{Stiavelli} {et~al.}(2005){Stiavelli}, {Djorgovski}, {Pavlovsky},
  {Scarlata}, {Stern}, {Mahabal}, {Thompson}, {Dickinson}, {Panagia}, \&
  {Meylan}}]{Stiavelli2005}
{Stiavelli}, M., {Djorgovski}, S.~G., {Pavlovsky}, C., {et~al.} 2005, \apjl,
  622, L1, \dodoi{10.1086/429406}

\bibitem[{{Suchyta} {et~al.}(2016){Suchyta}, {Huff}, {Aleksi{\'c}}, {Melchior},
  {Jouvel}, {MacCrann}, {Ross}, {Crocce}, {Gaztanaga}, {Honscheid}, {Leistedt},
  {Peiris}, {Rykoff}, {Sheldon}, {Abbott}, {Abdalla}, {Allam}, {Banerji},
  {Benoit-L{\'e}vy}, {Bertin}, {Brooks}, {Burke}, {Carnero Rosell}, {Carrasco
  Kind}, {Carretero}, {Cunha}, {D'Andrea}, {da Costa}, {DePoy}, {Desai},
  {Diehl}, {Dietrich}, {Doel}, {Eifler}, {Estrada}, {Evrard}, {Flaugher},
  {Fosalba}, {Frieman}, {Gerdes}, {Gruen}, {Gruendl}, {James}, {Jarvis},
  {Kuehn}, {Kuropatkin}, {Lahav}, {Lima}, {Maia}, {March}, {Marshall},
  {Miller}, {Miquel}, {Neilsen}, {Nichol}, {Nord}, {Ogando}, {Percival},
  {Reil}, {Roodman}, {Sako}, {Sanchez}, {Scarpine}, {Sevilla-Noarbe}, {Smith},
  {Soares-Santos}, {Sobreira}, {Swanson}, {Tarle}, {Thaler}, {Thomas},
  {Vikram}, {Walker}, {Wechsler}, {Zhang}, \& {DES
  Collaboration}}]{Suchyta2016}
{Suchyta}, E., {Huff}, E.~M., {Aleksi{\'c}}, J., {et~al.} 2016, \mnras, 457,
  786, \dodoi{10.1093/mnras/stv2953}

\bibitem[{Timlin {et~al.}(2018)Timlin, Ross, Richards, Myers, Pellegrino,
  Bauer, Lacy, Schneider, Wollack, \& Zakamska}]{Timlin_2018}
Timlin, J.~D., Ross, N.~P., Richards, G.~T., {et~al.} 2018, The Astrophysical
  Journal, 859, 20, \dodoi{10.3847/1538-4357/aab9ac}

\bibitem[{Tinker {et~al.}(2010)Tinker, Robertson, Kravtsov, Klypin, Warren,
  Yepes, \& Gottlöber}]{Tinker_2010}
Tinker, J.~L., Robertson, B.~E., Kravtsov, A.~V., {et~al.} 2010, The
  Astrophysical Journal, 724, 878, \dodoi{10.1088/0004-637x/724/2/878}

\bibitem[{{Tonry} {et~al.}(2012){Tonry}, {Stubbs}, {Lykke}, {Doherty},
  {Shivvers}, {Burgett}, {Chambers}, {Hodapp}, {Kaiser}, {Kudritzki},
  {Magnier}, {Morgan}, {Price}, \& {Wainscoat}}]{Tonry2012}
{Tonry}, J.~L., {Stubbs}, C.~W., {Lykke}, K.~R., {et~al.} 2012, \apj, 750, 99,
  \dodoi{10.1088/0004-637X/750/2/99}

\bibitem[{{Totsuji} \& {Kihara}(1969)}]{Totsuji1969}
{Totsuji}, H., \& {Kihara}, T. 1969, \pasj, 21, 221

\bibitem[{{Trainor} \& {Steidel}(2012)}]{Trainor2012}
{Trainor}, R.~F., \& {Steidel}, C.~C. 2012, \apj, 752, 39,
  \dodoi{10.1088/0004-637X/752/1/39}

\bibitem[{{Tremaine} {et~al.}(2002){Tremaine}, {Gebhardt}, {Bender}, {Bower},
  {Dressler}, {Faber}, {Filippenko}, {Green}, {Grillmair}, {Ho}, {Kormendy},
  {Lauer}, {Magorrian}, {Pinkney}, \& {Richstone}}]{Tremaine2002}
{Tremaine}, S., {Gebhardt}, K., {Bender}, R., {et~al.} 2002, \apj, 574, 740,
  \dodoi{10.1086/341002}

\bibitem[{{Uchiyama} {et~al.}(2018){Uchiyama}, {Toshikawa}, {Kashikawa},
  {Overzier}, {Chiang}, {Marinello}, {Tanaka}, {Niino}, {Ishikawa}, {Onoue},
  {Ichikawa}, {Akiyama}, {Coupon}, {Harikane}, {Imanishi}, {Kodama},
  {Komiyama}, {Lee}, {Lin}, {Miyazaki}, {Nagao}, {Nishizawa}, {Ono}, {Ouchi},
  \& {Wang}}]{Uchiyama2018}
{Uchiyama}, H., {Toshikawa}, J., {Kashikawa}, N., {et~al.} 2018, \pasj, 70,
  S32, \dodoi{10.1093/pasj/psx112}

\bibitem[{{Venemans} {et~al.}(2016){Venemans}, {Walter}, {Zschaechner},
  {Decarli}, {De Rosa}, {Findlay}, {McMahon}, \& {Sutherland}}]{Venemans2016}
{Venemans}, B.~P., {Walter}, F., {Zschaechner}, L., {et~al.} 2016, \apj, 816,
  37, \dodoi{10.3847/0004-637X/816/1/37}

\bibitem[{{Venemans} {et~al.}(2015){Venemans}, {Verdoes Kleijn}, {Mwebaze},
  {Valentijn}, {Ba{\~n}ados}, {Decarli}, {de Jong}, {Findlay}, {Kuijken}, {La
  Barbera}, {McFarland}, {McMahon}, {Napolitano}, {Sikkema}, \&
  {Sutherland}}]{Venemans2015}
{Venemans}, B.~P., {Verdoes Kleijn}, G.~A., {Mwebaze}, J., {et~al.} 2015,
  \mnras, 453, 2259, \dodoi{10.1093/mnras/stv1774}

\bibitem[{Virtanen {et~al.}(2020)Virtanen, Gommers, Oliphant, Haberland, Reddy,
  Cournapeau, Burovski, Peterson, Weckesser, Bright, {van der Walt}, Brett,
  Wilson, Millman, Mayorov, Nelson, Jones, Kern, Larson, Carey, Polat, Feng,
  Moore, {VanderPlas}, Laxalde, Perktold, Cimrman, Henriksen, Quintero, Harris,
  Archibald, Ribeiro, Pedregosa, {van Mulbregt}, \& {SciPy 1.0
  Contributors}}]{2020SciPy-NMeth}
Virtanen, P., Gommers, R., Oliphant, T.~E., {et~al.} 2020, Nature Methods, 17,
  261, \dodoi{10.1038/s41592-019-0686-2}

\bibitem[{Wang {et~al.}(2023)Wang, Yang, Hennawi, Fan, Sun, Champagne, Costa,
  Habouzit, Endsley, Li, Lin, Meyer, Schindler, Wu, Bañados, Barth, Bhowmick,
  Bieri, Blecha, Bosman, Cai, Colina, Connor, Davies, Decarli, Rosa, Drake,
  Egami, Eilers, Evans, Farina, Haiman, Jiang, Jin, Jun, Kakiichi, Khusanova,
  Kulkarni, Li, Liu, Loiacono, Lupi, Mazzucchelli, Onoue, Pudoka, Rojas-Ruiz,
  Shen, Strauss, Tee, Trakhtenbrot, Trebitsch, Venemans, Volonteri, Walter,
  Xie, Yue, Zhang, Zhang, \& Zou}]{Wang2023}
Wang, F., Yang, J., Hennawi, J.~F., {et~al.} 2023, A SPectroscopic survey of
  biased halos In the Reionization Era (ASPIRE): JWST Reveals a Filamentary
  Structure around a z=6.61 Quasar.
\newblock \doarXiv{2304.09894}

\bibitem[{{Wang} {et~al.}(2013){Wang}, {Wagg}, {Carilli}, {Walter}, {Lentati},
  {Fan}, {Riechers}, {Bertoldi}, {Narayanan}, {Strauss}, {Cox}, {Omont},
  {Menten}, {Knudsen}, {Neri}, \& {Jiang}}]{Wang2013}
{Wang}, R., {Wagg}, J., {Carilli}, C.~L., {et~al.} 2013, \apj, 773, 44,
  \dodoi{10.1088/0004-637X/773/1/44}

\bibitem[{{White} {et~al.}(2012){White}, {Myers}, {Ross}, {Schlegel},
  {Hennawi}, {Shen}, {McGreer}, {Strauss}, {Bolton}, {Bovy}, {Fan},
  {Miralda-Escude}, {Palanque-Delabrouille}, {Paris}, {Petitjean}, {Schneider},
  {Viel}, {Weinberg}, {Yeche}, {Zehavi}, {Pan}, {Snedden}, {Bizyaev},
  {Brewington}, {Brinkmann}, {Malanushenko}, {Malanushenko}, {Oravetz},
  {Simmons}, {Sheldon}, \& {Weaver}}]{White2012}
{White}, M., {Myers}, A.~D., {Ross}, N.~P., {et~al.} 2012, \mnras, 424, 933,
  \dodoi{10.1111/j.1365-2966.2012.21251.x}

\bibitem[{{White} \& {Rees}(1978)}]{White1978}
{White}, S.~D.~M., \& {Rees}, M.~J. 1978, \mnras, 183, 341,
  \dodoi{10.1093/mnras/183.3.341}

\bibitem[{{Willott} {et~al.}(2015{\natexlab{a}}){Willott}, {Bergeron}, \&
  {Omont}}]{Willott2015_surrogate}
{Willott}, C.~J., {Bergeron}, J., \& {Omont}, A. 2015{\natexlab{a}}, \apj, 801,
  123, \dodoi{10.1088/0004-637X/801/2/123}

\bibitem[{{Willott} {et~al.}(2017){Willott}, {Bergeron}, \&
  {Omont}}]{Willott2017}
---. 2017, \apj, 850, 108, \dodoi{10.3847/1538-4357/aa921b}

\bibitem[{{Willott} {et~al.}(2015{\natexlab{b}}){Willott}, {Carilli}, {Wagg},
  \& {Wang}}]{Willott2015}
{Willott}, C.~J., {Carilli}, C.~L., {Wagg}, J., \& {Wang}, R.
  2015{\natexlab{b}}, \apj, 807, 180, \dodoi{10.1088/0004-637X/807/2/180}

\bibitem[{{Willott} {et~al.}(2005){Willott}, {Percival}, {McLure}, {Crampton},
  {Hutchings}, {Jarvis}, {Sawicki}, \& {Simard}}]{Willott2005}
{Willott}, C.~J., {Percival}, W.~J., {McLure}, R.~J., {et~al.} 2005, \apj, 626,
  657, \dodoi{10.1086/430168}

\bibitem[{Willott {et~al.}(2009)Willott, Delorme, Reyl{\'{e}}, Albert,
  Bergeron, Crampton, Delfosse, Forveille, Hutchings, McLure, Omont, \&
  Schade}]{Willott_2009}
Willott, C.~J., Delorme, P., Reyl{\'{e}}, C., {et~al.} 2009, The Astronomical
  Journal, 137, 3541, \dodoi{10.1088/0004-6256/137/3/3541}

\bibitem[{Willott {et~al.}(2010)Willott, Albert, Arzoumanian, Bergeron,
  Crampton, Delorme, Hutchings, Omont, Reyl{\'{e}}, \& Schade}]{Willott_2010}
Willott, C.~J., Albert, L., Arzoumanian, D., {et~al.} 2010, The Astronomical
  Journal, 140, 546, \dodoi{10.1088/0004-6256/140/2/546}

\bibitem[{{Woods} \& {Fahlman}(1997)}]{Woods1997}
{Woods}, D., \& {Fahlman}, G.~G. 1997, \apj, 490, 11, \dodoi{10.1086/304843}

\bibitem[{{Yang} {et~al.}(2020){Yang}, {Wang}, {Fan}, {Hennawi}, {Davies},
  {Yue}, {Banados}, {Wu}, {Venemans}, {Barth}, {Bian}, {Boutsia}, {Decarli},
  {Farina}, {Green}, {Jiang}, {Li}, {Mazzucchelli}, \& {Walter}}]{Yang2020}
{Yang}, J., {Wang}, F., {Fan}, X., {et~al.} 2020, \apjl, 897, L14,
  \dodoi{10.3847/2041-8213/ab9c26}

\bibitem[{{Yang} {et~al.}(2021){Yang}, {Wang}, {Fan}, {Barth}, {Hennawi},
  {Nanni}, {Bian}, {Davies}, {Farina}, {Schindler}, {Ba{\~n}ados}, {Decarli},
  {Eilers}, {Green}, {Guo}, {Jiang}, {Li}, {Venemans}, {Walter}, {Wu}, \&
  {Yue}}]{Yang2021}
---. 2021, \apj, 923, 262, \dodoi{10.3847/1538-4357/ac2b32}

\bibitem[{Zehavi {et~al.}(2005)Zehavi, Zheng, Weinberg, Frieman, Berlind,
  Blanton, Scoccimarro, Sheth, Strauss, Kayo, Suto, Fukugita, Nakamura,
  Bahcall, Brinkmann, Gunn, Hennessy, Ivezi{\'{c}}, Knapp, Loveday, Meiksin,
  Schlegel, Schneider, Szapudi, Tegmark, Vogeley, \& and}]{Zehavi_2005}
Zehavi, I., Zheng, Z., Weinberg, D.~H., {et~al.} 2005, The Astrophysical
  Journal, 630, 1, \dodoi{10.1086/431891}

\bibitem[{Zeimann {et~al.}(2011)Zeimann, White, Becker, Hodge, Stanford, \&
  Richards}]{Zeimann_2011}
Zeimann, G.~R., White, R.~L., Becker, R.~H., {et~al.} 2011, The Astrophysical
  Journal, 736, 57, \dodoi{10.1088/0004-637x/736/1/57}

\end{thebibliography}
\bibliographystyle{aasjournal}

\end{document}